\documentclass[11pt,twoside]{article} 
\usepackage{times,fancyhdr}
\usepackage{graphicx}
\date{}

\title{High-Energy Astrophysics with Neutrino Telescopes} 
\author{T. Chiarusi and M. Spurio \\ 
Dipartimento di Fisica dell'Universit\`a di Bologna and INFN \\
Viale Berti Pichat 6/2 -– 40127 Bologna (Italy)}
\begin{document} 

\pagestyle{fancy}
\fancyhead{} 
\fancyhead[EC]{T. Chiarusi and M. Spurio}
\fancyhead[EL,OR]{\thepage}
\fancyhead[OC]{High-Energy Astrophysics with Neutrino Telescopes}
\fancyfoot{} 
\renewcommand\headrulewidth{0.5pt}
\addtolength{\headheight}{2pt} 

\maketitle 

\baselineskip=11.6pt

\begin{abstract}
Neutrino astrophysics offers new perspectives on the Universe investigation:
high energy neutrinos, produced by the most energetic phenomena in our Galaxy and in the Universe,  carry complementary (if not exclusive) information about the cosmos with respect to photons.
While the small interaction cross section of neutrinos allows them to come from the core of astrophysical objects, it is also a drawback, as their detection requires a large target mass. This is why it is convenient put huge cosmic neutrino detectors in natural locations, like deep underwater or under-ice sites. In order to supply for such  extremely hostile environmental conditions, new frontiers technologies are under development. 
The aim of this work is to review the motivations for high energy 
neutrino astrophysics, the present status of experimental results and the technologies used in underwater/ice {C}herenkov experiments, with a special focus on the efforts for the construction of a km$^3$ scale detector in the Mediterranean Sea.
\end{abstract}
\newpage

\section{Introduction}
\textit{Fiat lux.} It was written, and scientists never fail to observe new spectacular astrophysical discoveries when new experimental  techniques on new photons  wavelengths are available: from the Cosmic Microwave Background observation up to the TeV $\gamma$-ray astronomy using Imaging  Air- Cherenkov Technique. 

\textit{Fiat neutrinos}, it was never written, and Mr. Pauli itself has feared that this particle would never be discovered. 
Nevertheless,  observation of the solar neutrinos and of neutrinos from the supernova 1987A opened up a new observation field. High energy  neutrino astronomy  is  a young  discipline  derived from  the fundamental necessity  of extending conventional  astronomy beyond the usual electro-magnetic messengers.

One of  the main questions in  astroparticle physics is  the origin and
nature of  high-energy cosmic rays, CRs  (\S 2). It  was discovered at
the  beginning of the  last century  that energetic  charged particles
strike the  Earth and produce  showers of secondary particles  in the
atmosphere.   While the  energy spectrum  of  the cosmic  rays can  be
measured up to very high energies, their origin remains unclear. There
are many indications  of the galactic origin of the  CR bulk (protons
and other nuclei up to  $\sim 10^{15}\div 10^{16}$ eV), although it is
not  possible to  directly correlate  the CR  impinging  directions on
Earth to  astrophysical sources since  CRs are generally  deflected by
the  galactic magnetic  fields. 

Recent advances  on ground-based  $\gamma$-ray astronomy have  led to
the discovery of more than  80 sources of TeV gamma-rays, as described
in  \S \ref{gamma}.  
No definitive proof still exists that galactic CR originate from supernova explosions. Compelling evidences have been accumulated by TeV $\gamma$-ray telescopes on the possible CR acceleration by some peculiar galactic objects (\S \ref{nuastro}). 
Assuming that  at acceleration sites  a fraction of  the high-energy
CRs interact with the ambient matter or photon fields, TeV $\gamma$-rays are produced by the $\pi^0$ decay while neutrinos are produced by charged  pion decay. 
This is the so-called \textit{astrophysical hadronic model}, \S \ref{gammahadron}, which describes the mechanisms  which lay  behind the  production of neutrinos  and  high energy  photons  from  CR  interactions. 
In this framework, the  energy spectrum of secondary particles follows the same power law of the  progenitor CRs and it is possible to put constraints to the expected neutrino flux from sources where $\gamma$-rays are detected.
However, almost all observed objects emitting in the TeV $\gamma$-ray band are also sources of non-thermal X-rays, presumably of synchrotron origin, radiated by multi-TeV electrons. Since the same electrons can also radiate TeV $\gamma$-rays through inverse Compton scattering, this \textit{leptonic model} represents a competitive process for TeV radiation. Only the coincident measurement of neutrinos from the source would give a uncontroversial proof of the discovery of the galactic CR acceleration sites. 

The highest  energy  CRs are probably originated from extragalactic sources,  as indicated by recent measurements  (\S \ref{ankle}).  
Protons with $E> 10^{19}$ eV interact with  the cosmic  microwave  background.  This  effect,  known as  the Greisen-Zatsepin-Kuz\-min  cutoff,  restricts the  origin of  high  energy protons seen on Earth to a small fraction of the Universe, of the order of 100 Mpc. 
The prediction of high energy neutrino sources of extra-galactic origin is a direct consequence  of the UHE CR observations, \S \ref{extragalactic}. This connection between CRs, neutrinos and $\gamma$-rays can also be used (\S \ref{wb}) to put upper bounds on the expected neutrino flux from extragalactic sources, since the neutrino energy generation rate will never exceed the generation rate of high energy protons. These predictions impose that the scale of the neutrino detectors will be of the order of 1 km$^3$.

A  high-energy neutrino detector behaves as a \textit{telescope} when the neutrino direction is reconstructed with an angular precision of $1^o$ or better. This is the case for high energy charged current $\nu_\mu$ interactions.
 The accurate measurement of the $\nu_\mu$ direction (which can reach $0.2^o$ in water  for $\nu_\mu$ energies larger than few tens of TeV, \S \ref{waterice}) allows the association with (known) sources. It allows also the neutrino telescopes to face some of the most fundamental questions on HE physics beyond the standard model, \S \ref{particleph}: the nature of Dark Matter through the indirect search for WIMPs; the study of sub-dominant effects on neutrino oscillations, as those possibly induced by the violation of the Lorentz invariance; the study of relic particles (magnetic monopoles, nuclearites) in the cosmic radiation; the coincident neutrino emission with gravitational waves.

The small interaction cross section of neutrinos 
allows them to come from far away, but it is also a drawback, as their detection requires a large target mass. The idea of a neutrino telescope based on the detection of the secondary particles produced in neutrino interactions was first formulated in the 1960's by Markov \cite{markov}. He proposed \textit{to install detectors deep in a lake or  in the sea and to determine the direction of the charged particles with the help of  Cherenkov radiation}. 
As we will show in \S  \ref{toy}, starting from the Markov idea and from the present knowledge of TeV $\gamma$-rays sources, the challenge to detect galactic neutrinos is open for a kilo\-me\-ter-scale apparatus. We will use the fact that high-energy muons retain information on the direction of the incident neutrino and can pass through several kilometers of ice or water, \S \ref{nu_mu_det}. Along their trajectory, the muons emit  Cherenkov light. From the measured arrival time of the  Cherenkov light \S \ref{cherenkov}, the direction of the muon can be determined. This process is referred to as muon track reconstruction. We will also derive in a simple way that the number of optical sensors required to reconstruct muon tracks is of the order of 5000.

Neutrino production in astrophysical sites through $\pi$ or $K$ decay leads to a flavor ratio at sources of $\nu_e : \nu_\mu : \nu_\tau = 1: 2 : 0$, which is changed by the neutrino oscillation mechanism to $1 : 1 : 1$ on Earth. As discussed in \S \ref{nu_e_det} and \S \ref{nu_tau_det}, the measurements of showers induced by very and ultra high energy $\nu_e$ and $\nu_\tau$ is another very important challenge for large volume neutrino detectors, although the neutrino direction measurement is poorer for these flavors with respect to the $\nu_\mu$ channel. The extragalactic CRs-neutrinos connection \cite{halzen_nt} sets also the scale of the detectors to 1 km$^3$.

The properties of water and ice connected to the possibility of detecting high energy neutrinos are discussed in \S \ref{waterice}. 
The pioneering project for the construction of an underwater neutrino telescope was due to the DUMAND collaboration \cite{dumand}, which attempted to deploy a detector off the coast of Hawaii in the 1980s. At the time technology was not advanced enough to overcome these challenges and the project was cancelled. In parallel, the BAIKAL collaboration \cite{baikal1} started to work in order to realize a workable detector systems under the surface of the Baikal lake (\S \ref{early}). 

Regarding deep ice, a major step  towards the construction of a large neutrino detector (see \S  \ref{nutel}) is due to the AMANDA collaboration \cite{amanda1}. AMANDA deployed and operated the optical sensors in the ice layer of the Antarctic starting from 1993. After the completion of the detector in 2000, the AMAN\-DA collaboration proceeded with the construction of a much larger apparatus, IceCube. 59 of the 80 scheduled strings (April 2009) are already buried in the ice. Completion of this detector is expected to be in 2011. 

In water, the pioneering DUMAND experience is being continued in the Mediterranean Sea by the ANTARES \cite{antares1}, NE\-MO \cite{nemo1} and NESTOR \cite{nestor} collaborations, which demonstrated the detection technique (see \S  \ref{mediterranean}). The ANTARES collaboration has completed (May 2008) the construction of the largest neutrino telescope ($\sim 0.1$ km$^2$) in the Northern hemisphere. The ANTARES detector currently take data.
 These projects have lead to a common design study towards the construction of a km$^3$-scale detector in the Mediterranean Sea (\S  \ref{km3net}).
KM3NeT \cite{km3net} is an European deep-sea research infrastructure, which will host a neutrino telescope with a volume of at least one cubic kilo\-me\-ter at the bottom of the Mediterranean Sea that will open a new window on the Universe.

\section{The connection among primary Cosmic Rays, $\gamma$-rays and neutrino. Our Galaxy.}

\subsection{Primary Cosmic Rays} \label{PCR}
\begin{figure}[tbh]
\begin{center}
\vspace{11.0cm}
\includegraphics{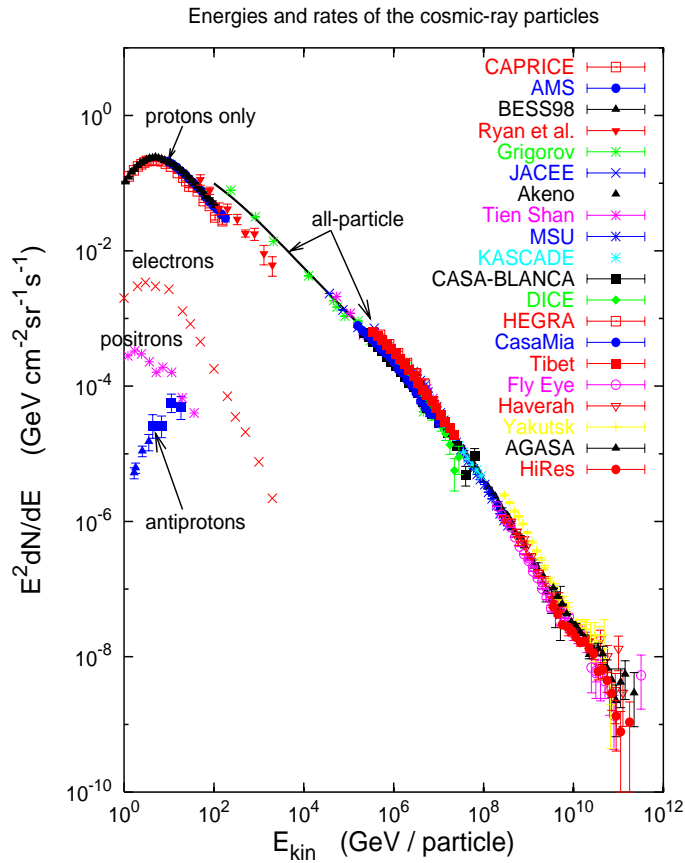}
  \caption{\it
Cosmic Ray spectrum from 10$^9$ to 10$^{21}$ eV as measured on Earth, from  \cite{hillascr}. Note that the vertical scale has been multiplied by $E^2$. On the low-energy domain, when the measurements are available, the contribution of protons, electrons, positrons and antiprotons it is also reported. See \cite{hillascr} for the reference to the experiments. 
\label{crspectrum} }
\end{center}
\end{figure}

Cosmic Rays (CRs) are mainly high energy protons (Fig. \ref{crspectrum}) and heavier nuclei which are constantly hitting the upper shells of the Earth's atmosphere. 
The energy spectrum spans from $\sim 10^9$ eV to more than $10^{20}$ eV, is of non-thermal origin and follows a broken power-law of the form:
\begin{equation}
\biggl[{dN_P\over dE}\biggr]_{obs} = K\cdot E^{-\alpha} \quad 
(\textrm{cm}^{-2} \textrm{sr}^{-1}\textrm{s}^{-1} \textrm{GeV}^{-1})
\label{cr}
\end{equation}

{\it Direct} or {\it indirect} techniques are used to measure the  CR spectrum. 
The measured power law spectrum of CRs (eq. \ref{cr}) is characterized by an index $\alpha=2.7$ up to energies of roughly $3 \times  10^{15}$ eV. 
Beyond  $3 \times  10^{15}$ eV, the index becomes $\alpha=3.1$.  This feature in the energy spectrum is known as the $knee$.
There is no consensus on a preferred accelerator model for energies above the $knee$ up to $10^{19}$ eV, where there is a flattening in the spectrum, denoted as the $ankle$.  

The highest CRs exceed even $10^{20}$ eV. After the $ankle$, it is generally assumed that CR sources are of extragalactic origin. The experimental search for sources of these ultra high energy CRs is recently entered a hot phase.
Detailed reviews of the theory and measurement of the primary CR spectrum can be found in  \cite{crflux,gaisserbook,greider}.

Up to energies of $\sim 10^{14}$ eV, the CR spectrum is directly measured above the atmosphere.  Stratospheric balloons or satellites have provided the most relevant information about the composition of CRs in the Galaxy and have contributed to establish the standard model of galactic CRs. Measurements show that $\sim$ 90\% are protons, $\sim$ 9\% are Helium nuclei and $\sim$ 1\% are heavier nuclei.

In this energy range, the mechanism responsible for the acceleration of particles is the Fermi mechanism  \cite{fermi1,fermi2}. This mechanism explains the particle acceleration by iterative scattering processes of charged particles in a shock-wave. These shock-waves are originated in environments of exceptional disruptive events, like stellar gravitational collapses. 
In each scattering process, a particle with energy $E$ gets an energy gain of $\Delta E \sim \beta E$, where $\beta \sim 10^{-2}$.
Due to the magnetic fields confinement, the scattered particles are trapped inside the acceleration region and they have a small probability to escape.
This iterative process of acceleration is a very appealing scheme for the origin of CRs, since it naturally explains the power law tendency in the spectrum. 

Supernova remnants (SNR) in the Galaxy are the most accredited site of acceleration of CRs up to the knee  \cite{ginzburg}, although this theory is not free from some difficulties  \cite{hillas}. 
The Fermi mechanism in the SNR \cite{snr}, predicts a power law differential energy spectrum $ \sim E^{-2}$
and fits correctly to the energy power involved in the galactic cosmic rays of $\sim 5\times 10^{40}$ erg/s.

The measured spectral index ($\alpha \sim 2.7$) is steeper than the expected spectrum near the sources, because of the energy dependence of the CR diffusion out of the Galaxy, as explained by the so called {\it leaky box} \cite{leakybox}. 
In the leaky box model, particles are confined by galactic magnetic fields ($\overline B\sim 3 \mu$G) and have a small probability to escape. The gyromagnetic radius for a particle with charge Z, energy E, in a magnetic field B is $R \simeq {E\over eZB} $.
During propagation, high energy particles (at a fixed value of $Ze$) have larger probability to escape from the Galaxy due to their larger gyromagnetic radii.
As a consequence, an energy-dependent diffusion probability $P$ can be defined. $P$ is experimentally estimated through the measurement of the ratio between light isotopes produced by spallation of heavier nuclei. 
It was found that $P(E)\sim E^{\alpha_D}$, with the diffusion exponent ${\alpha_D}\sim 0.6$  \cite{crflux}.  
The differential CR flux at the sources can be estimated as the convolution of the measured spectrum (\ref{cr}) and the CR escape probability $P$: 
\begin{equation}
\biggl[{dN_P \over dE}\biggr]_{sources} \propto \biggl[{dN_P \over dE}\biggr]_{obs}  \times P(E) \propto E^{-\alpha_{CR}}
\label{crsources} 
\end{equation}
\noindent with $\alpha_{CR}=\alpha-\alpha_D \sim 2$, as predicted by the Fermi model. 

The $knee$ of the CR spectrum is still an open question and different models have been proposed to explain this feature  \cite{knee}. 
Some models invoke astrophysical reasons: due to the iterative scattering processes involved in the acceleration sites, a maximum energy for the CRs is expected. This maximum energy depends on the nucleus charge $Ze$, and this leads to the prediction of a different energy cutoff for every nucleus type. As a consequence, CRs composition is expected to be proton-rich before the $knee$, and iron-rich after.
Other more exotic models try to explain the steepening in the CR flux, for instance the hypothesis of new particle processes in the atmosphere  \cite{kaza}.

Above $\sim 10^{14}$ eV, CR measurements are only accessible from ground detection infrastructures.
The showers of second\-da\-ry particles created by interaction of primary CRs in the atmosphere are distributed in a large area, enough to be detected by detector arrays. The energy region around the knee has been explored by different experiments, as for instance KASCADE \cite{kascade}.
Although the experimental techniques are very difficult and have poor resolution, observations of this region of the energy spectrum seem to indicate that the average mass of CRs increases when passing the knee. 
 
The SNR models cannot explain the CRs flux above $\sim 10^{16}$ eV, but there is no consensus on a preferred accelerator model up to $10^{19}$ eV.
CRs can be accelerated beyond the $knee$ if, for instance, the central core  of the supernova hosts a rotating neutron star. Already accelerated particles can also suffer additional acceleration due to the neutron star strong variable magnetic field. The maximum energy cannot exceed $\sim 10^{19}$ eV.

\subsection{High energy $\gamma$-rays} \label{gamma}
Some galactic accelerators must exist to explain the presence of CRs with energies up to the $ankle$. These sources can be potentially interesting for a neutrino telescope. 
Apart from details, it is expected that galactic accelerators are related to the final stage of the evolution of massive, bright and relatively short-lived stellar progenitors. 

Due to the influence of galactic magnetic fields, charged particles do not point to the sources. Neutral particles (gamma-rays and neutrinos) do not suffer the effect of magnetic fields: they represent the decay products of accelerated charged particles but cannot be directly accelerated.

Photons in the MeV-GeV energy range were detected by the Energetic Gamma-ray Experiment Telescope (EGRET) \cite{egret} on board of the CGRO satellite in the 1990s. 
The last EGRET catalogue contains 271 detections with high significance, from which 170 are not identified yet. 

Following its launch in June 2008 \cite{glast}, the Fermi Gamma-ray Space Telescope (Fermi) began a sky survey in August. Its Large Area Telescope (LAT) has  produced, in 3 months, a deeper and better-resolved map of the $\gamma$-ray sky than any previous space mission. The initial result for energies above 100 MeV \cite{Fermi_lat} regards the 205 most significant $\gamma$-ray sources, which are the best-characterized and best-localized ones. Most of them are in the galactic plane, and were associated with  known pulsars. 

\begin{figure*}
\vspace*{8.0cm}
\includegraphics{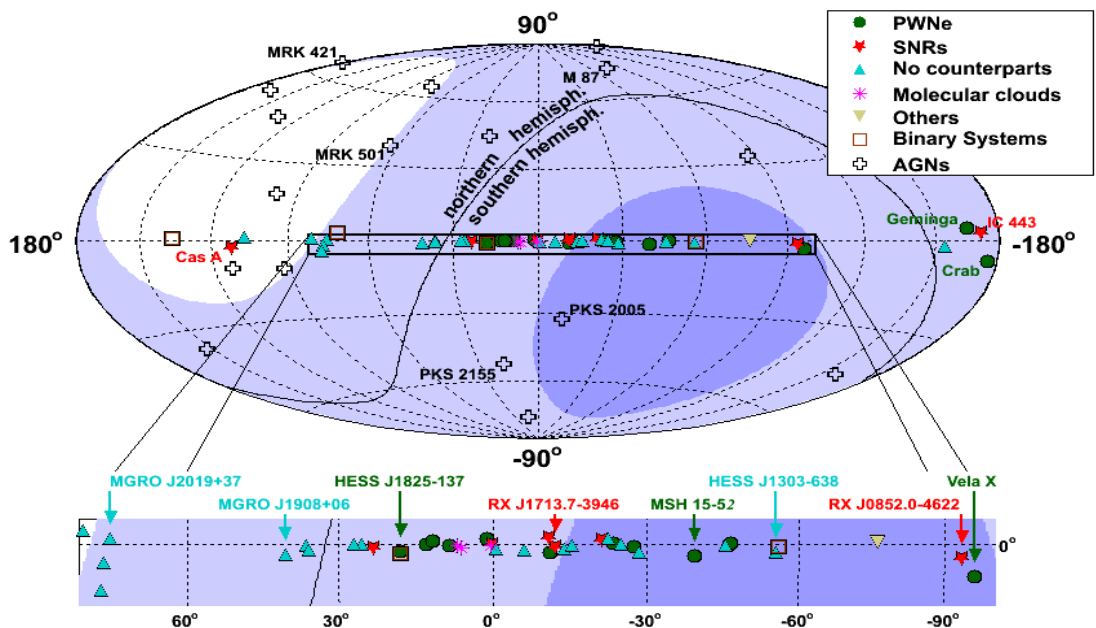}
  \caption{\it
Sky map of high energy $\gamma$-ray sources above 100 GeV. The shading indicates the visibility for a detector in the Mediterranean sea with $2\pi$ downward coverage; dark (light) areas are visible at least 75\% (25\%) of the time; the white area is not observable (from \cite{kappes}). 
    \label{gammasky} }
\end{figure*}

Gamma-rays above 100 GeV are detected on ground, using the Imaging Air-Cherenkov Technique (IACT). High-energy $\gamma$-rays are absorbed when reaching the Earth atmosphere, and the absorption process proceeds by creation of a cascade (shower) of high-energy relativistic secondary particles. These emit Che\-renkov radiation, at a characteristic angle in the visible and UV range, which  passes through the atmosphere.  As a result of Cherenkov light collection by a suitable mirror in a camera, the showers can be observed on the surface of the Earth. 

The pioneering ground based $\gamma$-ray experiment was built by the Whipple collaboration \cite{whipple}.  During the last decade, several ground-based $\gamma$-ray detectors were developed, both in the North \cite{hegra} and South \cite{veritas,cangaroo} Earth hemisphere. 
At present, the new generation apparatus are the H.E.S.S. \cite{hess} and VERITAS \cite{veritas_n} telescope arrays and  the MAGIC telescopes \cite{magic}. A full and detailed review of VHE astrophysics with the ground-based $\gamma$-ray detectors can be found in \cite{rnc,ropp}.
These IACT telescopes have produced a catalogue of TeV $\gamma$-ray sources. A (partial) sky map can be seen in Fig.  \ref{gammasky}.  Of particular interest (mainly for a neutrino detector placed in the North hemisphere) is the great population of new TeV $\gamma$-ray sources in the galactic centre region discovered by the H.E.S.S. telescope. A list of more than 70 galactic and extra-galactic sources is in \cite{rnc}. 

Both electrons and protons can be accelerated by astrophysical objects. We refer respectively to a {\it leptonic model} \cite{rnc,ropp} when electrons are accelerated, and to a {\it hadronic model} \cite{halzenRPP} when protons or other nuclei are accelerated.
The most important process which produces high energy $\gamma$-rays in the leptonic model is the Inverse Compton (IC) scattering. IC $\gamma$-rays are produced in the interactions of energetic electrons with ambient background photon fields: the CMB, and the diffuse galactic radiation of star light. This process is very efficient for producing $\gamma$-rays, since low energy photons are found in all astrophysical objects.
Multi-TeV electrons producing $\gamma$-rays of TeV energies via IC, produce also synchrotron radiation in the X-ray band  as well \cite{aharo}. Therefore, measurements of the synchrotron X-ray flux from a source is a signal that the accompanying $\gamma$-rays are likely produced by leptonic processes. 

Both models, the leptonic model and the hadronic model \cite{bere} could provide an adequate description of the  present experimental situation. If high energy photons are produced in the hadronic models, high energy neutrinos will be produced as well. 
Most of observed TeV $\gamma$-ray galactic sources have a power law energy spectrum $E^{-\alpha_\gamma}$, where $\alpha_\gamma \sim 2.0 \div 2.5$. The values of the spectral index are very close to the expected spectral index of CR sources, $\alpha_{CR}$. This lead to the conclusion that sources of TeV $\gamma$-rays can  also be the sources of galactic CRs.

Some of the most promising candidate neutrino sources in our Galaxy are extremely interesting, due to the recent results from $\gamma$-ray detectors. A neutrino telescope in the Northern hemisphere (as a detector in the Mediterranean sea) is looking at the same Southern field-of-view as the H.E.S.S. and CANGAROO Imaging Air Cherenkov  telescopes, while the neutrino telescope in the South Pole is looking at the Northern sky.

\subsection{TeV $\gamma$-rays and neutrinos from hadronic processes}\label{gammahadron}
The astrophysical production of high energy neutrinos is mainly supposed via the decay of charged pions in the beam dump of energetic protons in dense matter or photons field. 

Accelerated protons will interact in the surroundings of the CRs emitter with photons predominantly via the $\Delta^+$  resonance:
\begin{eqnarray}
p + \gamma \rightarrow \Delta^+ \rightarrow \pi^o + p \nonumber \\
p + \gamma \rightarrow \Delta^+ \rightarrow \pi^+ + n  
\label{delta}
\end{eqnarray}

Protons will interact also with ambient matter (protons, neutrons and nuclei), giving rise to the production of charged and neutral mesons.
The relationship between sources of VHE $\gamma$-ray ($E_\gamma > 100$ MeV) and neutrinos is the meson-decay channel. 
Neutral mesons decay in photons (observed at Earth as $\gamma$-rays):
\begin{equation}
\pi^o \rightarrow \gamma \gamma
\label{pio}
\end{equation}
\noindent while charged mesons decay in neutrinos:

\begin{eqnarray}
\pi^+ \rightarrow  \nu_\mu + \mu^+ \nonumber \\
&\hookrightarrow \mu^+ \rightarrow  \overline{\nu}_\mu + \nu_e + e^+ \nonumber  \\
\pi^- \rightarrow  \overline{\nu}_\mu  + \mu^-  \nonumber \\
&\hookrightarrow \mu^- \rightarrow  \nu_\mu + \overline{\nu}_e  + e^-   
\label{pipm}
\end{eqnarray}

Therefore, in the framework of the hadronic model and in the case of {\it transparent sources}, the energy escaping from the source is distributed between CRs, $\gamma$-rays and neutrinos. 
A transparent source is defined as a source of a much larger size that the proton mean free path, but smaller than the meson decay length. For these sources, protons have large probability of interacting once, and most secondary mesons can decay. 

Because the mechanisms that produce cosmic rays can produce also neutrinos and high-energy photons (from eqs. \ref{pio} and \ref{pipm}), candidates for neutrino sources are in general also $\gamma$-ray sources.
In the hadronic model there is a strong relationship between the spectral index of the CR energy spectrum $E^{-\alpha_{CR}}$, and the one of $\gamma$-rays and neutrinos. It is expected  \cite{kappes} that near the sources, 
the spectral index of secondary $\gamma$ and $\nu$ should be almost identical to that of parent primary CRs:  $\alpha_{CR} \sim \alpha_{\nu} \sim \alpha_{\gamma}$.  
Hence $\gamma$-ray measurements  give crucial information about primary CRs, and they constrain (see \S \ref{nuastro}) the expected neutrino flux.           

\subsection{Prediction of HE neutrino flux from astrophysical sources} \label{nuastro}

Here, we present some proposed mechanisms for the production of cosmic high energy neutrinos. Some of them seem to be  {\it guaranteed}, since complementary observations of  TeV $\gamma$-rays can hardly be explained by leptonic models alone. The expected neutrino fluxes at Earth, however, are uncertain and predictions differ in some cases up to orders of magnitude. 

\subsubsection{Shell-type supernova remnants}
Particles can be accelerated in the supernova remnants (SNRs) via the Fermi mechanism.
If the final product of the SN is a neutron star, already accelerated particles can gain additional energy, due to the neutron star strong variable magnetic field. Shell-type SNRs are considered to be the most likely sites of galactic CR acceleration, hypothesis supported by recent observations from the TeV $\gamma$-ray IACT. 

Of particular interest is the supernova remnant in the Vela Jr. (RX J0852.0-4622). This SNR is one of the brightest objects in the southern TeV –sky.
Recent observations of $\gamma$-rays exceeding 10 TeV in the spectrum of this SNR by H.E.S.S.  \cite{hessvela} have strengthened the hypothesis that the hadronic acceleration is the process that is needed to explain the hard and intense TeV $\gamma$-ray spectrum. 
H.E.S.S. has observed that the $\gamma$-ray TeV emission originates from several separated parts of a region of apparent size of $\sim 2^o$. 
The angular resolution of neutrino telescopes for the $\nu_\mu$ channel is much better than $2^o$ (\S  \ref{waterice}). The  expected neutrino-induced muon rate leads, in some calculations  \cite{guaranteed}, to encouraging results for a 1 km$^3$ detector in the Mediterranean sea. 
\begin{figure}[tbh]
\begin{center}
  \vspace{7.5cm}
  \includegraphics{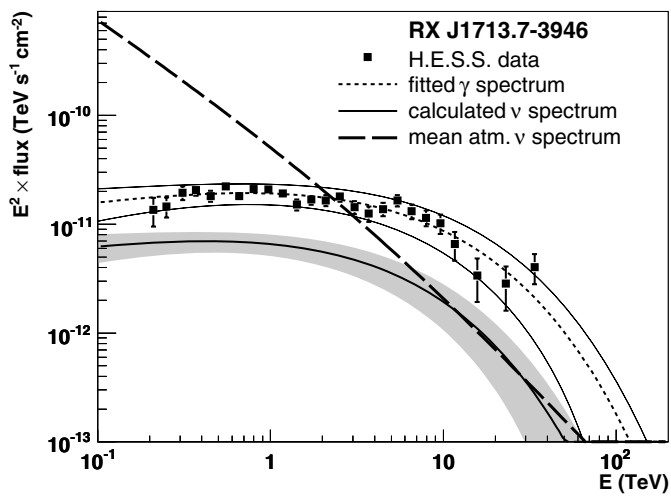}
  \caption{\it
Measured gamma-ray flux from RX J1713.7-3946 and estimated neutrino flux  \cite{kappes} with their error bands. The atmospheric neutrino flux (which represents the background) is integrated over the search window and averaged over one day.
    \label{gammatonu}  }
\end{center}
\end{figure}

A second important source is the SNR RX J1713.7-3946, which has been the subject of large debates about the nature of the process (leptonic or hadronic) that originates its gamma-ray spectrum  \cite{nohadro}.
RX J1713.7-3946 was first observed by the CANGAROO experiment which firstly claimed a leptonic origin  \cite{gangaroorx}. 
Successive observations with CANGAROO-II  \cite{enomoto} disfavour purely electromagnetic processes as the only source of the observed $\gamma$-ray spectrum. Neutrino flux calculations based on this result have predicted large event rates, also in neutrino telescopes with size smaller than 1 km$^3$   \cite{halzenalva,costantini}.
This source has successively been observed with higher statistics by the H.E.S.S. telescope \cite{hessmorfo}, supporting the hadronic origin.  
The measured spectrum deviates from a pure power law spectrum. It can be reasonably well described by a power law with an exponential
cutoff, $d\Phi_\gamma/dE_\gamma = I_0( E_\gamma/1\ TeV)^{\alpha_\gamma} \exp(-E_\gamma/E_c)$, where the cutoff parameter is $E_c= 12\pm 2$ TeV \cite{rnc}.

The neutrino flux calculation (shown in Fig. \ref{gammatonu})  based on the H.E.S.S. result, with the exponential cutoff in spectrum and other assumptions, lead to the prediction that the source should be marginally detectable in a kilometer-scale Mediterranean detector. This result strongly depends on the assumed cutoff value. Without cutoff, the event rate increases by a significant factor, making these sources easily accessible to neutrino telescopes. 

\subsubsection{Pulsar wind nebulae (PWNe)}
PWNe  are also called Crab-like remnants, since they resemble the Crab Nebula (\S \ref{candle}), which is the youngest and most energetic known object of this type. PWNe differ from the shell-type SNRs because there is a pulsar in the center which 
blows out equatorial winds and, in some cases,  jets of very fast-moving
material into the nebula.
The radio, optical and X-ray observations suggest a synchrotron origin for these emissions. H.E.S.S. has also detected TeV $\gamma$-ray emission from the Vela PWN, named Vela X. This emission is likely to be produced by the inverse Compton mechanism. The possibility of a hadronic origin for the observed $\gamma$-ray spectrum, with the consequent flux of neutrinos, was also considered  \cite{horns}.

The neutrino flux calculated for a few PWNe in the framework on a hadronic production of the observed TeV $\gamma$-rays (such as the Crab, the Vela X, the PWN around PSR1706-44 and the nebula surrounding PSR1509-58) agrees with the conclusions that all these PWNe could be detected by a kilometer-scale neutrino telescope. For instance, from 6 to 12 events are predicted in 1 y, with 1 background event due to the atmospheric neutrinos \cite{guetta}. Others \cite{kappes}, which assume an exponential cutoff in the energy spectrum, give more pessimistic results (10 events/5 y from the Vela X source, with 4.6 background events).

\subsubsection{The galactic Centre (GC)\label{gc}}
The galactic centre is probably the most interesting region of our Galaxy, also regarding the emission of neutrinos. It is specially appealing for a Mediterranean neutrino telescope since it is within the sky view of a telescope located at such latitude. The interest in it has increased after the recent discoveries of H.E.S.S..  

Early H.E.S.S. observations of the GC region showed a point-like source at the gravitational centre of the Galaxy (HESS J1745-290  \cite{hessj17}) coincident with the supermassive black hole Sagittarius A* and the SNR Sgr A East. In 2004, a more sensitive campaign revealed a second source, the PWN G 0.9+0.1  \cite{hesspwn}.

\begin{figure}[tbh]
\begin{center}
  \vspace{8.6cm}
  \includegraphics{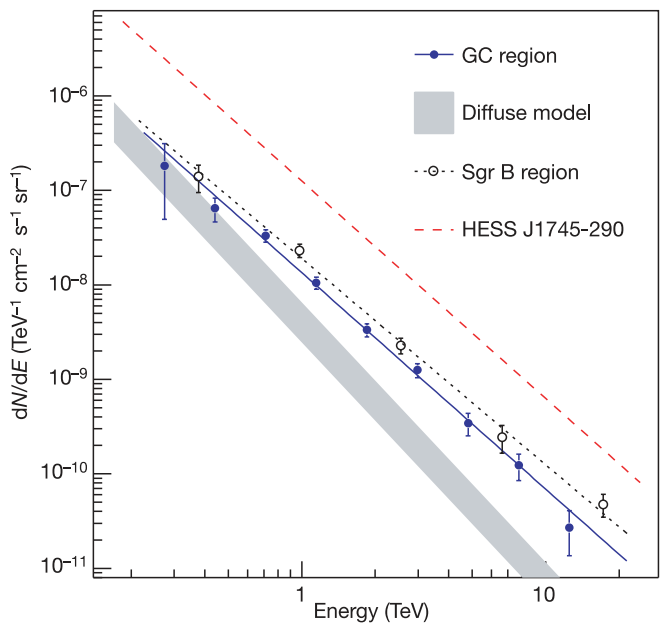}
  \caption{\it  H.E.S.S. \cite{hesscentre} measurement of $\gamma$-ray flux per unit solid angle in the Sgr B region, near the galactic center (open circle data points). 
The measured spectrum in the galactic region $|l| < 0.8^o, |b| < 0.3^o$ is shown using full circles. These data can be described by a power law with spectral index $\sim 2.3$.
In comparison, the  expected $\gamma$-ray flux assuming a CR spectrum as measured in the solar neighborhood is shown as a shaded region. 
The measured $\gamma$-ray flux ($>$1 TeV) implies a high-energy cosmic ray density which is 4 to 10 times higher than our solar neighborhood value. The spectrum of the source HESS J1745-290 is also shown for comparison. 
    \label{galacticcentre}  }\end{center}
\end{figure}

Thanks to the good sensitivity of the H.E.S.S. telescope, it is possible to subtract the GC sources and search for the diffuse $\gamma$- ray emission which spans the galactic coordinates $|l| < 0.8^o, |b| < 0.3^o$. This  diffuse emission of $\gamma$-ray with energies greater than 100 GeV is correlated with a complex of giant molecular clouds in the central 200 pc of the Milky Way  \cite{hesscentre}.
The measured $\gamma$-ray spectrum in the GC region is well described by a power law with index of $\sim 2.3$. The photon index of the $\gamma$-rays, which closely traces back the spectral index of the CR, indicates in the galactic centre a local CR spectrum that is much harder and  denser than that measured on Earth, as shown in Fig. \ref{galacticcentre}.
Thus it is likely that an additional component of the CR population is present in the galactic centre, above the diffuse CR concentration which fills the whole Galaxy. The fact that cosmic accelerators are very close to the  GC, and therefore the possibility of neglecting the CR diffusion loss due to propagation (see eq. \ref{crsources}), gives a natural explanation for the harder observed spectrum, which is closer to the intrinsic value of the CR spectral index. 
In  \cite{hesscentre} it is suggested that the central source HESS J1745-290 is likely to be the source of these CR protons (and thus of neutrinos), with two candidates for CR accelerations in its proximity: the SNR Sgr A East (estimated age around $10^4$ yrs), and the black hole Sgr A*.

\subsubsection{Microquasars}
Microquasars are galactic X-ray binary systems, which exhibit relativistic radio jets, observed in the radio band  \cite{radio}. The name is due to the fact that they result morphologically similar to the AGN, since the presence of jets makes them similar to small quasars. This resemblance could be more than morphological: the physical processes that govern the formation of the accretion disk and the plasma ejection in microquasars are probably the same ones as in large AGN. 

Microquasars have been proposed as galactic acceleration sites of charged particles up to $E\sim 10^{16} $ eV. The hypothesis was strengthened by the discovery of the presence of relativistic nuclei in microquasars jets like those of SS 433. This  was inferred from the observation of iron X-ray lines  \cite{iron}. 

Two microquasars, LS I +61 303 and  LS 5039, have been detected as $\gamma$-ray sources above 100 MeV and listed in the third EGRET Catalogue. They are also detected in the TeV energy range \cite{lsi61,ls5039}.

There is yet uncertainty as to what kind of compact object lies in LS I +61 303 (observed by the MAGIC telescope). 
Recently, a multiwavelength campaign including the MAGIC telescope, XMM-Newton, and Swift was conducted during 60\% of an orbit in 2007. A simultaneous outburst at X-ray and TeV $\gamma$-ray bands, with the peak at phase 0.62 and a similar shape at both wavelengths, gives conclusive indication of variability also in the $\gamma$-ray emission. The X-ray over TeV $\gamma$-ray flux ratio favors leptonic models \cite{magic_xray}. 
Because the source is located in the Northern sky, it is specially appealing for a neutrino telescope located in the Southern hemisphere as IceCube, which will be able to detect (or rule out) neutrinos coming from this source   \cite{icecubemicro}.

Microquasar LS 5039 (detected by H.E.S.S. in the Southern sky) has features similar to LS I +61 303, and the observed flux still does not allow an unequivocal conclusion about the variability of the source.
Different astrophysical scenarios have been proposed to explain the TeV $\gamma$-ray emission, which involve leptonic and/or hadronic interactions.
In particular, the leptonic model is strongly disfavored in  \cite{hadro5039}, and it is expected that LS 5039 could produce between $0.1\div 0.3$ events/year in a detector like ANTARES (see \S \ref{antares}). The event rate depends on the assumed neutrino spectrum (power law with index ranging from 1.5 to 2.0), and two energy cutoff ($E_{c} =$ 10 TeV and 100 TeV  \cite{hadro5039}). The expected rate is $\sim $ 25 times higher for a 1 km$^3$ detector in the Mediterranean sea.

Other microquasars were considered in  \cite{distefano}. The best candidates as neutrino sources are the steady microquasars SS433 and GX339-4. Assuming reasonable scenarios for TeV neutrino production, a 1 km$^3$-scale neutrino telescope in the Medi\-ter\-ranean sea could identify microquasars in a few years of data taking, with the possibility of a 5$\sigma$ level detection. 
In case of no-observation, the result would strongly constrain the neutrino production models and the source parameters.

\subsubsection{Neutrinos from the galactic plane}
In addition to stars, the   Galaxy  contains  interstellar  thermal  gas,  magnetic  fields  and  CRs which have roughly  the same energy density.
The  inhomogeneous  magnetic  fields confine  the CRs within  the  Galaxy. Hadronic  interactions of CRs with  the  interstellar  material  produce a  diffuse  flux  of  $\gamma$-rays  and  neutrinos (expected to be equal,  within  a  factor  of  $\sim$  2). The fluency at Earth is expected to be correlated to  the  gas  column  density  in  the  Galaxy:  the  largest  emission  is  expected  from  directions  along  the  line  of  sight  which  intersects  most  matter.  

Recently,  the  MILAGRO  collaboration  has  reported  the  detection  of  extended  multi-TeV  gamma emission  from  the  Cy\-gnus  region   \cite{milagrodif}, which is well correlated  to  the  gas  density and  strongly  supports the  hadronic  origin  of  the  radiation.  The  MILAGRO  observations  are  inconsistent  with  an  extrapolation  of  the  EGRET  flux  measured at  energies  of  tens  of  GeV.  This  supports the hypothesis   that  in  some  areas  of  the  galactic  disk  the  CR  spectrum  might  be  significantly  harder  that  the  local  one. 

With the assumption that  the  observed $\gamma$-ray  emission  comes  from  hadronic  processes,  it  is  possible  to  obtain  an  upper  limit  on  the  diffuse  flux of neutrinos from the galactic plane.  The KM3NeT consortium  \cite{km3} has made an  estimate  of  the  neutrino flux  from  the  inner  Galaxy,  assuming  that  the emission  is  equal  to  that  observed  from  the direction  of  the  Cygnus region. The  expected  signal  rate  for a  km$^3$ neutrino  telescope  located  in  the  Mediterranean  Sea is  between  4  and  9  events/year  for  the  soft (index $\alpha=2.55$)  and  hard ($\alpha=2.10$) spectrum  respectively,   with  an  atmospheric  neutrino  background  of  about  12  events  per  year.  

\subsubsection{Unknowns}
In addition to SNR, PWNe and microquasars, there are other theoretical environments in which hadronic acceleration processes could take place with production of a neutrino flux. For instance, neutron stars in binary systems and magnetars  \cite{magna}  might be sources of an observable neutrino flux. 

New improvements in the GeV- TeV scale $\gamma$-ray astronomy are expected in the next years. In particular practically all the IACT telescopes are improving their apparatus. 
News are also expected from the ARGO  \cite{argo} and MILAGRO  \cite{milagro} large field of view observatories. Finally, it is also worth remarking that a non-negligible number of VHE $\gamma$-ray sources detected by H.E.S.S. do not have a known counterpart in other wavelengths. The origin of such sources is a theoretical challenge in which neutrino astronomy may yield some insight.

	Although not certainly inspired by neutrino astronomy, it is interesting to quote this sentence from the former US Secretary of Defense, Donald Rumsfeld. The sentence is the exact words as taken from the official transcripts on the Defense Department Web site  \cite{slate}: 

\noindent \textbf{The Unknown.} \textit{As we know, /
There are known knowns. /
There are things we know we know. / 
We also know /
There are known unknowns. / 
That is to say /
We know there are some things/ 
We do not know. /
But there are also unknown unknowns, /
The ones we don't know /
We don't know. }

\section{The connection among extragalactic sources of primary Cosmic Rays, $\gamma$ rays and neutrino.}\label{3-extra}

\subsection{Measurements of the UHECR}\label{ankle}
The CR flux above $ 10^{19}$ eV, still dominated by protons and nuclei  \cite{halzencr}, is one particle per kilometer square per year per stereoradian. It has long been  assumed \cite{peters} that ultra high energy cosmic rays (UHECR) are extragalactic in origin  \cite{sokolsky}, and can be detected only by very large ground-based installations. Therefore, the structure in the CR spectrum above $\sim 10^{19}$ eV (the $ankle$) is usually associated with the appearance of this flatter contribution of extra-galactic CRs. In fact, above the $ankle$ the gyroradius of a proton in the galactic magnetic field exceeds the size of the Galaxy disk (300 pc). 
\begin{figure}[tbh]
\vspace{9.7cm}
\includegraphics{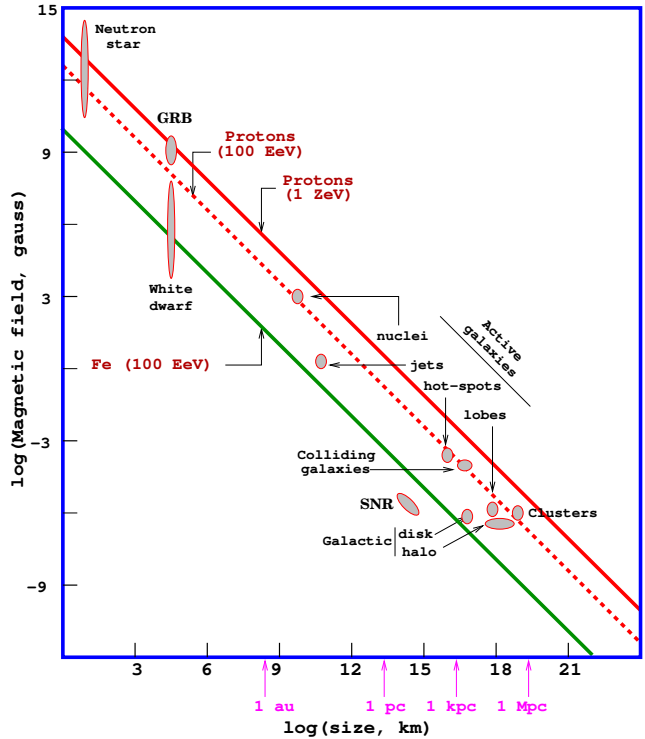}
  \caption{\it
The Hillas diagram (drawn by Murat Boratav). Acceleration of cosmic rays up to a given energy requires magnetic fields and sizes above the respective line. Some sources candidates are still controversial (1 EeV=$10^{18}eV$, 1 ZeV=$10^{21}eV$).  
    \label{fig:hillasplot}}
\end{figure}

Fig. \ref{fig:hillasplot}  \cite{ostro} shows a diagram first produced by Hillas (1984). Hillas derived the maximum energy which a particle of a given charge can reach, independently of the acceleration mechanism. It was obtained from the simple argument that the Larmor radius of the particle should be smaller than the size $R_{kpc}$ (in kpc) of the acceleration region. This energy $E$(EeV) (in units of $10^{18}$ eV) is given by:
\begin{equation}
E\textrm(EeV)\sim \beta Z B_{\mu G} R_{kpc}
\label{hillas}
\end{equation}
\noindent where $\beta$ is the velocity of the shock wave in the Fermi model or in any other acceleration mechanism. Fig. \ref{fig:hillasplot} gives the relation between the dimensions of the astrophysical objects and the magnetic fields needed to contain the accelerating particle, in order that protons can reach up to $10^{20}$ eV (dashed line)  or $10^{21}$ eV (upper full line),  and iron nuclei up to $10^{20}$  eV (lower full line). 
As can be seen from the Hillas plot, plausible acceleration sites may be the radio lobes or hot spots of powerful active galaxies.

The search for UHECR sources must take into account another effect, the Greisen-Zatsepin-Kuzmin cutoff (GZK)  \cite{gzk1,gzk2}, which imposes  a theoretical upper limit on the energy of cosmic rays from distant sources. 
Above a threshold of few $ 10^{19}$ eV, protons interact with the 2.7$^o$ K cosmic microwave background radiation (CMB) and lose energy through the resonant pion production of eq. \ref{delta}. Due to the GZK cutoff, protons above that threshold cannot travel distances further than few tens of Mpc. 

From the astrophysical point of view, this cutoff is very important because it limits the existence of standard astrophysical UHECR emitters inside our local super-cluster of galaxies. The GZK cutoff has stimulated important  debate, since there were two contradictory measurements in the region between $10^{19}\div 10^{20}$ eV, made by the 
AGASA \cite{agasa} and by the High Resolution Fly's Eye (HiRes)  \cite{hires} experiments. 

Nowadays, the largest experiment is the Auger Observatory  \cite{augerweb}, which combines the measurement of extensive air showers and light fluorescence detection. 
Auger has published  \cite{augerspectrum} the result of the first data set, rejecting the hypothesis that the cosmic ray spectrum continues in the form of a power law above $10^{19.6}$ eV with 6 sigma significance. 
In addition, Auger has reported the first hints of association of CRs with $E> 6\times 10^{19}$ eV and nearby (less than 100 Mpc) concentration of matter and AGN  \cite{augeragn}.  Although its statistical significance is still  limited, the results suggest that regions of matter with AGN can be the source candidates for UHECR acceleration. 

\subsection{Extra-galactic CR and $\nu$ sources}\label{extragalactic}

The prediction of high energy neutrino sources of extra-galactic origin is a direct consequence  of the CR observations. 
As for the origin of UHE Cosmic Rays, Active Galactic Nuclei (AGN) are the principal candidates as neutrino sources. Other potentially promising particle accelerators are $\gamma$-ray bursts (GRBs). Here we consider these two astrophysical classes of objects, with a particular attention to the possible neutrino production mechanism. 
Finally, radio observation of starburst galaxies has motivated the idea of the existence of $hidden$ sources of CR. These sources can represent pure high energy neutrino injectors, and some predictions are presented.

Extra-galactic  sources are very far and the possibility of a individual discovery in a km$^3$ scale neutrino telescope is expected only in particular theoretical models, or using the {\it source stacking} methods: it is a 
combined analysis for different classes of objects which enhance the neutrinos detection probability \cite{amanda_sta}. 

An alternative way to prove the existence of extragalactic neutrino sources is through the measurement of the {\it cumulative flux } in the whole sky. 
The only way to detect this {\it diffuse flux of high energy neutrinos} is looking for an excess of high energy events in the measured energy spectrum induced by  atmospheric neutrinos.

Theoretical models constrain the neutrino diffuse flux, \S \ref{wb}.  These upper bounds are derived from the observation of the diffuse fluxes of $\gamma$-rays and UHECR. One of them (the Waxman-Bahcall, shortened as W\&B) is used as the reference limit to the predicted neutrino flux coming from different extra-galactic sources.

In addition to neutrinos generated by high energy cosmic accelerators, there are high energy neutrinos induced by the propagation of CRs in the local Universe \cite{nuuhe}. The subsequent pions decay will produce a neutrino flux (called $GZK$ or \textit{cosmological} neutrinos) similar to the W\&B bound above  $5\times 10^{18}$ eV  \cite{waxuhe}, since neutrinos carry approximately 5\% of the proton energy.

\subsubsection{Active Galactic Nuclei (AGN)}\label{agn}
Active Galactic Nuclei (or AGN) are galaxies with a very bright core of emission embedded in their centre, where a supermassive black hole ($10^6 \div 10^9 $ solar masses) is probably present. As outlined in \S \ref{ankle}, the Auger observatory has reported the first hints of correlation
between CR directions and nearby concentrations of matter in which AGN are present. This measurement (although still controversial) suggests that AGN are the most promising candidates for UHECR emission.

The supermassive black hole in the centre of AGN would attract material onto it, releasing a large amount of gravitational energy. 
According to some models  \cite{dinamo}, the energy rate generated with this mechanism by the brightest AGNs can be $L > 10^{47}$ erg s$^{-1}$. Early models  \cite{mannold,halzenold,protheold},  postulating the hadronic acceleration in the AGN cores, predicted a production of secondary neutrinos well above the W\&B upper limit, and the prediction from some of these models has been experimentally disproved by AMANDA  \cite{icecubediffuse}. 
More recent models  \cite{blazar} predict fluxes close to the W\&B bound. 
For instance, a prediction has recently been carried out for the Centaurus A Galaxy, which is only 3 Mpc away. In  \cite{halmul} the estimate neutrino flux from hadronic process is $E^2d\Phi_\nu/dE \le 5\times 10^{-13}$ TeV$^{-1}$ cm$^{-2}$s$^{-1}$. By varying the power-law indices between -2.0 and -3.0, they obtained between 0.8 and 0.02
events/year for a generic neutrino detector of effective muon area (\S \ref{effarea}) of $\sim$ 1 km$^2$.

A particular class of  AGN (called $blazars$) has  their  jet  axis  aligned  close  to  the  line  of  sight  of  the  observer. Blazars  present  the  best  chance  of  detecting  AGNs  as  individual  point  sources  of  neutrinos because of a  significant  flux  enhancement in the jet through  Doppler  broadening.  Blazars  exhibit  non-thermal continuum  emission  from  radio  to  VHE  frequencies and  are  highly  variable,  with  fluxes  varying  by  factors  of  around  10  over  timescales  from  less  than  1  hour  to  months.  
The third EGRET catalog \cite{egret3} contains a list of 66 blazars, plus  27 additional candidates, and 119 are in the recent Fermi LAT bright gamma-rays source list \cite{Fermi_lat}; 
an  increasing  population  of  TeV  blazars  at  higher  redshifts  is   being  detected  by  the  latest  generation  of  $\gamma$-ray  IACT; so far  18  blazars  have  been  discovered  over  a  range  of  red-shifts  from  0.03  to  $>$0.3  \cite{ropp}.  

In  hadronic  blazar  models,  the  TeV  radiation  is  produced 
by highly relativistic baryons in jets interacting with radiation fields and the ambient matter.  Owing to the low matter density in relativistic jets, photo-production of pions is commonly believed to be the most important energy loss channel, followed by proton synchrotron radiation. The $\gamma$ rays from neutral pion decay induce electromagnetic cascades, disrupting the strict neutrino-to-gamma-ray ratio of pion decay kinematics for the emerging radiation.
Another important effect to take into account is that the  observed  TeV $\gamma$-ray  spectrum  from  extragalactic  sources  is  steepened  due  to  absorption  by  the  Extragalactic  Background  Light  (EBL).  In  the  case  of  a  distant  blazar,  such  as  1ES1101  at  z=0.186,  the  observed  spectral  index  of 2.9  is estimated to correspond  to  a spectral  index  as  hard  as  1.5  near the source \cite{1es11}.  
Neutrinos,  however,  are  unaffected  by  the  EBL. As a consequence of the effective hardening of the spectrum, some TeV-$\gamma$ bright  blazars, in some models, are expected to  produce  $\nu_\mu$  fluxes  exceeding  the  atmospheric  neutrino  background  in  a cubic kilometer neutrino telescope.
H.E.S.S.  recently  reported also highly  variable  emission  from  the  blazar  PKS2155-304   \cite{pks2155}.  A  two  order  of  magnitude  flux  increase,  reaching 10  Crab  Units (C.U., defined in \S \ref{candle})  was  observed  during  a  one  hour  period.  Such  flaring  episodes  are  interesting  targets  of  opportunity  for  neutrino  telescopes.  Assuming  that  half  of  the  $\gamma$-rays  are  accompanied  by  the  production  of  neutrinos,  a  flare  of  10  C.U.  lasting   around  2.5  days  would  result  in  a  neutrino  detection  at  the  significance  level  of  3  sigma  \cite{km3}.

\subsubsection{Gamma ray bursts (GRBs)}\label{grb}
GRBs are short flashes of $\gamma$-rays, lasting typically from milliseconds to tens  of seconds, and carrying most of their energy in photons of  MeV scale. 
The likely origin of the GRBs with duration of tens of seconds is the collapse of massive stars to black holes. Observations suggest that the formation of the central compact object is associated with Ib/c type supernovae  \cite{grb1,grb2,grb3}.

GRBs also produce X-ray, optical and radio emission subsequent to the initial burst (the so called $afterglow$ of the GRB).  The detection of the afterglow is performed with sensitive instruments that detect photons at wavelengths smaller than MeV $\gamma$-rays.
In 1997 the Beppo-Sax  \cite{sax} satellite obtained for the first time high-resolution X-ray images of the GRB970228 afterglow, followed by successive observations in optical and longer wavelengths with an angular resolution of arcminute.   
This accurate angular resolution allowed the redshift measurement and the identification of the host galaxy. It was the first step to demonstrate the cosmological origin of GRBs. 

Leading models assume that a $fireball$, produced in the collapse, expands with an highly relativistic velocity (Lorentz factor $\Gamma \sim 10^{2.5}$) powered by radiation pressure. 
Protons accelerated in the fireball internal shocks lose energy through photo-meson interaction with ambient photons (the same process of eq. \ref{delta}). 
In the observer frame, the condition required to the resonant production of the $\Delta^+$ is
$E_\gamma E_p = 0.2$ GeV$^2 \Gamma^2$.
For the production of gamma-rays with $E_\gamma \sim 1$ MeV the characteristic proton energy required is $E_p = 10^{16} $ eV, if  $\Gamma \sim 10^{2.5}$.
The interaction rate between photons and protons is high due to the high density of ambient photons and yields a significant production of pions. The charged ones decay in neutrinos, typically   carrying 5\% of the proton energy. Hence, neutrinos with $E_\nu \sim 10^{14}$ eV are expected  \cite{wbgrb}. 
Other neutrinos with lower energies can also be produced in different regions or stages where GRB $\gamma$-rays are originated. 
Depending on models, a different contribution of neutrinos is expected at every time stage of the GRB. For instance, the neutrino emission from early afterglows of GRBs, due to dissipation made by the external shock with the surrounding medium or by the shock internal dissipation, was discussed in \cite{murase07}. Here, the implications of recent Swift \cite{swift} satellite observations concerning the possible neutrino signals in a neutrino telescope were also considered in detail. 

\begin{figure}[ht]
  \vspace{8.cm}
  \includegraphics{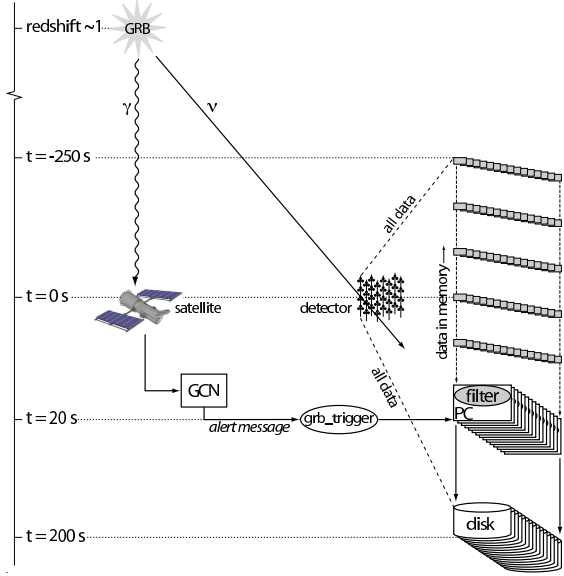}
  \caption{\it 
Sketch of an external alert from a GRB satellite to the ANTARES neutrino telescope. When an alert from the GRB Coordinates Network (GCN, which include active satellites) is received by ANTARES, all raw data in memory covering a few minutes are saved to disk. Any possible neutrino signal from the GRB (before, during, and after the photon detection by the satellite) is stored on disk \cite{mieke}.    
\label{grbanta}  }
\end{figure}

Some calculations of the neutrino flux  \cite{guettagrb}  from GRB show that a kilometer-scale neutrino telescope can be sufficient to allow detection. The average energy of these neutrinos (100 TeV) corresponds to a value for which neutrino telescopes are highly efficient. Nevertheless, being transient sources, GRBs detection has the advantage of being  practically background free, since neutrino events  are correlated both in time and direction with $\gamma$-rays. As for the case of the ANTA\-RES  detector \cite{mikePHD,antaresdaq}, unfiltered data can also be stored in the occurrence of a GRB alert from a satellite or a ground\-based telescope (see Fig. \ref{grbanta}  \cite{mieke}). The analysis of collected data around a GRB alert can be carried out some time later, with the advantage of using very precise astronomical data, improved by later observations of the afterglow with optical telescopes.

\subsubsection{Starburst or neutrino factories}\label{sec:starburst}
Radio observations have motivated the idea of the existence of regions with an abnormally high rate of star formation, in the so-called \textit{starburst galaxies}, which are common throughout the Universe. These regions of massive bursts of star-formation can dramatically alter the structure of the galaxy and can input large amounts of energy and mass into the intergalactic medium. Supernovae explosions are expected to enrich the dense star forming region with relativistic protons and electrons  \cite{sb1,sb2}. 
These relativistic charged particles, injected into the starburst interstellar medium, would lose energy through pion production. Part of the proton energy would  be converted into neutrinos by charged meson decays and part into $\gamma$-rays by neutral meson decays. Very recently, the (relatively) nearby NGC 253 galaxy in the southern hemisphere and the M82 galaxy in the northern hemisphere were identified as starburst galaxies by, respectively, the H.E.S.S. \cite{hess_burst} and  VERITAS \cite{veritas_burst} telescopes. The $\gamma$-ray flux above 220 GeV measured from NGC 253 imply a cosmic-ray density about three orders of magnitude larger than that in the center of our Galaxy.
Such hidden accelerators of CRs are thus intense neutrino sources, since mainly neutrinos would be able to escape from these dense regions. 
A cumulative flux of GeV neutrinos from starburst galaxies was calculated in  \cite{loeb} as $E^2_\nu\Phi_\nu \simeq 10^{-7} $ GeV cm$^{-2}$ s$^{-1}$sr$^{-1}$, a level which can be detected by a km$^3$-scale neutrino detector.

\subsection{The upper limits for transparent sources}\label{wb}
The observation of diffuse flux of gamma-rays and of UHE CRs can be used to set theoretical upper bounds on the total flux of neutrino from extragalactic sources (diffuse neutrino flux). High energy $\gamma$-rays can be produced in astrophysical acceleration sites by decay of the neutral pion (eq. \ref{pio}). Neutrinos will be produced in parallel from decay of the charged pions and they will escape from the source without further interactions, due to their low cross section. High-energy photons from $\pi^o$ decay, on the contrary, will develop electromagnetic cascades when interacting with the intergalactic radiation field. Most of the $\gamma$-ray energy will be released in the 1 MeV-100 GeV range. Therefore, the observable neutrino flux (within a factor of two due to the branching ratios and kinematics at production of charged and neutral pions) is limited by the bolometric observation of the gamma-ray flux in this energy band.

The diffuse gamma-ray background spectrum above 30 MeV was measured by the EGRET experiment as  \cite{egretdiffuse}: 
\begin{equation}
E^2 I_\gamma (E) = (1.37\pm 0.06)\times 10^{-6} \ 
(\textrm{GeV}\  \textrm{cm}^{-2} \textrm{sr}^{-1}\textrm{s}^{-1})
\label{eq:boundgamma}
\end{equation}

If nucleons escape from a cosmic source, a similar bound can be derived from the measured flux of CR from extragalactic  origin. Fermi acceleration mechanism  can take place when protons are magnetically confined near the source. Neutrons produced by photo-production interactions of protons with radiation fields (eq. \ref{delta}) can escape from transparent sources and decay into cosmic protons \underline{outside} the region of the magnetic field of the host accelerator. 

\begin{figure}[ht]
  \vspace{7.cm}
  \includegraphics{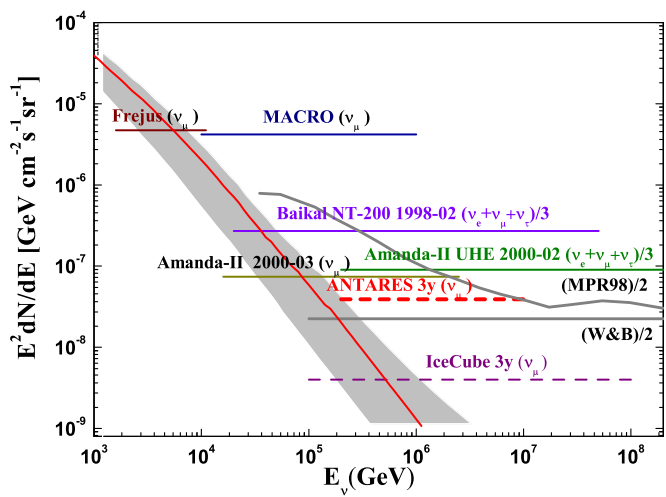}
  \caption{\it 
Sensitivities and upper limits for a $E^{-2}$ diffuse flux of high energy neutrinos of one flavor, see text. Experimental upper limits are indicated as solid lines, ANTARES and IceCube 90\% C.L. sensitivities with dashed lines. 
Upper limits obtained from all-flavor analyses (Baikal and Amanda-II  UHE 2000-02) are not directly comparable to the $\nu_\mu$ upper limits. However, for the assumed astrophysical neutrino production models and for a wide range of neutrino oscillation parameters (\S \ref{oscillations}), the flavor flux ratio at Earth  can be assumed to be $\nu_e : \nu_\mu : \nu_\tau = 1\ :\ 1\ :\ 1$. In that case, either a single flavor limit can be multiplied by three (and compared to an all-flavor result) or an all-flavor limit can be divided by three and compared to a single-flavor result, as we did in the figure.
For reference, the W\&B  and MPR98 limits for transparent sources are also shown. Both upper bounds are divided by two, to take into account the neutrino oscillation effects.  
    \label{diffuse}  }
\end{figure}

Some additional factors have to be considered before establishing a relationship between CR and neutrino fluxes. These factors take into account the production kinematics, the opacity of the source to neutrons and the effect of propagation. This last factor is the subject to the larger uncertainties, because it has a strong dependence on galactic evolution and on the poorly-known magnetic fields in the Universe. There is some controversy about how to use relationships to constrain the neutrino flux limit. There are however two relevant predictions:
\begin{itemize}
\item{\bf The Waxman-Bahcall upper bound.}
Following Mannheim \cite{mannold}, the upper bound proposed by Waxman-Bahcall  \cite{wb} (W\&B) takes the cos\-mic-ray observations at $E_{CR}\sim 10^{19}$ eV to constrain neutrino flux. With a simple inspection of Fig. \ref{crspectrum}, we can see that  $E^2 dN/dE \sim 10^{-8}$ GeV cm$^{-2}$s$^{-1}$sr$^{-1}$ at $10^{19}$ eV. This flux is two orders of magnitude lower than the limit provided by the extragalactic MeV-GeV $\gamma$-ray background (eq. \ref{eq:boundgamma}). 

In the computation of the upper bound, several hypothesis are made: it is assumed that neutrinos are produced by interaction of protons with ambient radiation or matter; that the sources are transparent to high energy neutrons; that the  $10^{19}$ eV CRs produced by neutron decay are not deflected by magnetic fields; finally (and most important) that the spectral shape of CRs up to the GZK cutoff is $dN/dE \propto E^{-2}$, as typically expected from the Fermi mechanism. The upper limit that they obtain is:
\begin{equation}
E^2_\nu d\Phi/dE_\nu < 4.5 \times 10^{-8} \
(\textrm{GeV}\  \textrm{cm}^{-2} \textrm{sr}^{-1}\textrm{s}^{-1})
\label{eq:wb}
\end{equation}

Although this limit may be surpassed by hidden or optically thick sources for protons to p$\gamma$ or pp(n) interactions, it represents the ``reference'' threshold  to be reached by large volume neutrino detectors (see Fig. \ref{diffuse}). \vskip 0.2cm
\item{\bf Mannheim-Protheroe-Rachen (MPR) upper bound.}
The W\&B limit was criticized as not completely model-inde\-pen\-dent. In particular, the main observation was about the choice of the spectral index $\alpha =2 $. In   \cite{mpr} a new upper bound was derived using as a constraint not only the CRs observed on Earth, but also the observed gamma-ray diffuse flux.
 The two cases of sources $opaque$ or $transparent$ to neutrons are considered; the intermediate case of source partially transparent to neutrons give intermediate limits. 

The limit for sources $opaque$ to neutrons is:
\begin{equation}
E^2_\nu d\Phi/dE_\nu < 2 \times 10^{-6}\
(\textrm{ GeV}\  \textrm{cm}^{-2} \textrm{sr}^{-1}\textrm{s}^{-1})
\label{eq:mpr}
\end{equation}
This is two orders of magnitude higher than the W\&B limit and similar to the EGRET limit on diffuse gamma rays (eq. \ref{eq:boundgamma}), because a source opaque to neutrons produces very few CRs (neutrons cannot escape and cannot decay outside the source), but it is transparent to neutrinos and $\gamma$-rays. This limit was already excluded in a wide energy range by the AMANDA-II experiment, as shown in Fig. \ref{diffuse}.

The limit for sources $transparent$ to neutrons decreases from the value of eq. \ref{eq:mpr} at $E_\nu \sim  10^6$ GeV to the value of eq. \ref{eq:wb} at $E_\nu \sim  10^9$ GeV. Above this energy, the limit increases again due to poor observational information.  
\end{itemize}

The W\&B and the MPR limits for neutrino of one flavor are reported in Fig. \ref{diffuse}. The original values are divided by two, to take into account the neutrino oscillations from the source to the Earth (see \S \ref{oscillations}). 
Experimental upper limits are indicated as solid lines, ANTARES \cite{zornozaPHD} and IceCube 90\% C.L. sensitivities for 3 years with dashed lines. Frejus \cite{rhode}, MACRO \cite{macro_diffuse}, Amanda-II  2000-03 \cite{icecubediffuse} limits refer to muon neutrinos. 
Baikal \cite{baik3} and Amanda-II  UHE 2000-02 \cite{ice23} refer to neutrinos of all-flavors. In this case, the original upper limits are divided by three. In fact, due to neutrino oscillations, we expect at Earth a flux of cosmic neutrinos of all flavors in the same proportion. The red line inside the shadowed band represents the Bartol \cite{bartol} atmospheric neutrino flux. The lowest limit of the band represents the flux from the vertical direction, with a  negligible contribution from prompt neutrinos.
The upper limit of the band represents the flux from the horizontal direction, with one of the prompt model which gives the maximum contribution \cite{fiore}.

Extragalactic neutrino sources at the MPR-limit above 100 TeV cannot be  experimentally excluded yet owing to the Earth occultation.  If the main neutrino production mechanism is the photo-meson production due to the interaction of protons with radiation fields, the neutrino output becomes maximal at  energies above 100 TeV. Due to the increasing neutrino interaction cross-section, the Earth  becomes optically thick (see \S \ref{ref_flux_rate}) to neutrinos above this energy. Ultra High Energy neutrino events must thus be looked for near the horizontal direction.
As we will discuss in \S \ref{effarea}, the experimental neutrino detection probability is expressed by the \textit{effective neutrino area}. This quantity strongly depends from the background suppression capability of the detector for horizontal or downward-going events.

\section{Particle and fundamental physics with neutrino telescopes.}
\label{particleph}
Neutrino detectors can contribute to the multi-messenger astronomy, to solve some of the outstanding problems of high energy astrophysics described in the previous sections. In addition, these experiment will address some of the fundamental questions of high energy physics beyond the standard model: search for relic particles  in the cosmic radiation; what is the nature of the Dark Matter; neutrino oscillations through the ``standard''  mass-flavors  mechanism and with possible subdominant effects, as those induced by the violation of the Lorentz invariance or of the equivalence principle. 

\subsection{Relic particles in the cosmic radiation}\label{relic}

\subsubsection{Magnetic Monopoles}\label{monopoles}
Most of the Grand Unified Theories (GUTs) predict the creation of magnetic monopoles (MM) in the early Universe \cite{thooft,polyakov}.
MM are topologically stable and carry a magnetic charge defined as a multiple integer of the Dirac charge $g_D = \hbar c/2e$, where $e$ is the elementary electric charge, $c$ the speed of light in vacuum and $\hbar$ the Planck constant. Depending on the GUT group, the masses inferred
for magnetic monopoles can take range over many orders of magnitude, from $10^8$ to $10^{17}$ GeV.

MM are stable particles and they would have survived until now, diluted in the Universe, as predicted by theoretical studies which set limits on their fluxes, like the Parker flux limit \cite{parker}. Stringent limits over a very wide MM velocity range were set by the underground MACRO experiment \cite{macroMM}. 
\begin{figure}[tbh]
\vspace{6.5cm}
\includegraphics{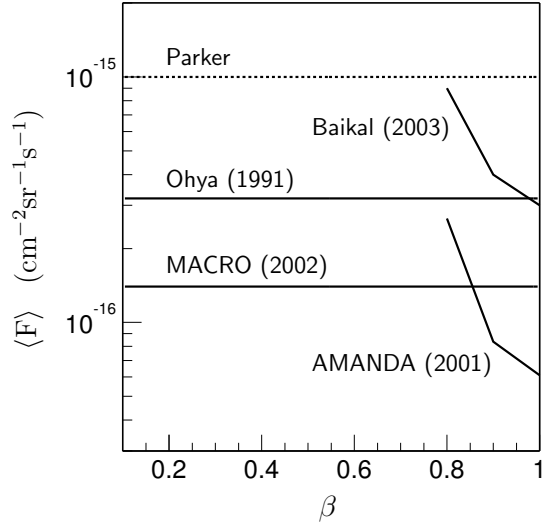}
  \caption{\it
Experimental upper limits (90\% C.L.) on the MM flux as a function of the MM velocity $\beta c$ (solid lines). The Parker \cite{parker} bound is also shown (dashed line). From \cite{bvr}.
    \label{fig:mm}}
\end{figure}

Neutrino detectors can be used to search for fast MM. Since fast MM have a large interaction with matter, they can lose large amounts of energy in the terrestrial environment. The total energy loss of a relativistic MM with one Dirac charge is of the order
of $10^{11}$ GeV \cite{derkaoui} after having crossed the full Earth diameter. Because MM are expected to be accelerated in the galactic coherent magnetic field domain to energies of about $10^{15}$ GeV \cite{wick}, some could be able to cross the Earth and produce 
an upgoing signals in a neutrino detector. 

The monopole magnetic charge $g = m g_D$ can be expressed as an equivalent electric charge $g = 68.5\ me$, where $m$ is an integer. In a medium with a refractive index $n$, a MM with $\beta=v/c > 1/n$ emits a factor $(gn/ze)^2$ more Cherenkov light than an electrical charge $ze$ with the same velocity. Thus relativistic MM with $\beta \ge 0.74$ carrying one Dirac charge will emit a large amount of direct Cherenkov light when traveling through the neutrino detector; in water ($n=1.35$) it gives rise to $\sim $ 8500 more intense light than a muon. Fig. \ref{fig:mm} show the limits on MM from the Baikal \cite{baikalMM}  and AMANDA \cite{amandaMM} Cherenkov neutrino telescope, together with the best limits from other experiments \cite{macroMM,ohya}.

\subsubsection{Nuclearites}\label{nuclearites}
Nuclearites are hypothetical nuggets of strange quark matter that could be present in cosmic radiation. Their origin is related to energetic astrophysical phenomena. Down-going nuclearites could reach a neutrino telescope with velocities $\sim$ 300 km/s, emitting blackbody radiation at visible wavelengths while traversing water/ice. Nuclearites with $M \ge 10^{10}$ GeV would be electrically neutral; the small positive electric charge of the quark core would be neutralized by electrons forming an electronic cloud. 

The relevant energy loss mechanism is represented by the elastic collisions with the atoms of the traversed media \cite{derujolanuc}.
Nuclearites moving into the water/ice could be detected because of the black-body radiation emitted by the expanding cylindrical thermal shock wave. The luminous efficiency (defined as the fraction of dissipated energy appearing as light) was estimated, in the case of
water, to be $\sim 3 \times 10^{-5}$ \cite{derujolanuc}. The best limits on the search for nuclearites in the cosmic radiation are from the MACRO experiment \cite{macroNUCL}.  
Because nuclearites are slowly moving particles, light coming from the passage of a nuclearite on a km-scale neutrino detector may span into intervals from tens of $\mu $s up to 1 ms. Preliminary results were presented by the ANTARES collaboration \cite{antaresNUC}.

\subsection{Indirect dark matter searches}\label{wimps}
The existence of nonbaryonic Dark Matter (DM) in our Universe is supported by strong cosmological observational evidences \cite{dm1}.  The presence of non-visible mass in galaxies is also motivated by the fact that dark matter halos seem to help stabilize spiral disk structure. Nonbaryonic  DM may consist of Weakly Interacting Massive Particles (WIMPs). A long list of nonbaryonic cold (i.e. non-relativistic) dark matter candidates has been suggested, among which the supersymmetric (SUSY) neutralino and the axion seem to be the most promising. SUSY postulates a symmetry between bosons and fermions predicting SUSY partners \cite{dm2}. In theories where R parity is conserved there exists a stable lightest supersymmetric particle (LSP). If the neutralino is the LSP, it is a natural WIMP candidate: it is a weakly interacting particle with a mass between roughly a GeV and a TeV and would be expected to have a significant relic density.

From the experimental point of view, ''direct'' and ''indirect'' methods \cite{dm3} exist for detecting the WIMPs in the galactic halo. The methods can probe complementary regions of the supersymmetric parameter space, even when more extensive LHC results will become available. Direct methods detect Weakly Interacting Particles via the elastic scattering of the WIMP with a nucleus: the energy deposited in a low-background detector  can be measured. Indirect methods look for by-products of WIMP decay or annihilation such as neutrinos resulting from the annihilation of WIMPs. Neutrino telescopes can perform indirect WIMP searches looking for high energy neutrinos from WIMP annihilation in the core of the Earth or the Sun. 

Dark matter WIMPs existing in the galactic halo can be captured in a celestial body by losing energy through elastic collisions and becoming gravitationally trapped. As the WIMP density increases in the core of the body, the WIMP annihilation rate increases until equilibrium is achieved between capture and annihilation. High energy neutrinos are produced via the hadronization and decay of the annihilation products (mostly
fermion-antifermion pairs, weak and Higgs bosons) and may be detected as upward-going muons in neutrino telescopes. 

The capture rate for an astrophysical body depends, apart from the mass of the celestial body and from the escape velocity, on several poorly known factors: the WIMP mean halo velocity, the WIMP local density and the WIMP scattering cross section.  The WIMP may scatter from nuclei with spin (hydrogen in the Sun) via an axial-vector spin-dependent interaction. In this case, the WIMP couples to the spin of the nucleus or via a scalar interaction in which the WIMP couples to the nuclear mass.
Elastic scattering is most efficient when the mass of the WIMP is similar to the mass of the scattered nucleus. Hence, the heavy nuclei in the Earth core make it very efficient in capturing WIMPs with $m_X \le 100$ GeV. Nuclei in the Sun, in contrast, have a smaller average mass; the Sun is nevertheless efficient in capturing WIMPs due to the larger value of the  escape velocity.

As studied by underground experiment like Super-Kamio\-kande \cite{SKDM}, MACRO \cite{macroDM} and others, WIMP annihilation signal would appear in a neutrino telescope as a statistically significant excess of upward-going muon events from the direction of the Sun or of the Earth among the background of atmospheric neutrino-induced upward-going muons.
The precise direction measurement allows a restriction of the search for WIMP annihilation neutrinos to a narrow cone pointing from the Earth center or from the Sun, greatly reducing the background. This detection method achieves an increasingly better signal to noise ratio for high WIMP masses, because of the increase in neutrino cross section with energy and longer range of high energy muons.

At present AMANDA \cite{icecubeDM} and Baikal \cite{BaikalDM} have published results on indirect search of WIMPs, while ANTARES \cite{antaresDM} has presented preliminary results. Instead of the general supersymmetry scenario, in AN\-TA\-RES the more constrained approach of minimal supergravity (mSUGRA) was used. mSU\-GRA models are characterized by four free parameters and a sign: $m_{1/2}$, $m_0$, $A_0$, $\tan(\beta)$ and sgn$(\mu)$. They investigated mSUGRA models that possess a relic neutralino density that is compatible with the cold dark matter density $\Omega_X$ as measured by the WMAP experiment.
No excess is found, and the sensitivity has been sufficient to put constraints on parts of the mSUGRA parameter space. 

Both the IceCube and a cubic kilometer experiment in the Mediterranean sea would be sensitive to a wide part of the mSUGRA parameter phase-space. 

\subsection{Neutrino oscillations}\label{oscillations}

In recent years, neutrino oscillation became a well known phenomenon, which plays also an important role on determining the flavor on Earth of   neutrinos of cosmic origin. Neutrino oscillations were observed in atmospheric neutrinos, in solar neutrino experiments and on Earth based accelerator and reactor experiments. A complete review about neutrino oscillations can be found in \cite{nuosci}.

As already mentioned, high energy neutrinos are produced in astrophysical sources mainly through the decay of charged pions, in $p \gamma$, pp, pn interactions (eq. \ref{pipm}). Therefore, neutrino fluxes of different flavors are expected to
be at the source in the ratio:
\begin{equation}
\nu_e : \nu_\mu : \nu_\tau = 1\ :\ 2\ :\ 0
\label{nusurce}
\end{equation}

Neutrino oscillations will induce flavor changes while neutrinos propagate through the Universe. One has to consider {\it mass
eigenstates} $\nu_m=\nu_1,~\nu_2,~\nu_3$ in the propagation, instead of {\it weak flavor eigenstates} $\nu_l=\nu_e,~\nu_\mu,~\nu_\tau$. 
The weak flavor eigenstates $\nu_l$ are linear combinations of the mass 
eigenstates $\nu_m$ through the elements of the mixing matrix $U_{lm}$:
\begin{equation}
\nu_l = \sum_{m=1}^3 U_{lm}\ \nu_m
\end{equation}
Because mixing angles are large, the flavor eigenstates are well separated from those of mass.
The oscillation probability in the simple case of only two flavors, for instance $(\nu_\mu,~\nu_\tau)$ and one mixing angle $\theta_{23}$, is:
\begin{equation}
\left\{ \begin{array}{ll}
      \nu_\mu =~\nu_2 \cos\ \theta_{23} + \nu_3 \sin\ \theta_{23} \\
      \nu_\tau=-\nu_2\sin\ \theta_{23} + \nu_3\cos\ \theta_{23} 
\end{array} 
\right. 
\end{equation} 

The survival probability for a pure a $\nu_\mu$ beam:
\begin{equation}
P(\nu_\mu \rightarrow \nu_\mu) = 
1- \sin^2 2\theta_{23}~\sin^2 \left( { 
{1.27 \Delta m^2 \cdot L}\over {E_\nu}} \right)
\label{eq:posci}
\end{equation}

\noindent where $\Delta m^2=m^2_3-m^2_2$ (eV$^2$), $L$ (km) is the distance travelled by the neutrino from production to detection and $E_\nu$ (GeV) the neutrino energy. $\theta_{23}$ and $\Delta m^2$ may be experimentally determined from the variation of $ P(\nu_\mu \rightarrow \nu_\mu)$ as a function of the zenith angle or from the variation in $L/E_\nu$.

With three neutrino flavors, three mass differences can be defined (two linearly independent). The mass difference measured with atmospheric neutrinos is
$\Delta m^2_{atm}\simeq$ $\Delta m^2_{23}$ $= \pm 2.5\times 10^{-3}$ eV$^2$. For the mixing angle, $\theta_{23} \simeq 45^o$ (that correspond to maximal mixing), while  $\theta_{13}$ is small. 
The values of $\Delta m^2_{12}$ and of the other mixing angle $\theta_{12}$ are determined by the solar neutrino experiments and KamLand. The most recent data favor \cite{pdg}  very clearly the solution with a best fit: 
$\Delta m^2_{sol}\simeq \Delta m^2_{12} = 7.6\times 10^{-5} eV^2$ and $\sin^2\theta_{12}=0.32 \ (\theta_{12}=34.4^o)$.

According to these neutrino oscillation parameters, the ratio of fluxes of neutrinos from astrophysical origin (i.e. very large baseline $L$) in eq. \ref{nusurce}  changes to a flux ratio at Earth  \cite{learpak} as:
\begin{equation}
\nu_e : \nu_\mu : \nu_\tau = 1\ :\ 1\ :\ 1
\label{nuearth}
\end{equation}

Theoretical predictions which does not take into account neutrino oscillations must be corrected to include this effect (as we did in Fig.    \ref{diffuse} ). In particular, muon neutrinos are reduced at Earth by a factor of two.

Neutrino telescopes can detect thousand of atmospheric $\nu_\mu$ per year. For instance, the ANTARES detector is efficient for $E_\nu >$ few tens of GeV, were atmospheric upward-going $\nu_\mu$ are still suppressed by flavor oscillations.  
Neutrino telescopes can test also non-standard oscillations. If the standard mass-induced oscillation is assumed as the leading process for flavor changes, other  mechanisms can be tested for  flavor
transitions as a subdominant effect. As an example, we discuss in the following the case of the Lorentz invariance \cite{lore1}.

\subsection{Violation of the Lorentz invariance}\label{lorentz}
Quantum gravity theories assume that the space–time take a foa\-my
nature \cite{lore2}. Interactions with this space–time foam may lead to the breaking of CPT symmetry, leading to the violation of Lorentz invariance \cite{lore3}. In addition, some theories of quantum gravity predict that there is a minimum length scale, of order the Planck length ($10^{-35}$ m): the existence of a fundamental length scale may
also induce the violation of Lorentz invariance (VLI).

Lorentz invariance violation may manifest itself in many different
ways and can be tested with many different experimental systems, in particular with  a modified neutrino oscillation length\footnote{In the literature, neutrino oscillations induced by the simplest models of VLI and violation of the equivalence principle (VEP) are described within the same formalism. In the following we will mention only this VLI formalism for simplicity. We refers to \cite{lore2} for the more general case.}. 

The VLI subdominant effect can be studied using atmospheric neutrino data collected by neutrino telescopes, after correction for the known mass-induced neutrino oscillation. In this scenario \cite{battistoni}, neutrinos can be described in terms of three distinct bases: flavor eigenstates, mass eigenstates and velocity eigenstates, the latter being characterized by different maximum attainable velocities (MAVs) in the limit of infinite momentum. Here, both mass-induced oscillations and VLI transitions are treated in the two-family approximation. It is also assumed that mass and velocity mixings occur inside the same families (e.g., $\nu_2$ and $\nu_3$). In the case of mass-flavor oscillation, the survival probability of muon neutrinos at a distance $L$ from production is given by eq. \ref{eq:posci}. 
In the VLI case, the $\nu_\mu$ survival probability is:
\begin{equation}
P(\nu_\mu \rightarrow \nu_\mu) = 
1- \sin^2 2\theta_{v}~\sin^2 \left(  
{2.54\times 10^{18} \Delta v L {E_\nu}} \right)
\label{eq:osciVLI}
\end{equation}
\noindent where $\Delta v = (v_{\nu_3} - v_{\nu_2}$) is the neutrino MAV difference in units of c and $\theta_v$ is the mixing angle. Notice that neutrino flavor oscillations induced by VLI are characterized by an $L E_\nu$ dependence of the oscillation probability (eq. \ref{eq:osciVLI}), to be compared with the $L/E_\nu$ behavior of mass-induced oscillations (eq. \ref{eq:posci}).

When both mass-induced transitions and VLI induced transitions are considered simultaneously, the muon neutrino survival probability can be expressed as 
\begin{equation}
P(\nu_\mu \rightarrow \nu_\mu) = 
1- \sin^2 2\Theta~\sin^2 \Omega  
\label{eq:osciMA-VLI}
\end{equation}
\noindent where $2\Theta =atan(a_1/a_2)$ and $\Omega = \sqrt{a_1^2+a_2^2}$, with $a_1$ and $a_2$ which depend from $\Delta m^2, \theta_{23}, \Delta v, \theta_v, L$ and $E_\nu$ \cite{battistoni}.

The same formalism also applies to violation of the equivalence principle, after substituting $\Delta v/2$ with the adimensional product $\mid \phi\mid \Delta \gamma$ ; $\Delta \gamma$ is the difference of
the coupling constants for neutrinos of different types to the gravitational potential $\phi$ \cite{lore4}.

The most conservative bounds from underground experiments were obtained by Super-Kamiokande \cite{lore1}  and MACRO \cite{battistoni}. In particular, the 90\% confidence level limits obtained by MACRO are  $|\Delta v| < 6 \times 10^{-24}$ at sin$2\theta_v=0$ and $|\Delta v| < 2.5-5 \times 10^{-26}$ at sin$2\theta_v=\pm 1$. 
The typical energy scale of upward thoroughgoing neutrino-induced muons measured from these experiments is of the order of $E_\nu \sim 10\div 100$ GeV.
Neutrino telescopes are sensitive to much higher neutrino energies. The larger number of atmospheric neutrino events and the greater average energy allows neutrino telescopes to be much more sensible. 
The AMANDA-II expected sensitivity (90\% C.L.) for maximal mixing and six years of simulated data is $|\Delta v| < 2.1 \times 10^{-27}$ \cite{amandaLI}.
A null observation would be able to place very stringent bounds on quantum decoherence effects, and on VLI parameters which modify the dispersion relation for massive neutrinos. 

\subsection{HE neutrinos in coincidence with gravitational waves}\label{gw}
High-energy neutrinos and gravitational waves (GW), contrarily to high-energy photons and charged cosmic rays may escape from dense astrophysical regions and travel over large distances without being absorbed, pointing back to their emitter.

It is expected that many astrophysical sources produce both gravitational waves and HE neutrinos \cite{elewyck}. A possible coincident detection will then provide important information on the processes at work in the astrophysical accelerators. 
Furthermore, if a mechanism (as for instance the gamma-ray bursts) allows a precise measurement of the time delay between neutrinos and GW signals, some quantum-gravity effects can be tested with the possibility to constrain some dark energy models \cite{gw1}. 

The search for coincident detection is motivated by the advent, in association of neutrino telescopes, of a new generation of GW detectors VIRGO \cite{virgo} and LIGO \cite{ligo} (which are now part of the same experimental collaboration). 
The neutrino/GW association requires the measurement of the event time, arrival direction and associated angular uncertainties. Each candidate is obtained by the combination of reconstruction algorithms specific to
each experiment, and quality cuts used to optimize the signal-to-background ratio. A joint neutrino/GW analysis program is planned for the ANTARES neutrino telescope and VIRGO+ LIGO \cite{elewyck}. 
The coincident (non-)obser\-va\-tion shall play a critical role in our understanding of the most energetic sources of cosmic radiation and in constraining existing models. They could also reveal new, \textit{hidden} sources unobserved so far by conventional photon astronomy.

\section{Neutrino detection principle}\label{detection}

The basic idea for a neutrino telescope is to build a matrix of light detectors inside a transparent medium. This medium, such as deep ice or water:
\begin{itemize}
\item offers large volume of free target for neutrino interactions;
\item provide shielding against secondary particles produced by CRs;
\item allows transmission of Cherenkov photons emitted by relativistic particles produced by the neutrino interaction.
\end{itemize}

Other possibilities, such as detecting acoustic or radio signals generated by EeV ($10^{18}$ eV) neutrinos in a huge volume of water or ice are not considered in this review \cite{sapric}. 

High energy neutrino interact with a nucleon $N$ of the nucleus, via either charged current (CC) weak interactions
\begin{equation}
\nu_l + N \rightarrow l+X
\label{cc}\end{equation}
\noindent or neutral current  (NC) weak interactions
\begin{equation}
\nu_l + N \rightarrow \nu_l +X \ .
\label{nc}\end{equation}

\begin{figure}[t]
\vspace{6.5cm}
\includegraphics{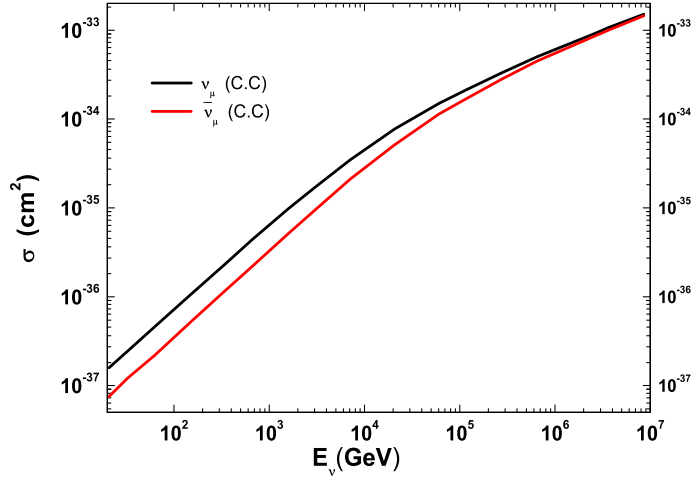}
  \caption{\it
    Cross section for $\nu_\mu$ and $\overline\nu_\mu$  as a function of the (anti)-neutrino energy according to CTEQ6-DIS  \cite{crossunc} parton distributions. 
    \label{nusigma} }
\end{figure}

At energies of interest for neutrino astronomy, the leading order differential cross section for the $\nu_l N \rightarrow l X$ CC interactions is given by  \cite{gandi}

\begin{eqnarray}
{d^2\sigma_{\nu N} \over dxdy} &=& {2G_F^2 m_N E_\nu \over \pi}
{M^4_W \over (Q^2+ M^2_W)^2} \nonumber \\
&\times & [xq(x,Q^2)+x(1-y)^2 \overline q(x,Q^2)]
\end{eqnarray}

\noindent where $x = Q^2/2m_N(E_\nu - E_l)$ and $y = (E_\nu-E_l)/E_\nu$ are the so-called scale variables or Fenyman-Bjorken variables, $Q^2$ is the square of the momentum transferred between the neutrino and the lepton, $m_N$ is the nucleon mass, 
$M_W$ is the mass of the W boson, and $G_F$ is the Fermi coupling constant.
The functions $q(x,Q^2)$ and $\overline{q}(x,Q^2)$ are the parton distributions for quarks and antiquarks. 
Fig. \ref{nusigma} shows the $\nu_\mu$ and $\overline \nu_\mu$ cross sections as a function of the neutrino energy. As can be seen, at low energies the neutrino cross section rises linearly with $E_\nu$ up to $\sim 10^4$ GeV. For higher energies, the invariant mass $Q^2 = 2m_NE_\nu xy$ could be larger than the W-boson rest mass, reducing the increase of the total cross section. Since there is not data which constrain the structure functions at very small $x$, outside the range measured with high precision at the HERA collider, some uncertainties are estimated on the total cross section at large energies  \cite{crossunc}. Computer libraries \cite{pdf} provide a collection of parton distribution function (PDF) to model the neutrino cross section also at very high energies.

\begin{figure}[tbh]
 \vspace{8.5cm}
 \includegraphics{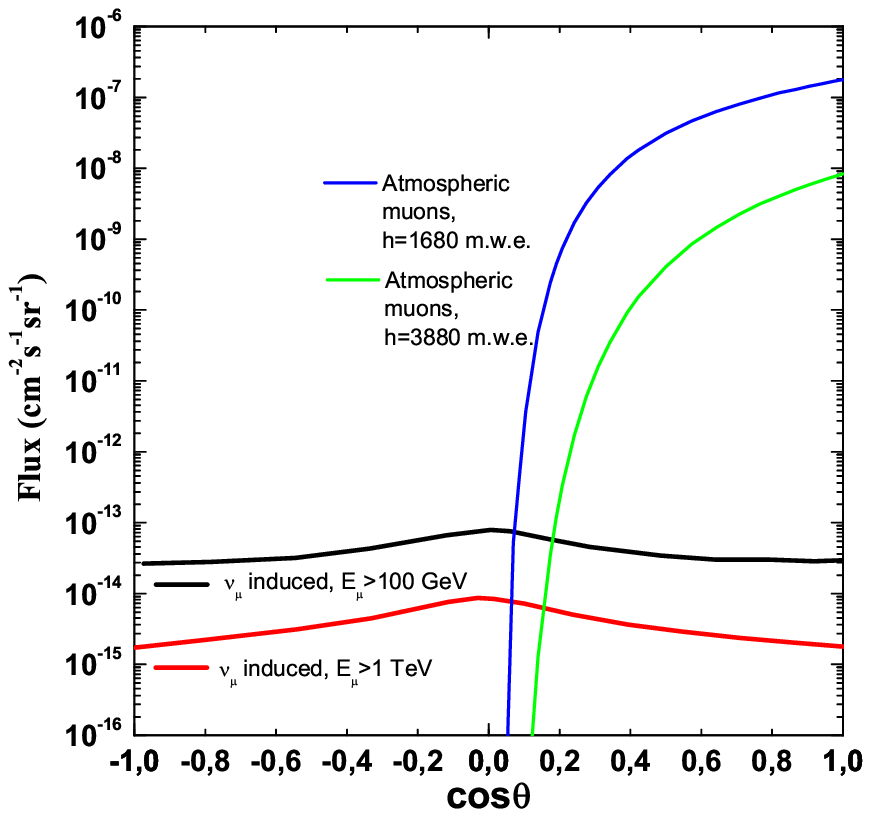}
  \caption{\it
Different contributions (as a function of the cosine of the zenith angle) of  the: $i)$ atmospheric muons (computed according to \cite{mupage0}) for two different depths; ii) atmospheric neutrino induced muons (from \cite{bartol}), for two different muon energy thresholds. 
    \label{atmunu} }
\end{figure}

Cosmic neutrino detectors are not background free. Showers induced by interactions of cosmic rays with the Earth's atmosphere produce the so-called {\it atmospheric muons} and {\it atmospheric neutrinos}. Atmospheric muons can penetrate the atmosphere and up to several kilometers of ice/water. Neutrino detectors must be located deeply under a large amount of shielding in order to reduce the background. 
The flux of down-going atmospheric muons exceeds the flux induced by atmospheric neutrino interactions by many orders of magnitude, decreasing with increasing detector depth, as is shown in Fig. \ref{atmunu}.
The previous generation of experiments which had looked  also for astrophysical neutrinos (MACRO \cite{macro}, Super-Kamiokande \cite{sk}) was located under mountains, and has reached almost the maximum possible size for underground detectors. 


Charged particles travel through the medium until they either decay or interact. The mean length of the distance travelled is called the \textit{path length} of the particle and it depends on  its energy loss in the medium. If the path length exceeds the spatial resolution of the detector, so that the trajectory of the particle can be resolved, one have a \textit{track}. In a high energy neutrino detector, one can distinguish between two main event classes:  events with a track, and events without a track (showers). 

Relativistic charged particles emit Cherenkov radiation in the transparent medium. A detector must measure with high precision the number and arrival time of these photons on a three-dimensional array of Photo Multiplier Tubes (PMTs), from which some of the properties of the neutrino (flavor, direction, energy) can be inferred. 

In order to behave as a neutrino \textit{telescope}, a neutrino \textit{detector} must be able to point at a specific celestial region if a signal excess over the background is found.
 Neutrino telescopes must have the same peculiarities of GeV-TeV $\gamma$-ray experiments (satellites, imaging Cherenkov) to associate some of the signal excesses to objects known in other electromagnetic bands. In order to achieve  an angular resolution of a fraction of degree, only the CC $\nu_\mu$ interaction can be used. The angular resolution for other flavors and for NC is so poor that there is no possibility to perform associations. For the same reason, the particle physics and general physics open questions which can be afforded with a neutrino telescope largely rely on the $\nu_\mu$ channel.

On the other hand, a high energy neutrino detector is motivated by discovery and must be designed to detect neutrinos of all flavors over a wide energy range and with the best energy resolution. This is of particular interest for the case of the neutrino diffuse flux from extragalactic sources.
In addition, the neutrino oscillation changes the source admixture from $\nu_e : \nu_\mu : \nu_\tau = 1: 2 : 0$ to $1 : 1 : 1$.  While above hundreds of TeV muon and electron neutrinos become absorbed by the Earth, the tau neutrino is \textit{regenerated} \cite{halzen_tau}: high energy $\nu_\tau$ will produce  a secondary $\nu_\tau$ of lower energy, lowering its energy down to $10^{15}$ eV, where the Earth is transparent.

\begin{figure}[tbh]
  \vspace{5.0cm}
   \includegraphics{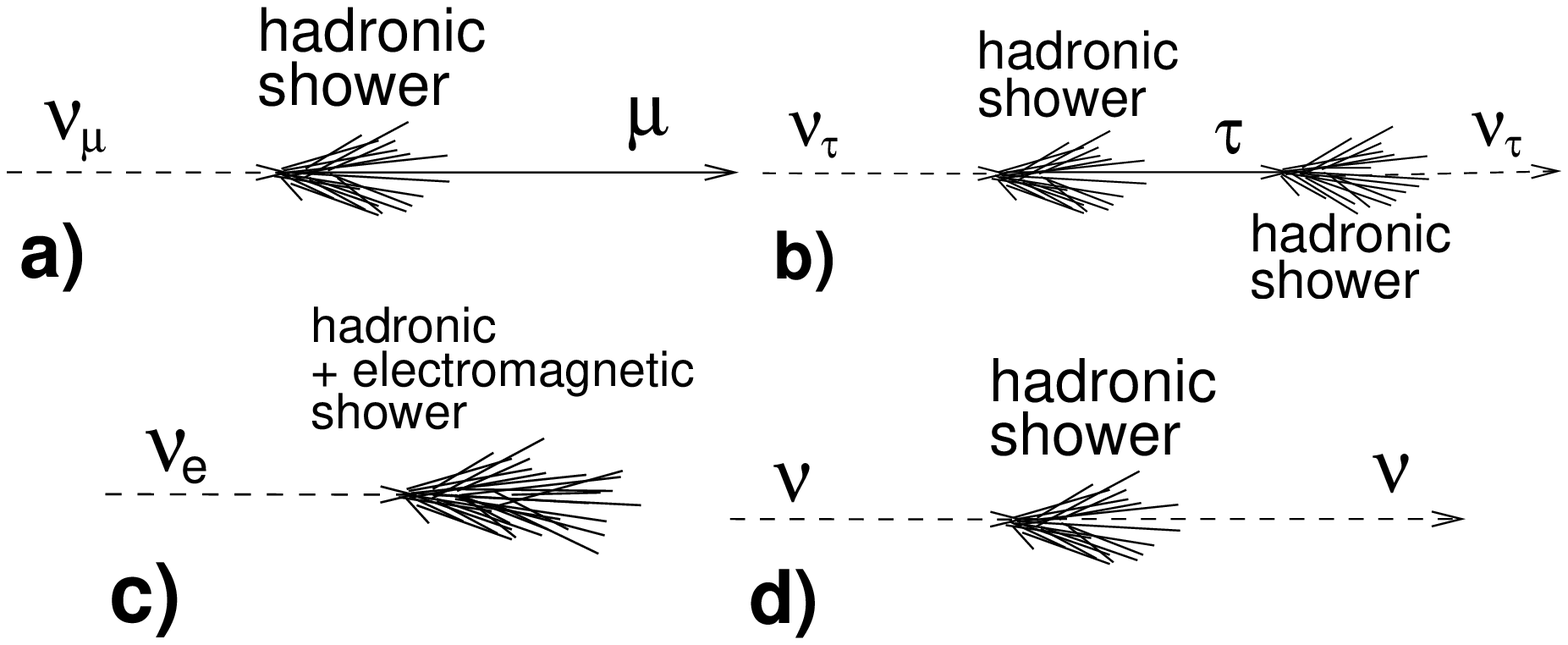}
  \caption{\it
Some event signature topologies for different neutrino flavors and interactions: 
a) CC interaction of a $\nu_\mu$ produces a muon and a hadronic shower; 
b) CC interaction of a $\nu_\tau$ produces a $\tau$ that decays into a $\nu_\tau$, tracing the double bang event signature. 
c) CC interaction of  $\nu_e$ produces both an EM and a hadronic shower; d) a NC interaction produces a hadronic shower.
Particles and anti-particles cannot be distinguish in neutrino telescopes. From  \cite{bettinathesis}
    \label{sketch} }
\end{figure}

Schematic views of a $\nu_e,\nu_\mu$ and $\nu_\tau$ CC events and of a NC event are shown in Fig. \ref{sketch}. Neutrino and anti-neutrino reactions are not distinguishable; thus, no separation between particles and anti-particles can be made. Showers occur in all event categories shown in Figure. However, for CC $\nu_\mu$, often only the muon track is detected, as the path length of a muon in water exceeds that of a shower by more than 3 orders of magnitude for energies above 2 TeV. Therefore, such an event might very well be detected even if the interaction has taken place several km outside the instrumented volume, provided that the muon traverses the detector. 

Neutrino telescopes, at the contrary of usual optical telescopes, are 'looking downward'. Up-going muons can only be produced by interactions of (up-going) neutrinos. From the bottom hemisphere, the neutrino signal is almost background-free.
Only atmospheric neutrinos that have traversed the Earth, represent the irreducible background for the study of cosmic neutrinos. The rejection of this background depends upon the pointing capability of the telescope and its possibility to estimate the parent neutrino energy. 
As we will discuss in \S \ref{waterice}, either water or ice is used as media.
A deep sea-water telescope has some advantages over ice and lake-water experiments due to the better optical properties of the medium. However, serious technological challenges must be overcome to deploy and operate a detector in deep sea, as we will discuss in \S  \ref{nutel}.

\subsection{Electron neutrino detection}\label{nu_e_det}

A high energy electron resulting from a charged current $\nu_e$  interaction has a high probability to radiate a photon via brems\-strah\-lung after few tens of cm of water/ice (the water radiation lenght is $\sim$ 36 cm). The following process of $e^+e^-$ pair productions, and subsequent brems\-strah\-lung, rapidly  produce an electromagnetic (EM) shower until the energy of the constituents falls below the critical energy $E_c$ and the shower production stops; the remaining energy is then dissipated by ionization and excitation.

The Cherenkov light from EM showers is emitted isotropically in azimuth with respect to the shower axis. The lateral extension of a EM shower is of the order of 10 cm \cite{pdg}  and therefore negligible compared to the longitudinal one. Thus, the EM shower is described by the longitudinal shower profile (a parametric formula is given in \cite{pdg}). 
The longitudinal profiles are used to parameterize the total shower length $L$ as a function of the initial shower energy. The shower length is defined as the distance within which 95\% of the total shower energy has been deposited. For salt water \cite{bettinathesis} it is found that 
$L \textrm{[m]} = 3.04 + 1.09 \log_{10}(E \textrm{[GeV]})$. For a 10 TeV electron  $L$ is 7.4 m 

A showers size of order of 10 m is small compared to the spacing of the PMTs in any existing or proposed detector. EM showers represent, to a good approximation, a point source of Cherenkov photons. Pointing  accuracy for showers is inferior to that can be achieved for the $\nu_\mu$ channel. Reconstruction of Monte Carlo simulated events performed  in the framework of the IceCube and ANTARES collaborations shows a precision of the order of $\sim 10^o$, with the possibility to reduce it to few degrees for a small subsample of events. 

Finally, we should mention (only for completeness)  two effects which has some consequences on EM shower: 

\noindent - the Glashow resonance \cite{glas_res}, which affects the $\overline \nu_e$ through the resonant process 
$\overline \nu_e + e^- \rightarrow W^- \rightarrow q + \overline q', (\overline \nu_\ell + \ell)$. The  resonance peak is for a $\overline \nu_e$ energy of 6.3 PeV. This resonant channel constitutes only a small portion to the overall cross section in the energy range between 100 GeV and  100 PeV and it must be taken into account in Mont Carlo simulations; 

\noindent - the Landau- Pomeranchuk - Migdal (LPM) effect \cite{bib@LandPom,bib@Mig}, which  has an influence on showers of ultra high energies ($E > 10^{16}$ eV). The LPM effect suppresses the radiative energy losses of the particles in the shower; in this case, the longitudinal development of  EM or hadronic cascades can be largely enhanced. 

\begin{figure}[tbh]
  \vspace{7.0cm}
   \includegraphics{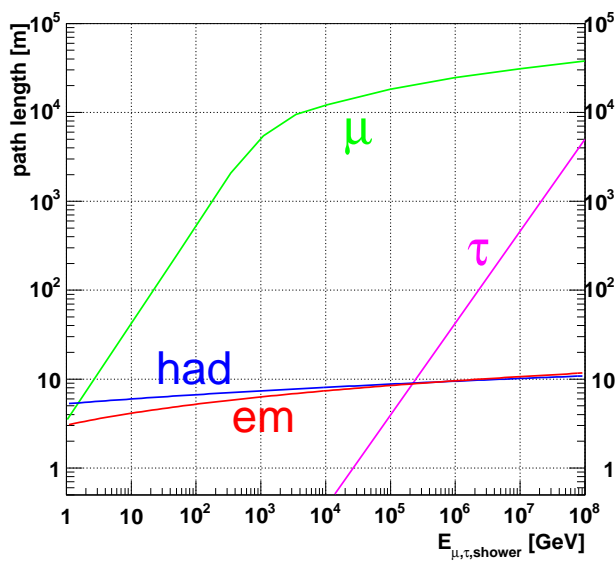}
  \caption{\it
Path length of particles produced by neutrino interactions in water: muons, taus, electromagnetic and hadronic showers, versus their respective energy. The shower lengths are calculated using a shower profile parameterization as described in  \cite{bettinathesis}.
    \label{nu_path} }
\end{figure}

\subsubsection{Neutral currents interactions}\label{nu_nc_det}
The NC channel gives the same signature for all neutrino flavors. In this channel, a part of the interaction energy is always carried away unobserved by the outgoing neutrino, and therefore the error on the reconstructed energy of the primary neutrino increases accordingly. 
Even though EM and hadronic showers are different from each 
other in principle, the $\nu_e$ CC and the $\nu_x$ NC channels are not 
distinguishable in reality, because any proposed detector is too sparsely instrumented. 

Hadronic cascades suffer event-to-event fluctuations which are much more important with respect to the EM ones \cite{bettinathesis}. The dominant secondary particles in a hadronic shower are pions; kaons, protons or neutrons occur in variable fractions. Muons (from pions decay) can be present as well: as these usually leave the shower producing long tracks, they contribute significantly to the fluctuations. 

Monte Carlo simulations (for instance, from the ANTARES collaboration \cite{brunner}) shows that above 1 TeV of shower energy, the largest part of the Cherenkov light  is generated by EM sub-showers. 
Referring to Fig. \ref{nu_path}, the EM shower has a shorter length than the hadronic shower below 100 TeV. The electromagnetic component (produced by the $\pi^0$ decay) in the hadronic shower increases with increasing shower energy. The longitudinal profile of hadronic showers can be parameterized in the same way as the EM one. The shower length $L_H$ is again defined as the distance within which 95\% of the total shower energy is deposited. It can be described as: 
$L_H \textrm{[m]} = 5.28 + 0.70 \log_{10} (E \textrm{[GeV]})$.  
For what concerns the measurement of the incoming neutrino direction,  the angular difference between the shower and the neutrino falls below $2^o$ for $E_\nu$ above $\sim$ 1 TeV. It is thus negligible with respect to the precision of the shower direction measurement. 

\subsection{Tau neutrino detection}\label{nu_tau_det}

For $\nu_\tau$ CC interactions, the produced $\tau$-lepton travels some distance (depending on its energy) before it decays and produces a second shower. The Cherenkov light emitted by the charged particles in the showers can be detected if both the $\nu_\tau$ interaction and the $\tau$ decay occur inside the instrumented volume of the detector. 
Below 1 PeV, also the $\nu_\tau$ CC channels (except for the case where the $\tau$ produces a muon) belong to the class of events without a track, because the $\tau$ track cannot be resolved.

The $\tau$ lepton has a short lifetime, and in the energy range of interest it travels (depending on the Lorentz factor $E_\tau/m_\tau c^2$) from a few meters up to a few kilometers before it decays (see Fig. \ref{nu_path}). 
Radiation losses, on the other hand, play a much smaller role than for the muon, because of its 17 times larger mass. Most of the possible $\tau$  decay modes include the generation of a hadronic or an electromagnetic cascade. Thus, if the track of the $\tau$  is long enough to distinguish between the primary interaction of the $\nu_\tau$   and the decay of the tau (typically for $\tau$ energies above  1 PeV, see 
Fig. \ref{nu_path}), the expected signatures for the $\nu_\tau$ CC events are that of a shower, plus a track, plus another shower. This signature is called \textit{double bang event}, if the Cherenkov light emitted by the charged particles in the first shower can be detected and separated from the light emitted by the particles produced in the $\tau$ decay.

Alternatively, if the $\tau$ starts or ends outside the instrumented volume, a track plus a shower can be detected. This signature is called \textit{lollipop event}. In small size neutrino detector (like ANTARES) 
the expected \textit{lollipop} event rate above 1 PeV is far below 1 event per year. For larger detector (IceCube, the 1 km$^3$ Mediterranean sea telescope) the optimal $\nu_\tau$ energy value for \textit{double bang} events is around $\sim 10^{16}$ eV, because the tau path length rapidly exceeds the dimensions of the detectors for increasing energies. 
If the $\tau$ decays into a muon, the event is presumably not distinguishable from an original $\nu_\mu$ CC interaction.  

\subsection{Muon neutrino detection}\label{nu_mu_det}

\begin{figure}[tbh]
\vspace{7.0cm}
\includegraphics{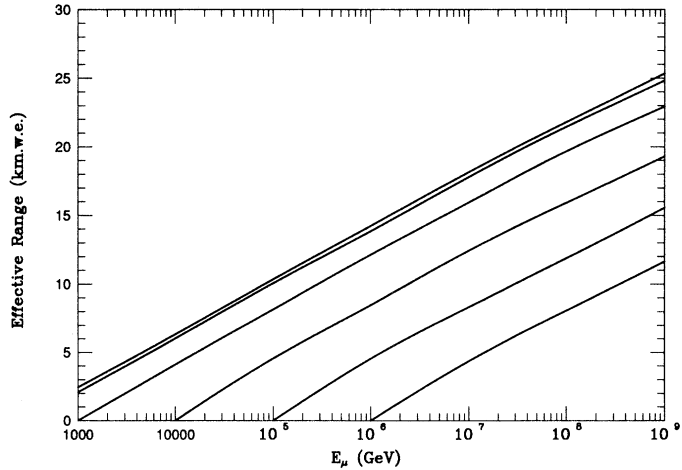}
  \caption{\it
Effective muon range as a function of the initial energy $E_\mu$. Curves correspond (from top to bottom) to different threshold energies $E_{thr}^\mu$ of the muon arriving at the detector ($E_{thr}^\mu =1,10^2,10^3,10^4,10^5,10^6$ GeV). From  \cite{lista}.
    \label{murange} }
\end{figure}

Muon neutrinos are especially interesting in a search for cosmic point sources of neutrinos with energies larger than $\sim$ 1 TeV. In this energy range, $\nu_\mu$ interaction can occur outside the detector volume, while in most cases muons are energetic enough to completely traverse the detector. This gives a clean experimental signal which allows accurate reconstruction of muon direction, closely correlated with the neutrino direction.

The relation between neutrino and muon directions is essential for the concept of a neutrino telescope. Since neutrinos are not deflected by (extra-) galactic magnetic fields, it is possible to trace the muon back to the neutrino source. This is equivalent to traditional astronomy where photons point back to their source.
The average angle $\theta_{\nu \mu}$ between the incident neutrino and the outgoing muon can be approximated by:
\begin{equation}
{\theta_{\nu \mu}} \le {0.6^o \over \sqrt{E_\nu (TeV)}}
\label{thetanumu}
\end{equation}
\noindent where $E_\nu$  is the neutrino energy (see Fig. \ref{fig:nu_mu}). 

Muon energy losses are due to several processes such as ionization, pair production, bremsstrahlung and photonuclear interactions \cite{murange}. The total energy loss per unit length can be written in a parameterized formula as:
\begin{equation}
dE_\mu/dx= \alpha (E_\mu) + \beta (E_\mu) \cdot E_\mu
\end{equation}
\noindent where $\alpha(E_\mu)$ is an almost constant term that accounts for ionization, and $\beta(E_\mu)$ takes into account the radiative losses. 

Fig. \ref{murange} shows the effective muon range $ R_{eff}$ in water. $ R_{eff}$ represents the range after which a muon of initial energy $E_\mu$ has still a residual energy $E_{thr}^\mu$ at the detector. 
As an example, a muon with initial energy $E_\mu=10$ TeV travels more than 4 km in water and arrives with more than 1 TeV of residual energy. The event will be detected even if the neutrino interaction vertex is outside the instrumented detector volume of the telescope.

A muon travelling through rock, ice or water is subject to multiple scattering. The deviation of the muon direction due to this process after travelling a distance $x$  is given by  \cite{pdg} :
\begin{equation}
{\theta_{ms}} = {13.6 (MeV) \over E_\mu} \sqrt{x/X_0}[1+0.0038\ln(x/X_0)]
\label{thetams}
\end{equation}
\noindent where $X_0$ is the radiation length of the medium. At energies and distances that concern us, $\theta_{ms}$  is smaller than $\theta_{\nu \mu}$ and the effect can be neglected.
Muon direction can be measured with an intrinsic resolution which depends on many factors, and in particular on the propagation medium. From Monte Carlo simulations, the precision is of the order of less than $1^o$ in ice, and $\sim 0.2^o$ in water (see \S \ref{waterice}).

\subsection{Cherenkov radiation}\label{cherenkov}
Any operating or proposed neutrino telescope in the TeV-PeV range is working by collecting the optical photons produced by the Cherenkov effect of relativistic particles. The light is collected by a three-dimensional array of photomultiplier tubes (PMTs). The information provided by the number of photons detected and their arrival times are used to infer the neutrino flavor, direction and energy. 
\begin{figure*}
\vspace{6.0cm}
\includegraphics{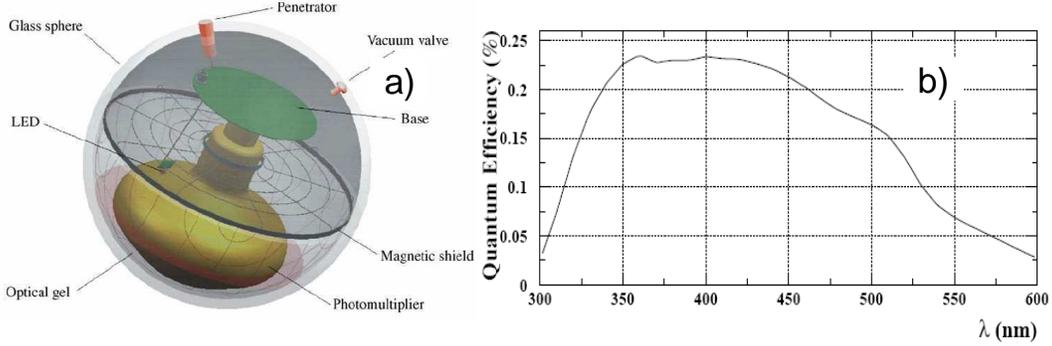}
  \caption{\it (a) Sketch of an ANTARES optical module (OM). Those used by IceCube, NEMO and NESTOR experiments are similar. A large hemispherical (10 inches in diameter) photomultiplier (PMT) is protected by a pressure-resistant glass sphere. The outer diameter of the sphere is 43.2 cm.  A mu-metal cage protects the PMT from the Earth  magnetic field. An internal LED is used for the calibration. (b) The quantum efficiency for PMTs commonly used in ice or water (from Hamamatsu). }
    \label{fig:om}
\end{figure*}

Cherenkov radiation is emitted by charged particles crossing an insulator medium with  speed exceeding that of light in the medium  \cite{chere}. The charged particle polarizes the molecules along the particle trajectory, but only when the particle moves faster than the speed of light in the medium, an overall dipole moment is present. Light is emitted when the  electrons of the insulator restore themselves to equilibrium after the disruption has passed, creating a coherent radiation emitted in a cone with a characteristic angle $\theta_C$ given by 
\begin{equation}
cos\theta_C= {c/n \over \beta c}={1 \over \beta n}
\end{equation}
\noindent where $n$ is the refracting index of the medium and $\beta$ is the particle speed in units of $c$. For relativistic particles ($\beta \simeq 1$) in seawater ($n \simeq$ 1.364) the Cherenkov angle is $\theta_C \simeq 43^o$. 

The number of Cherenkov photons, $N_\gamma$, emitted per unit wavelength interval, $d\lambda$ and unit distance travelled, $dx$, by a charged particle of charge $e$ is given by
\begin{equation}
{d^2N \over dxd\lambda} = {2\pi \over 137 \lambda^2} \bigl(1-{1\over n^2\beta^2}\bigr)
\label{cphot}
\end{equation}

\noindent where $\lambda$ is the wavelength of the radiation. From this formula it can be seen that shorter wavelengths contribute more significantly to the Cherenkov radiation. The light absorption by water/ice will strongly suppress some wavelengths, see \S  \ref{waterice}. 

Fig. \ref{fig:om} shows one example of an optical module used in ice and water experiments (see \S \ref{nutel} and \S \ref{mediterranean}). The PMT  quantum efficiency is large in the wavelength range between 300-600 nm, matching very well the region in which ice/water are transparent to light. Typically, in the wavelength range between 300-600 nm, the number of Cherenkov photons emitted per meter is about $3.5 \times 10^4$.

\section{Why a km$^3$ telescope}\label{toy}

In this section we develop a {\it toy model} for a neutrino telescope, dedicate to non-expert readers. Our aim is to derive, using a simple calculation, why a cubic kilometer scale detector is needed, and what is the number of optical sensors required to detect neutrinos in the instrumented volume. 
In \S \ref{nuastro} and \S \ref{extragalactic} we have presented some theoretical models of, respectively, galactic and extragalactic neutrino sources. Each model is characterized by the differential neutrino energy spectrum ${d\Phi_\nu \over dE_\nu}$ (TeV$^{-1}$ cm$^{-2}$s$^{-1})$. The rate of {\it detectable} events is simply derived using the {\it neutrino effective area}. This quantity (\S \ref{effarea}) is computed by each experiment, and can be used to compare the sensitivity of the different experiments.

\subsection{The reference flux}\label{candle}
As an \textit{ingredient} in our {toy model}, we need a reference flux. TeV $\gamma$-rays astronomers found convenient to use the Crab flux as {\it reference flux} or {\it standard candle}  for their measurements. For practical reasons, this is useful also for neutrino astronomy. 

The Crab nebula\footnote{The Crab pulsar, with a rotational period of 33 ms and a spin-down luminosity $L = 5\times 10^{38}$ erg/second, is a particular important source for high energy astrophysics. In fact, the pulsar powers a surrounding synchrotron nebula which has been detected in radio and X-ray wavelengths \cite{crablepto}. It is believed that the rotational energy of the pulsar is mostly carried away by a relativistic wind of electrons and positrons. The interaction of this wind with the surrounding medium creates a relativistic shock wave, where the leptons are thought to be accelerated to high energies \cite{crabwind}.}
 was discovered at TeV energies in 1989 \cite{weekes} and it is conventionally used as a reference source of TeV $\gamma$-rays, due to its relative stability and high intensity.
A reference flux equal to 1 C.U. (Crab Unit) is defined as the flux similar to that of the Crab, assuming no energy cutoff and a spectral index $\Gamma   =2$ :
\begin{equation}
E_\gamma^2{d\Phi_\gamma \over dE_\gamma } = 10^{-11} \ 
\textrm{cm}^{-2} \textrm{s}^{-1} \textrm{TeV} = \textrm{1 C.U.} 
\label{crab}
\end{equation}
The recent Crab TeV $\gamma$-rays spectrum measured by different experiments shows a steepening; a better fit of the data is provided by a power law (energy $E$ in TeV) with an exponential cutoff: 
${d\Phi_\gamma \over dE_{\gamma}} = I_0 E^{-\Gamma} e^{(-E/E_c)}$. This formula, with $\Gamma = 2.39\pm 0.03_{stat} $ and a cutoff 
energy $E_c = (14.3\pm 2.1_{stat})$ TeV, gives a differential
flux normalization at 1 TeV of \cite{crabhess}: 

\noindent $I_0 = (3.76\pm~0.07_{stat})\times~10^{-11}$ cm$^{-2}$s$^{-1}$TeV$^{-1}$.

Regarding neutrino telescopes, the Crab is in a particular sky position (see Fig. \ref{gammasky}), because it can  be seen by both telescopes located in the Northern and in the Southern Earth hemisphere. 
In order to perform a back-of-the-envelope calculation, we assume a reference  neutrino flux equal to that of the gamma flux from the Crab, $  E_\nu^2{d\Phi_\nu \over dE_\nu }\simeq E_\gamma^2{d\Phi_\gamma \over dE_\gamma }$. This is only true if hadronic processes are involved in the Crab TeV $\gamma$-ray emission (and for this particular source it seems to be disfavored with respect to the leptonic model). A detailed prediction of the event rate in a neutrino telescope as big as the IceCube experiment is $N(E_\mu~>~1$ TeV) $\sim$ 2.8 yr$^{-1}$ \cite{guaranteed}. Our simple calculation in the next section can be compared with this result.

\subsection{Event rate for the reference flux}\label{ref_flux_rate}

The $\nu_\mu$ CC interactions give the  possibility of a measurement of the neutrino direction within $1^o$ or better (it can reach up to $\sim 0.2^o$ in water). The $\nu_\mu$ channel allows also an enhancement  of the {\it effective volume} of the detector. The effective volume corresponds to the product of the detector effective area and the muon range $R_\mu$, it is larger than the {\it instrumented volume} and it increases with increasing neutrino energy.

The event rate in a neutrino telescope can be expressed in terms of:
\begin{eqnarray}
{N_\mu(E_{thr}^\mu) \over T} & = & 
\int dE_\nu \cdot {d\Phi_\nu \over dE_\nu}(E_\nu)\cdot A \cdot P_{\nu\mu}(E_\nu, E_{thr}^\mu) \nonumber \\
& \cdot & e^{-\sigma (E_\nu) \rho N_A Z(\theta)}
\label{rate}
\end{eqnarray}

\noindent where $A$ and $T$ are the detector area and observation time, respectively. $P_{\nu \mu}$ is the probability that the neutrino gives an observable muon (see below). The {\it effective neutrino area} is the quantity:
\begin{equation}
A^{eff}_\nu(E_\nu)  = P_{\nu\mu}(E_\nu, E_{thr}^\mu) \cdot A \cdot e^{-\sigma (E_\nu) \rho N_A Z(\theta)}
\end{equation}
The term $ e^{-\sigma (E_\nu) \rho N_A Z(\theta)}$, where $\sigma (E_\nu)$ is the total neutrino cross section, $N_A$ the Avogadro number, $(\rho N_A)$ the target nucleon density and $\theta$ the neutrino direction with respect to the nadir,
takes into account the absorption of neutrinos along the Earth path $Z(\theta)$. From the nadir $Z(0)=6.4\times 10^8$ cm, the absorption becomes not negligible for $\sigma  > 10^{-34}$ cm$^2$ (or equivalently $E_\nu > 50$ TeV, see Fig. \ref{nusigma}), when $e^{-\sigma  \rho N_A Z} \simeq e^{-0.05}\simeq 0.95$.

\begin{figure}[tbh]
  \vspace{8.0cm}
  \includegraphics{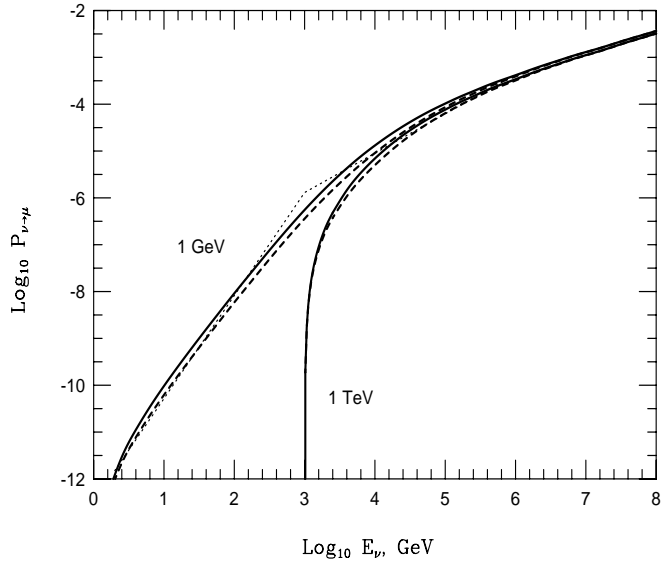}
  \caption{\it
$ P_{\nu\mu}$  \cite{gaisser} for two values of the muon threshold energy: 1 GeV and 1 TeV. The solid line are neutrinos and the dashed line antineutrinos. The dotted lines show a power law approximation.
    \label{figpnumu} }
\end{figure}

$ P_{\nu\mu}(E_\nu, E_{thr}^\mu)$ represents the probability that a neutrino with energy $E_\nu$ produces a muon of energy $E_\mu$ which survives with energy $> E_{thr}^\mu$ after the propagation from the interaction point to the detector. It can be expressed in terms of:
\begin{eqnarray}
P_{\nu\mu}(E_\nu, E_{thr}^\mu) &=&   N_A \int_0^{E_\nu} {d\sigma_\nu\over dE_\mu} (E_\mu , E_\nu)   \nonumber \\
&\times & R_{eff}(E_\mu, E_{thr}^\mu) dE_\mu 
\label{pnumu}
\end{eqnarray}

\noindent where $d\sigma_\nu / dE_\mu$ is the differential neutrino cross section to produce a muon of energy $E_\mu$, and $ R_{eff}(E_\mu, E_{thr}^\mu)$ is the effective muon range.
One can tabulate $P_{\nu\mu}$ for a given muon energy threshold. Fig. \ref{figpnumu} shows $P_{\nu\mu}$  for two values of muon threshold energy: 1 GeV (which was the characteristic muon threshold for large area underground detector, like MACRO) and 1 TeV (which is the characteristic value for neutrino telescopes). 

Note the dependence of $P_{\nu\mu}$ for $ E_{thr}^\mu> 1$ GeV on neutrino energy  \cite{gaisserbook}: it can be approximated with  $ P_{\nu\mu} \propto E_\nu^2$ for $E_\nu < 1$ TeV, and 
$ P_{\nu\mu} \propto E_\nu$ for $1<E_\nu < 10^3$ TeV. The two energy regimes reflect the energy dependence of the neutrino cross section (which depends almost linearly on energy) and effective muon range (which depends linearly on muon energy up to $\sim 1$ TeV, when muon radiative losses become dominant). 

To solve analytically eq. \ref{rate} in our simplified model, we use the approximation for neutrino energies larger than 1 TeV: $P_{\nu\mu}(E_\nu, E_{thr}^\mu) \simeq P_o E_\nu^{0.8}= 10^{-6} E_\nu ^{0.8}$ ($E$ in TeV)  \cite{gaisser}. With  this approximation, the event rate for a neutrino source equivalent to 1 C.U. (eq. \ref{crab}) can be analytically computed in the range between $1\div 10^3$ TeV, neglecting the Earth's absorption: 
\begin{eqnarray}
{N_\mu(E_{thr}^\mu) \over T} & = & 
\int_{1\ TeV}^{10^3\ TeV} dE_\nu \cdot  (KE_\nu ^{-2})\cdot A \cdot (P_o E_\nu ^{0.8})  = \nonumber \\
&=& 5\times 10^{-19}\cdot A\ cm^{-2} s^{-1} 
\label{rate1km}
\end{eqnarray}
\noindent where $K= 10^{-11}$ TeV$^{-1}$ cm$^{-2}$ s$^{-1} $. 
Here, the area A is the surface surrounding the instrumented volume where muons of energy $E_{thr}^\mu$ can be detected. To give a rough energy-inde\-pen\-dent estimation, a 1 TeV muon has a range (in water) of $R^\mu_{1\ TeV}$ = 2.42 km \cite{murange}. The cross sectional area of a sphere of diameter  $R^\mu_{1\ TeV}$ is $A \sim 5$ km$^2$. Assuming this value in eq. \ref{rate1km}, the number of expected events is $\sim$ 1/year, in a rough agreement with the more detailed computation of \cite{guaranteed}.

\subsection{The neutrino effective area} \label{effarea}
The energy-dependent effective area $ A^{eff}_\nu (E_\nu)$ must be computed using Monte Carlo (MC) simulations. Because it depends from many factors, including the detector characteristics, $ A^{eff}_\nu$  is strongly detector-dependent. The effective area allows a more detailed computation of the rate of expected  events for a given neutrino flux and the  sensitivity of each experiment can be expressed in terms of the effective area \cite{bailey}. In the MC calculation, a flux of neutrino is generated, which interacts in a huge volume $V_{gen}$ of material surrounding the instrumented volume. The  effective neutrino area is then computed as:
\begin{equation}
A^{eff}_\nu  = {N_{x}\over N_{gen} } \times V_{gen} \times (\rho N_A) \times \sigma (E_\nu) \times e^{-\sigma (E_\nu) \rho N_A Z(\theta)}
\label{nuareamc}
\end{equation}
\begin{figure*}[tbh]
  \vspace{9.0cm}
  \includegraphics{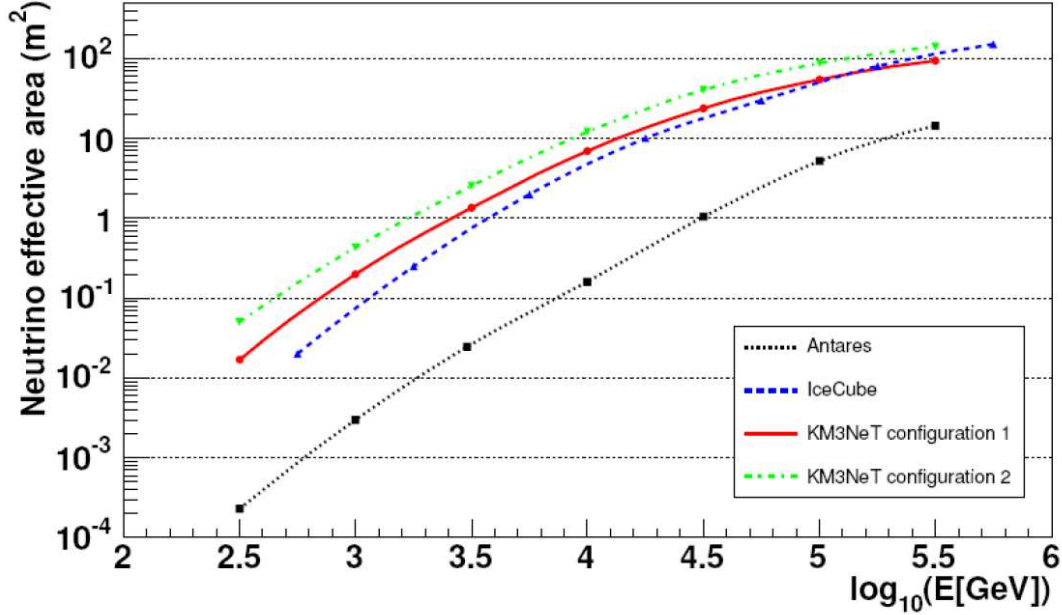}
  \caption{\it Neutrino effective area as a function of the true simulated neutrino energy. Two different versions of a km$^3$ (\S \ref{km3net}) detector in the Mediterranean sea are compared to the IceCube (\S \ref{icecube}) detector in the Antarctic South Pole. The ANTARES (\S \ref{antares}) neutrino effective area is shown for comparison.}
    \label{km3effarea}
\end{figure*}

\noindent where $N_x$ is the number of detected (triggered, reconstructed) events and  $N_{gen}$ is the number of generated events. The other quantities have been already defined. Fig.  \ref{km3effarea} shows the  effective neutrino area for IceCube, ANTARES and for two different configurations of a cubic kilometer detector in the Mediterranean sea, presented in the KM3Net design report  \cite{km3}. 
The effective area represents a useful tool to compare different experiments, because for a given source it is straightforward to compute the number of expected events as:
\begin{equation}
{N_\mu \over T} = \int dE \cdot  {d\Phi_\nu \over dE_\nu} \cdot  A^{eff}_\nu (E_\nu)   
\label{rateanu}
\end{equation}

In eq. \ref{nuareamc}, $N_{x}$ depends on neutrino energy and direction. This has two consequences:

\noindent {\it i)} sources with a similar fluency (number of neutrino per unit of area and unit of time) but different spectral index produce a different response in  neutrino telescopes (harder is the spectral index, better the source is seen);

\noindent {\it ii}) due to the Earth motion, the position in the detector frame of a given source in the sky changes with daytime. The effective area must be computed for each declination, by averaging over the local coordinates (zenith and azimuth angle).

\subsection{Number of optical sensors in a neutrino telescope}
\label{num_pmt}
The instrumented volume for a neutrino telescope must be at least of the order of a km$^3$. How many optical sensor (PMTs) are needed? This is one of the major impact  factor on the cost of  an experiment.

In this computation, we assume that the telescope use a PMT with a 10" diameter, detection area $A_{pmt}\sim 0.05\ m^2$ and quantum efficiency\footnote{New 10" PMTs with higher quantum efficiency, $\epsilon_{pmt}\simeq 0.35$, are also considered in the KM3NeT consortium} $\epsilon_{pmt}\simeq 0.25$, see Fig. \ref{fig:om}. Similar PMTs, which have the advantage to fit inside commercial pressure-resistant glass spheres (optical module, OM), have been chosen by the IceCube, ANTARES, NEMO and NESTOR collaborations. The overall efficiency of the OM is somewhat reduced with respect to that of the PMT, due to the presence of the glass, glue between glass and the PMT, and mu-metal cage for magnetic shield: $\epsilon_{om}\simeq 0.8 \epsilon_{pmt}=0.2$.

As we will discuss in \S \ref{water}, ice or seawater absorption length $\lambda_{abs}$ is larger than 50 m for light in the 400-500 nm range. 
A photon falling inside the effective PMT volume $V_{pmt}$= $A_{pmt} \times  \lambda_{abs} \simeq 2.5\ m^3$ can produce a photoelectron (p.e.) with a probability $\epsilon_{om}\simeq 0.2$.

Let us call $N_{pmt}$ the number of optical sensors inside the instrumented volume (it is the number to be determined). The rate R between the effective PMT volume of $ N_{pmt}$ and the instrumented volume is:
\begin{equation}
R = { V_{pmt}\times N_{pmt} \over 10^9 \ m^3} = 2.5\times 10^{-9} N_{pmt} 
\label{npmt}
\end{equation}

The total number of Cherenkov photons emitted by a 1 km length muon track in the wavelength range of PMTs sensitivity (\S  \ref{cherenkov}) is $N_\gamma \simeq 3.5 \times 10^7$. The fraction converted into photoelectrons which gives a signal is:
\begin{eqnarray}
N_{p.e.} & = &  N_\gamma \times R \times \epsilon_{om}  \nonumber  \\
 & \simeq & (3.5\times 10^7) \cdot (2.5\times 10^{-9} N_{pmt})\cdot \epsilon_{om}  \nonumber \\ 
 & = & 1.8\times 10^{-2} N_{pmt} 
\label{npe}
\end{eqnarray}

The number of PMTs needed to reconstruct a muon track is of the order of a few tens. 
We must take into account that in most cases many photons arrive on the same PMT during the integration window of the electronics (which is of the order of 20-50 ns).  For this reason, on average around $ N_{p.e.} \sim$ 100 p.e. are necessary to affect few tens of OM. 
The number of optical sensors in a neutrino detector follow straightforward  from eq. \ref{npe}:
\begin{equation}
N_{pmt} \simeq 100/ 1.8\times 10^{-2} \simeq 5000
\label{npmtmin}
\end{equation}

As we will shown in \S \ref{icecube}, the IceCube collaboration is in an advanced status to bury under the ice 4800 OMs; the neutrino telescope in the Mediterranean sea, \S \ref{km3net}, plan to deploy between 5000 and 10000 OMs, depending on the financial budget.

\section{Water and Ice properties}\label{waterice}

The  effects of the medium (water or ice) on light propagation are absorption and scattering of photons.  These  affect the reconstruction  capabilities of  the telescope.   In  fact, absorption reduces  the amplitude  of  the  Cherenkov  wavefront,  i.e. the  total amount of light on  PMTs. Scattering changes the direction of propagation of the Cherenkov photons and the distribution of their arrival time on the PMTs; this degrades the measurement of the direction of the incoming neutrino.
We define \textit{direct photons} those which arrive on a PMT in the   Cherenkov  wavefront, without be scattered; otherwise, we define them \textit{indirect photons}.
\begin{figure}[ht]
  \vspace{7.cm}
  \includegraphics{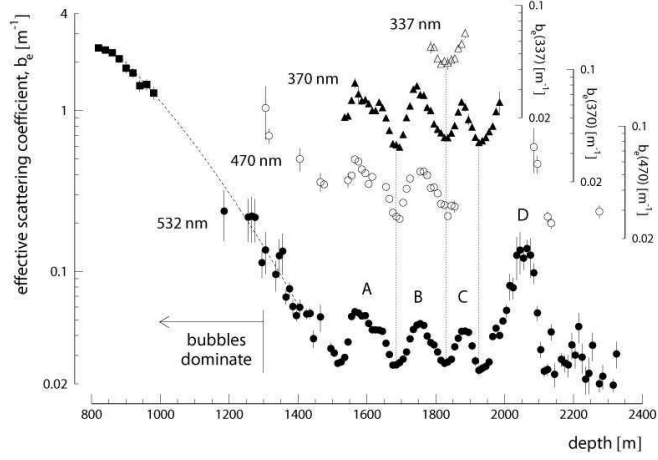}
\caption{\it Depth dependence of scattering coefficient $b_e(\lambda)=1/ L_b^{eff}(\lambda)$ as measured by the IceCube collaboration  \cite{iceprop} for 4 different wavelengths. }
    \label{iceprop} 
\end{figure}

The propagation of light in a transparent medium is quantified for a given wavelength $\lambda$,  by  the  medium  inherent  optical  properties:  absorption  $a(\lambda)$,   scattering  $b(\lambda)$  and  attenuation
$c(\lambda) = a(\lambda)  + b(\lambda)$ coefficients, or, alternately,
absorption    $L_a(\lambda)   =   a(\lambda)^{-1}$,   scattering
$L_b(\lambda)  =  b(\lambda)^{-1}$  and  attenuation  $L_c(\lambda)  =
c(\lambda)^{-1}$ lengths.   
Each of  these lengths represents  the path after which a beam on  initial intensity $I_0$ at wavelength $\lambda$ is reduced in intensity by  a factor of 1/e through absorption and scattering, according to the following relation:
\begin{equation}
I_{i}(x,\lambda)=I_0(\lambda)e^{-x/L_{i}(\lambda)}; \quad i=a,b, c
\end{equation}
where  $x$ (in meters) is the  optical path  traversed by  the light.   

A complete description of  light scattering would  require, in addition to the geometric scattering length $L_b(\lambda)$, the knowledge of the scattering angular  distribution.
Gustav Mie developed (1908) an analytical solution of the Maxwell equations for scattering of electromagnetic radiation by spherical particles, which is appropriate for modeling light scattering in transparent media. In particular for ice the predominant scattering centers are sub-millimeter sized air bubbles and micron sized dust particles. 

Generally, light can be scattered multiple times before it reaches an optical sensor. 
The average cosine of the light field of photons that have undergone multiple ($=n$ times) scattering obeys a simple relationship:
\begin{equation}
 \langle \cos\theta \rangle_n =\langle \cos\theta \rangle^n 
\end{equation}
On average, per step, a photon advances at an angle of $\langle \cos\theta \rangle $ a distance of  $L_b(\lambda)$  between each scatter. Hence
after $n$ scatters, a photon has moved in the incident direction: 
\begin{equation}
L_b^{eff}(\lambda) = L_b(\lambda) \sum_{i=0}^{n} \langle \cos\theta \rangle^i
\simeq { L_b(\lambda)\over {1- \langle \cos\theta \rangle} } \end{equation}
Experimental measurements are generally expressed in terms of the \textit{effective} light scattering length $L_b^{eff}(\lambda)$, instead of the (strongly correlated) values of average scattering angle $\langle \cos\theta \rangle$ and geometric scattering length  $L_b(\lambda)$.

In the following discussion, we will point out that seawater has a smaller value of $L_a(\lambda)$ with respect to ice (which is more transparent). By referring to the discussion in \S \ref{effarea}, the same instrumented volume of ice corresponds to a larger effective volume with respect to seawater. On the other hand, the effective scattering length $L_b^{eff}$ for ice is smaller than water. This is a cause of a larger degradation of the angular resolution of the detected neutrino-induced muons in ice with respect to the water (see \S \ref{track}).

Another difference between ice and water is that optical modules in seawater suffer some background from the natural radioactivity of elements (mainly $^{40}K$) and from luminescence produced by organisms living in the deep sea (\S \ref{optbck}). Ice is (almost) background-free.

\subsection{Ice properties}\label{ice}
\begin{figure}[tbh]
  \vspace{9.cm}
  \includegraphics{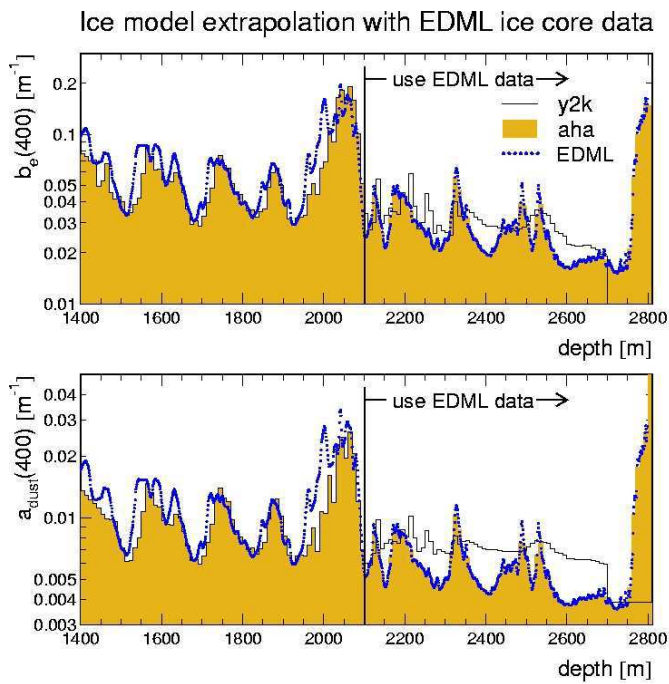}
\caption{\it Depth dependence of scattering coefficient $b_e(\lambda=400\ nm)$ and  absorption coefficient $a_{dust}(\lambda=400\ nm)$ as used in Monte Carlo simulations in the Antarctica experiments  \cite{icethesis}.  }
    \label{deepice} 
\end{figure}

The ice in which AMANDA and IceCube (see \S \ref{icecube}) are embedded has optical properties that vary significantly with depth and that need to be accurately modeled. Impurities trapped in the ice depend on the quality of the air present in the snows: Antarctica ice is laid down through a process of snowfall, hence trapping bubbles of air as it compacts itself. This happened over roughly the last 10$^5$ years. Because of variations in the long-term dust level in the atmosphere during this period, as well as occasional volcanic eruptions, impurity concentrations are depth dependent. 
IceCube and AMANDA detectors  have both pulsed and steady light sources located at different depths under the ice. These sources are used to measure both the attenuation length and scattering length. This is done by measuring the arrival time distributions of photons at different distances from a light source. 
\begin{figure*}[ht]
\begin{center}
\includegraphics[width = 14. cm]{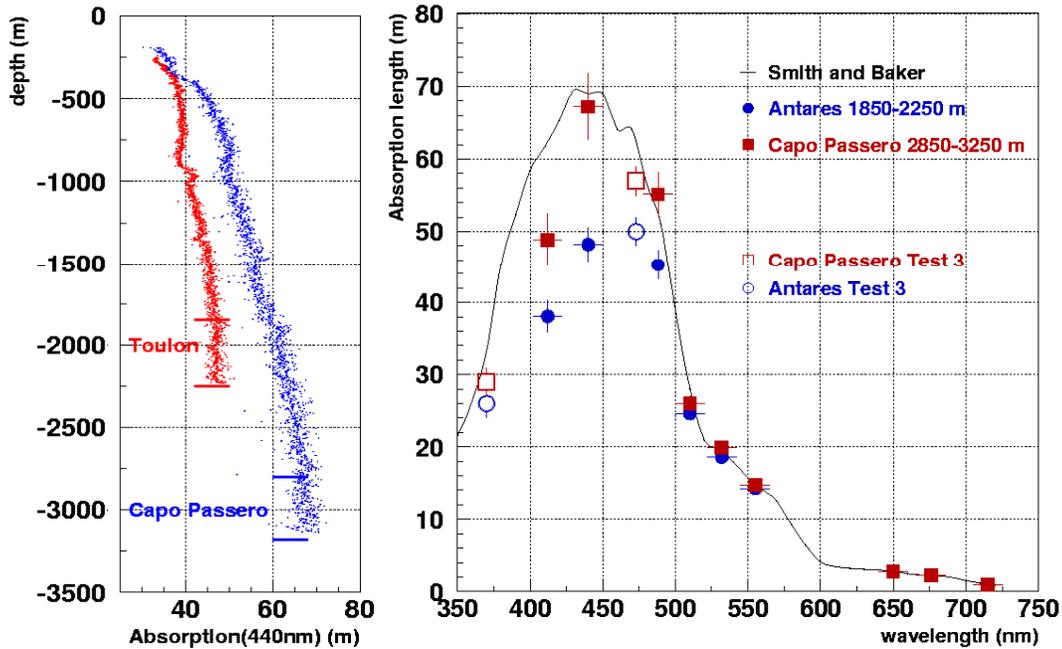}
\caption{\label{fig@absoNvsA} \it The absorption  length measured  in the
underwater sites of  Toulon (blue) and Capo Passero (red), of  the ANTARES (\S \ref{antares}) and NEMO (see \S \ref{nemo}) experiments, respectively. Left: $L_a(\lambda)$  with $\lambda = $ 440
nm  as function  of  the  depth.  Right:  $L_a(\lambda)$  as  function of  the wavelength, compared to  the behaviour  of {\it pure}  seawater (solid  line).}
\end{center}
\end{figure*}

The scattering centers for light propagation in IceCube are air bubbles and  dust particles of various types \cite{iceprop}. 

Air bubbles play a major role in light scattering  down to depths of approximately 1250 m below the surface. Then, the pressure of ice layers above compact these air bubbles into air hydrate crystals, which have an index of refraction nearly identical to that of ice. 

Four main components of the dust were individuated: insoluble mineral grains, sea salt crystals, liquid acid drops and soot. Sea salt crystals and liquid acid drops contribute negligibly to absorption, sea salt being the strongest scattering component. Insoluble mineral grains are the most common component, and contribute to both absorption and scattering, while soot contributes mainly to absorption.
The relative abundance of each of these components was derived from ice core measurements  \cite{icedust}.

Fig. \ref{iceprop} shows the effective scattering coefficient $b_e(\lambda)$ measured in Antarctica. The strong drop off from depths of around 1250 m is due to the transition from the region where air bubbles are dominant to the region where the four main dust peaks are present in the ice.

The effects of ice properties on photon propagation and arrival times on PMTs are evaluated by the IceCube collaboration through Monte Carlo (MC) simulations. MC parameters are adjusted until an agreement is met between simulation and real data for photon timing distributions. Some models were developed inside the AMANDA/ IceCube collaborations  \cite{icethesis}. The depths below AMANDA, which are now included in the active volume of IceCube, were described with a model corrected using the measurements in ice cores taken at other sites in Antarctica, in particular at East Dronning Maud Land (EDML). Absorption (due to dust) and scattering coefficients obtained with different simulations and compared with the measurements are shown in Fig. \ref{deepice}.

\subsection{Optical properties of water}\label{water}

Optical  properties   of  water depend on many factors. Environmental parameters  such as  water temperature and  salinity are indicators of the aggregation state of H$_2$O molecules, which biases the  diffusion of light.   Water absorption and  scattering depend also on  the density and the  size of the  floating particulate, which affects the telescope response also  in terms of detector aging: due to bio-fouling  and  sediments  sticking   on  the  optical  modules,  efficiency of the photon detection can be compromised.

For these reasons, together with the strong necessity of reducing the atmospheric muon background,  a site for a neutrino  telescope must be located at  great  depth.  In this  condition, high  pressure and extremely slow water currents make the site characteristics stable.
\begin{figure}[ht]
\begin{center}
\includegraphics[width = 8. cm]{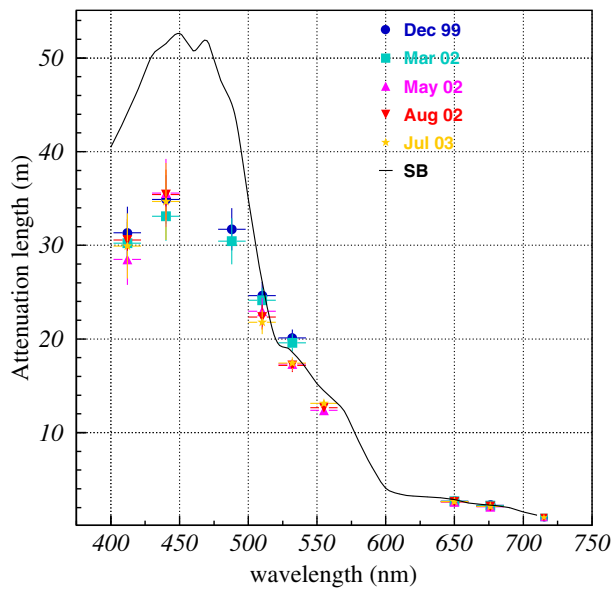}
\caption{\label{fig@attenN}\it Average  attenuation length  measured with
the AC9 in the Capo Passero  site \cite{bib@giorgio}, at depth 2850-3250 m from year 1999 to  2003.  Only statistical  errors  are  plotted. The  solid  line
indicates  the values  of $L_c(\lambda)$  for optically  pure seawater
reported by Smith and Baker  \cite{bib@SmithBaker}.}
\end{center}
\end{figure} 

Apart from  the BAIKAL  experiment (\S \ref{baikal}), situated  in the Siberian lake Baikal at a depth of approximately 1 km, submarine sites have  been   preferred  in  order  to  reach   deeper  locations.  The preference for undersea sites  is not free from drawbacks: because of  the salts into  the water,  they  present  an irreducible  optical  background  due  to  the  radioactive  decay  of $^{40}$K and to the  bioluminescence, which strongly depends on environmental factors.

In order to  minimize the bias induced by  external agents, the  telescope sites  must be in addition far  enough from continental shelf breaks and river  estuaries, which can  induce turbulent  currents and  spoil the purity  of water. At  the same  time, the  neutrino telescope  should be close to scientific and  logistic infrastructures on shore.  With such requirements,  the Mediterranean  Sea  offers optimal  condition on  a worldwide scale.
In water the absorption and attenuation coefficients  $a(\lambda)$ and $c(\lambda)$ are directly measured by means of  dedicated instruments, like the AC9  manufactured by WETLabs  \cite{bib@ac9}.  Water  optical  properties  are  strongly  dependent  on the  wavelength:  light transmission   is  extremely   favored  in   the  range   350-550  nm  \cite{bib@ceff},   where  the   photomultipliers   used  to detect   Cherenkov radiation  reach the  highest quantum efficiency (see Fig. \ref{fig:om}).

In  natural seawater,  optical  properties are  also function  of  water temperature, salinity and dissolved particulate  \cite{bib@PopeFry,bib@KouLabChy}.   Measurements of the profiles  of temperature,  salinity,  attenuation and  absorption lengths  performed by the  NEMO collaboration (\S \ref{nemo}) in the  site named  {\it  Capo Passero}, 100  km off  shore from  the  coast of Sicily, during various sea campaigns from year 1999 to the end of year
2003 show that such quantities are  stable and constant at depths greater than 1500$\div$2000 m  \cite{bib@giorgio} (see Fig. \ref{locationNT} for the geographical location of sites). 
\begin{figure*}[htb]
\begin{center}
\includegraphics[width = 14.5 cm]{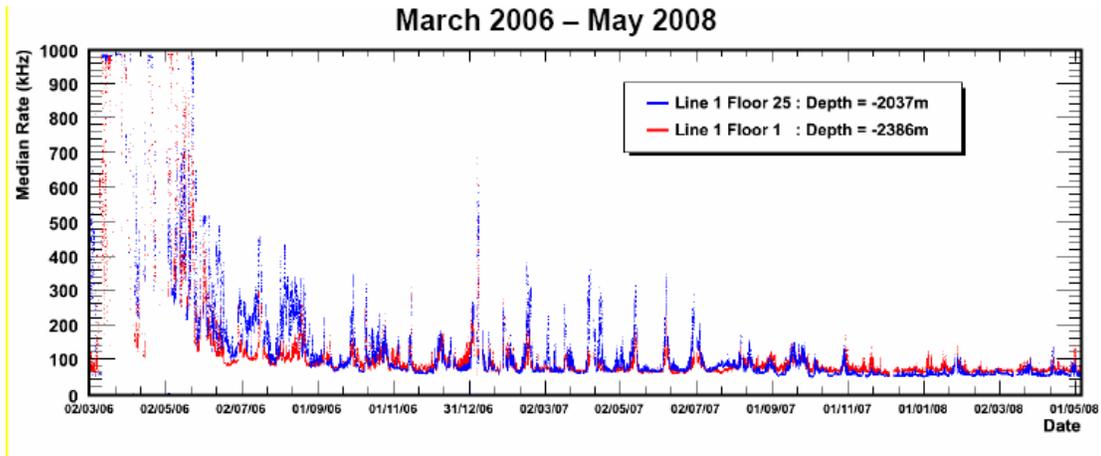}
\caption{ \label{rates_anta} \it Median rates (in kHz) measured with the 10" PMTs of the ANTARES experiment, on optical modules at two different depths (2037 m and 2386 m). Data from March 2006 up to May 2008 \cite{circe}. The contribution of the $^{40}$K decay is evaluated to be almost constant to $\sim 30-40$ kHz.}
\end{center}
\end{figure*}

The  nature of  particulate, either organic  or inorganic,  its dimension and  concentration, affect light  propagation.  All   these  environmental  parameters  may  vary significantly,  for each  marine  site,  as a  function  of depth  and
time. Moreover, it is known that seasonal effects like the increase of surface biological activity (typically during spring)  or the precipitation of sediments
transported  by flooding rivers,  enlarge the  amount of  dissolved and
suspended particulate, worsening the water transparency.

Fig.  \ref{fig@absoNvsA} (left) shows the absorption  length $L_a(\lambda =  440$ nm)  as a function  of depth measured in the site of  the ANTARES experiment (near Toulon, France) and in the Capo Passero site.  
The right side of the figure shows the measured absorption length
(in the  range of  350 nm $\leq\lambda\leq $700 nm, circle and square dots) in the two sites, compared to the model of {\it pure} seawater reported by Smith and Baker   \cite{bib@SmithBaker} (solid black line). 

Fig. \ref{fig@attenN} shows the mean attenuation length $L_c(\lambda)$ measured in different seasons in the Capo Passero site; such  averages are obtained with measurements from 2850  m to 3250 m depth. 

\subsection{Optical background in water} \label{optbck} 
The background in seawater has two main natural contributions:   the decay of  radioactive  elements in  water,  and  the  luminescence produced by organisms, the so called bioluminescence. 

The $^{40}$K is by far the dominant of all radioactive isotopes present in natural seawater. $^{40}$K decay channels are:
\begin{eqnarray}    
^{40}K &\rightarrow & ^{40}Ca\;+\;e^-\;+\;\bar{\nu_e} \; (\mbox{BR = 80.3}\%)\nonumber\\
^{40}K\;+\;e^-&\rightarrow & ^{40}Ar\;+\;\nu_e\;+\;\gamma \; \ (\mbox{BR = 10.7}\%)\nonumber
\end{eqnarray}
and both contribute  to the  production of  optical  noise. A  large fraction  of electrons  produced  in the  first  reaction  is  above the threshold  for  Cherenkov  light  production. The  photon  originating  in the  second reaction  has an energy  of 1.46  MeV and  can therefore  lead (through Compton  scattering) to  electrons  above the Cherenkov threshold.

The  intensity of   Cherenkov  light from  $^{40}$K radioactive  decays depends mostly on  the $^{40}$K concentration in sea  water. Since salinity in  the Mediterranean  Sea has small  geographical variation, this  Cherenkov light intensity is largely site independent.

Bioluminescence is ubiquitous in oceans and there are  two sources in deep sea:  steady glow  of  bacteria and  flashes produced  by animals. These can  give rise to an optical  background up to several orders of magnitude more intense than the one due to $^{40}$K (see Fig. \ref{rates_anta}).
The two components of optical background described above are clearly visible. 
Bursts observed in the counting rates are probably due to the passage of light emitting organisms close to the detector.
\begin{figure}[ht]
\begin{center}
\includegraphics[height = 9. cm]{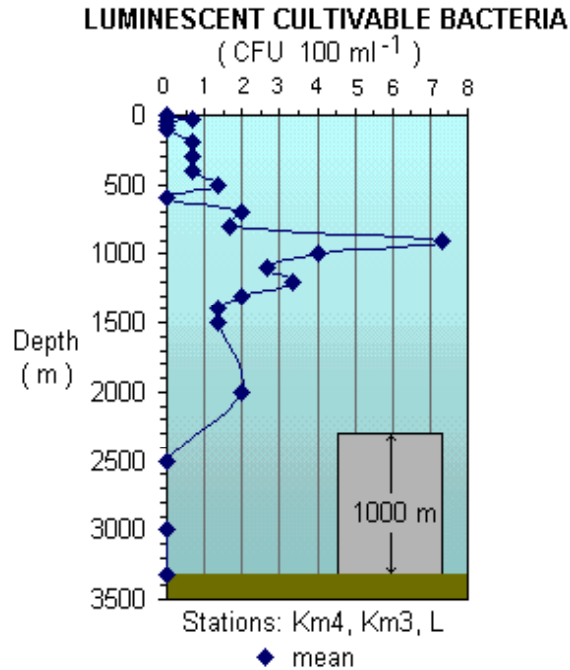}
\caption{\label{fig@bacteria} \it The amount of luminescent bacteria per unit of seawater volume, sampled in the Capo Passero site at different depths. The bacteria were  cultived at atmospheric pressure.}
\end{center}
\end{figure} 
\begin{figure}[htb]
\begin{center}
\includegraphics[width = 9. cm]{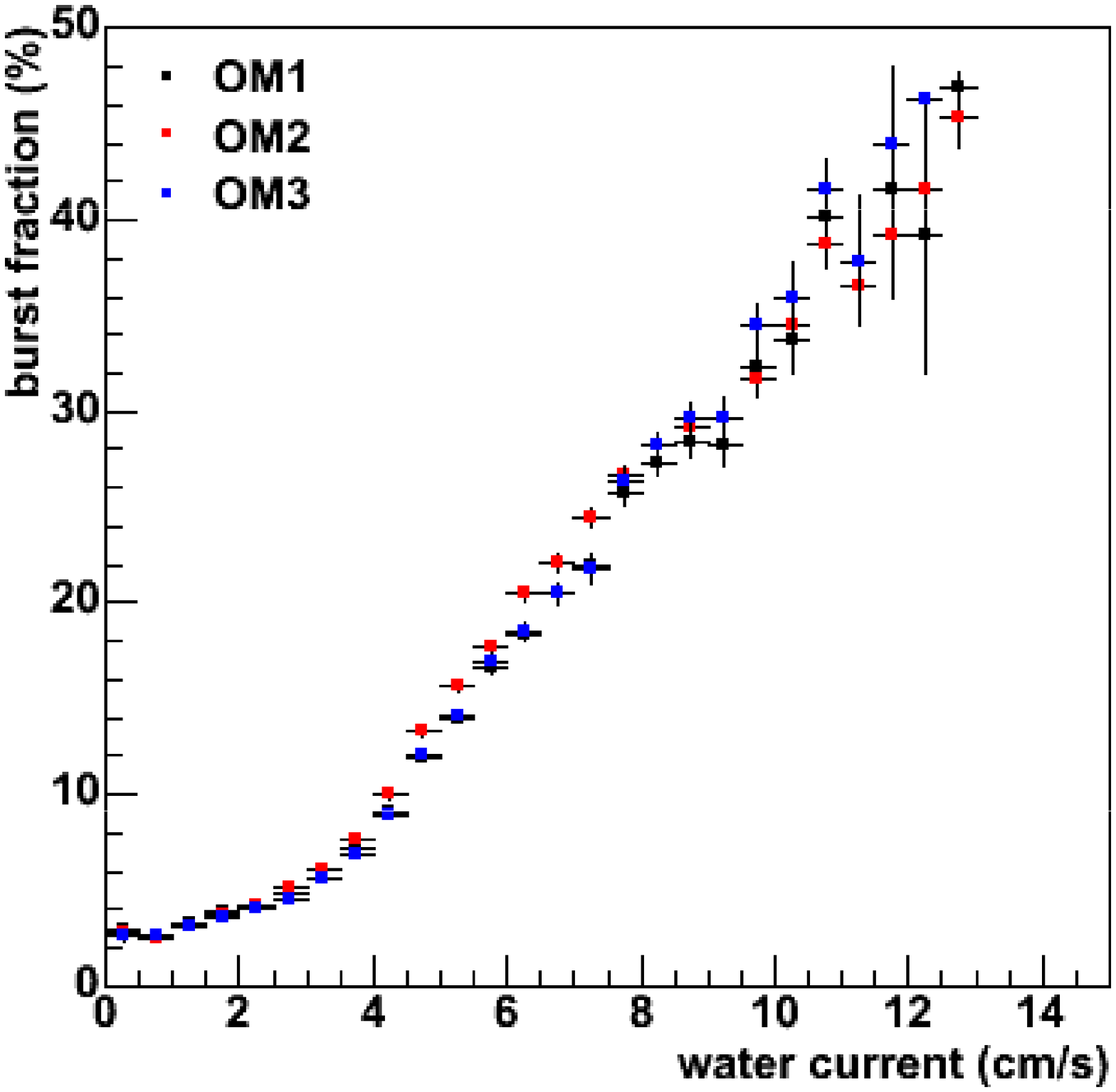}
\caption{\label{vel_vs_bact}\it Correlation between the burst fraction and the seawater current velocity as measured by the ANTARES detector. The burst fraction is the fraction of time with count rates on OMs exceeding 120\% of the baseline rate. }
\end{center}
\end{figure}

The  phenomena of  bioluminescence are  not yet fully  understood. The typical spectrum  of bioluminescence  light is centered  around 470-480 nm \cite{bib@biolum1,bib@biolum2},  the   wavelength  of  maximal transparency  of water.  The distribution of luminescent organisms in deep-sea varies with location, depth,  and time but there is a general pattern of decrease in abundance with depth. Fig. \ref{fig@bacteria} shows the amount  of luminescent cultivable bacteria as  a function of depth, measured in the Capo Passero site. Such measurements show a bioluminescence that is significantly lower with respect to what discovered at similar depth in the Atlantic Ocean  \cite{bib@aberdeen1,bib@aberdeen2}.

Deep sea currents were monitored at the ANTARES, NE\-MO and Nestor sites for  long time  periods. ANTARES discovered that the baseline component is neither correlated with sea current, nor with burst frequency; however, long-term variations of the baseline were observed. Periods of high burst activity are not correlated with variations of the baseline component, suggesting that each of the two contributions is caused by a different population. Moreover, a strong correlation is observed between bioluminescence phenomena and the sea current velocity, as shown in Fig. \ref{vel_vs_bact}.

\subsection{Track reconstruction in water and ice}\label{track}

In a 1 km$^3$ scale detector, most of the high energy muons produce tracks which are visible over more than 1 km. This long lever arm allows for good directional reconstruction, depending on the medium (water or ice), number and orientation of the optical sensors. 
A rough estimate of the muon energies it is also possible, either by the length of their tracks, or by measuring the specific energy loss; at energies above 1 TeV, muon energy loss $(dE/dx)$ is proportional to muon energy.

Muon reconstruction is done by maximum likelihood methods. The fitter finds the likelihood for different track positions and directions, and, optionally, energy. To do this, it uses functions which model the light propagation, giving the Probability Distribution Function for a photon, radiated from a track with a given orientation, to reach a PMT at a given distance and orientation as a function of time. Usually, these functions are pre-calculated using a simulation that tracks photons through the medium  \cite{tr-ice2,tr-ice1,aart}. 
\begin{figure*}[htb]
\begin{center}
\includegraphics[width = 14.5 cm]{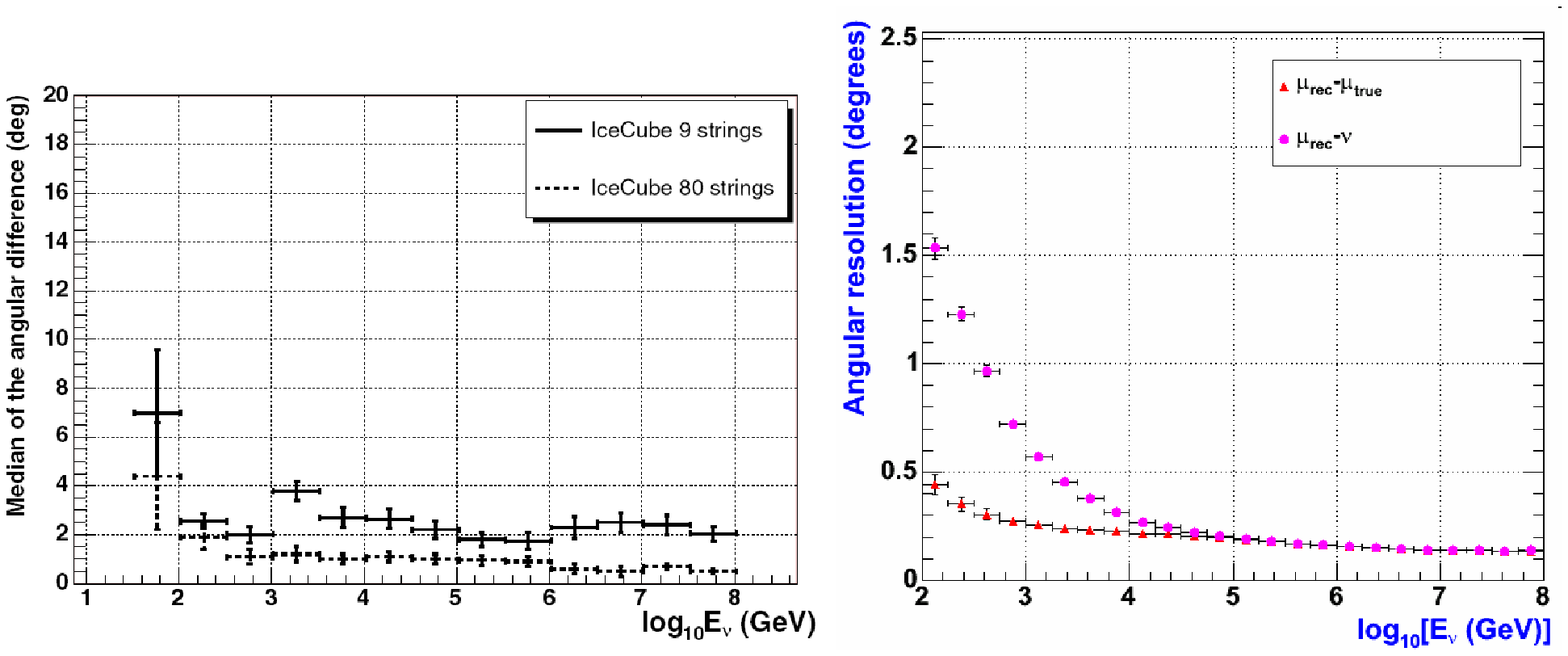}
\caption{\label{ang_resolution}\it (Left) Angular resolution (evaluated with MC) for 9 strings and for the full IceCube array as a function of event energy \cite{tr-ice0}. It is shown here the differences between true and reconstructed muon track.  (Right) The same, for the underwater ANTARES detector. In this case, it is also shown the difference with respect to neutrino direction.}
\label{fig:nu_mu}
\end{center}
\end{figure*}

Fig. \ref{ang_resolution} shows the angular resolution in ice and water resulted from MC calculations. For neutrino-induced muons (up-going) with $E_\mu> 1$ TeV, the  directional resolution is of the order of few tenths of degree  in water and around 1 degree in ice. 
$\nu_e$ and low-energy $\nu_\tau$ (below 1 PeV) interactions, or neutral current interactions, occurring inside the detector instrumented volume produce a shower, a compact deposition of energy, \S \ref{nu_e_det}, \S \ref{nu_tau_det}.

\subsubsection{Direct and indirect photons}\label{direct_indirect_ph}
\begin{figure}[tbh]
\vspace{6.0cm}
\includegraphics{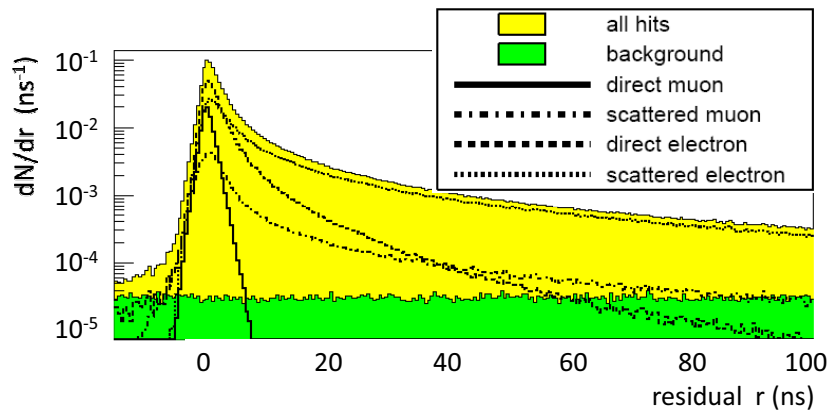}
  \caption{\it
Distribution of the difference r between the expected and measured photon arrival times for 1 TeV muon in water. A measurement of the light intensity and arrival time in one PMT is denoted as ''hit''. Contributions are shown for scattered and direct photons originating from the muon itself and from secondary electrons and positrons. All hits occurring within a distance of 100 m from the track are included in the figure. The distribution for background photons corresponds to a flat background rate of 60 kHz. Adapted from \cite{aart}.
    \label{dir_indir} }
\end{figure}

The effect on the tracking algorithms due to the arrival of \textit{direct} and \textit{indirect} Cherenkov photons is presented in Fig. \ref{dir_indir} (made for a simulation in water using 1 TeV muons, \cite{aart}).  Direct photons are emitted by a muon almost exactly at the Cherenkov angle and they arrive at the PMT without be scattered by the propagation medium. They carry the most precise timing information: the arrival time is only smeared by the dispersion and the transit time spread of the PMTs. 
A muon reconstruction algorithm which can ideally use only direct photons, is able to obtain the particle direction with the highest precision. This can be proved with Monte Carlo simulations, where direct and indirect photons can be tagged. The difference between the expected and the measured arrival time (the time residual $r$) for these photons has a Gaussian shape distribution (full line in Fig. \ref{dir_indir}).

Photons that originate from secondary electrons or that have scattered, are often delayed with respect to this time. However, also for these photons the distribution of time residuals peaks at zero, which means that they can still be used in the reconstruction process.
This is the reason why the measurement of the direction of electromagnetic and hadronic showers in $\nu_e$ CC and in $\nu_x$ NC interactions is much less precise with respect to the $\nu_\mu$ CC interaction channel.

\subsubsection{The background of atmospheric muons}
The bulk (see Fig. \ref{atmunu}) of reconstructed events in any neutrino telescope are downward going muons produced in cosmic-ray air showers \cite{bazzottiPHD}. 
Atmospheric muons can be used for a real-time monitoring of the detector status, of the time variation of the PMTs efficiencies, and for detector calibration.
Atmospheric muons can also be used for the study of the telescope pointing capability through the measurement of the moon shadow. On the other side, muons (especially in bundles) are a major background source: downward going particles wrongly reconstructed as upward going and simultaneous muons produced by different cosmic ray primaries could mimic high energy neutrino interactions.

Muons in bundle are very common: they represent the majority of the events which trigger a neutrino telescope. All the muons in a bundle are almost parallel, and propagate in a plane perpendicular to the shower axis at the same time. Because they are produced in the decay of secondary mesons, they follow the transverse and longitudinal momentum distribution of the parent mesons. As a consequence, the most energetic muons are expected to be closer to the axis shower. Most of the muons in the bundle have a radial distance with respect to the shower axis smaller than 10 meters \cite{mupage0,mupage1}, i.e. much smaller than the grid size of the detectors. For this reason, usually muon reconstruction algorithms find a \textit{track} which is in reality due to the sum of the signals induced by the muons in the shower. The muon multiplicity in the bundle is not measured (although works are in progress inside collaborations to obtain a rough estimate of the bundle muon multiplicity).

Because of the high rate of downward going muons, to distinguish $\nu$-induced events it is not enough to select events with the most likely reconstruction as upward going. Fairly stringent cuts must be applied to eliminate tracks with reasonable likelihoods for being downward going. This can be done by cutting on estimated errors from the likelihood fit, or using other quality estimators (see for instance Fig. \ref{zenazi}). The exact cuts depend on the medium (water or ice); cuts are also analysis-dependent, since different analyses are interested in signals from different energy ranges and zenith angles.

Another important possible background source is due to random coincident muon events. This happens  when two muon bundles from independent cosmic-ray air showers traverse the detector in the same time-window (few $\mu s$) when the detector is read out. The effect depends strongly from the depth and size of the detectors.  For instance, the upper level of IceCube (\S \ref{icecube}) is at the depth of $\sim 1450$ m and the frequency of these coincident events is relevant. For the smaller-sized AMANDA-II detector, the estimated trigger rate of these events is 2-3 Hz  \cite{icecubediffuse}, compared with the total trigger rate of 80 Hz. 
Running (or proposed) underwater experiments are deeper and this effect is largely reduced \cite{mupage2}.
In IceCube, specific algorithms have been developed to find and reject coincident events, by separating hits from the two tracks based on their separation in space and/or time.

\section{The pioneers: DUMAND and Lake Baikal experiments}\label{early}

\subsection{The prototype: DUMAND}
The project to realize the Markov idea started in 1973 during the Cosmic Ray Conference in Denver. The Deep Underwater Muon And Neutrino Detection (DUMAND) project \cite{dum1,dum2} born in 1976 and existed through 1995. The goal was the construction of the first deep ocean neutrino detector, to be placed at a 4800 m depth in the Pacific Ocean off Keahole Point on the Big Island of Hawaii. Many preliminary studies were carried out, from technology to ocean water properties. A prototype vertical string of instruments suspended from a special ship was employed to demonstrate the technology, and to measure the cosmic ray muon flux at  various depths (2000-4000 m, in steps of 500 m) in deep ocean \cite{dumand}. 

A major operation took place in December 1993, when one string of photo-detectors, a string of environmental instruments and a junction box were placed on the ocean bottom and cabled to shore. While the cable laying was successful, short circuits soon developed in the instruments and it was no longer possible to communicate with the installed apparatus. In 1995 the US DOE cancelled further efforts on DUMAND.

All subsequent designs for underwater experiments have taken  advantage  of this experience. Some reasons for the long DUMAND development time were: $i)$ huge depth of the chosen site; $ii)$ lack of advanced fibre-optics technology for data transmission; $iii)$ lack of reliable pressure-resistant underwater connectors; $iv)$ lack of Remotely Operated Vehicle (ROV) for underwater connections; $v)$ limited funding.

The Baikal group has been working in Lake Baikal in Siberia for about the same time as the DUMAND group in Hawaii. Initially, in the mid-1970's, the groups worked together. However, political problems developed after the Soviet invasion of Afghanistan, and the US DUMAND group was told by its government (Reagan administration) that no funding would be available to collaborate with the Soviet groups. Hence the teams reluctantly took separate paths.

\subsection{The experiment in Lake Baikal}\label{baikal}
The possibility to build a neutrino telescope in the Russian Lake Baikal was born with the basic idea of using the winter ice cover as a platform for assembly and deployment of instruments, instead of using a ship \cite{baik1}. After initial small size tests, in 1984-90 single-string arrays equipped with 12 - 36 PMTs were deployed and operated via a shore cable. During this period, underwater and ice technologies were developed, optical properties of the Baikal water as well as the long-term variations of the water luminescence were investigated in great detail. Deep Baikal water is characterized by an absorption length of $L_{a}$(480nm) = $20\div  24$ m, an effective scattering length of $L_b^{eff}= 30\div 70$ m and a strongly anisotropic scattering function with a mean cosine of scattering angle $\langle \cos\theta \rangle =0.85 \div 0.9$. 

\begin{figure}[htb]
\begin{center}
\includegraphics[width = 9. cm]{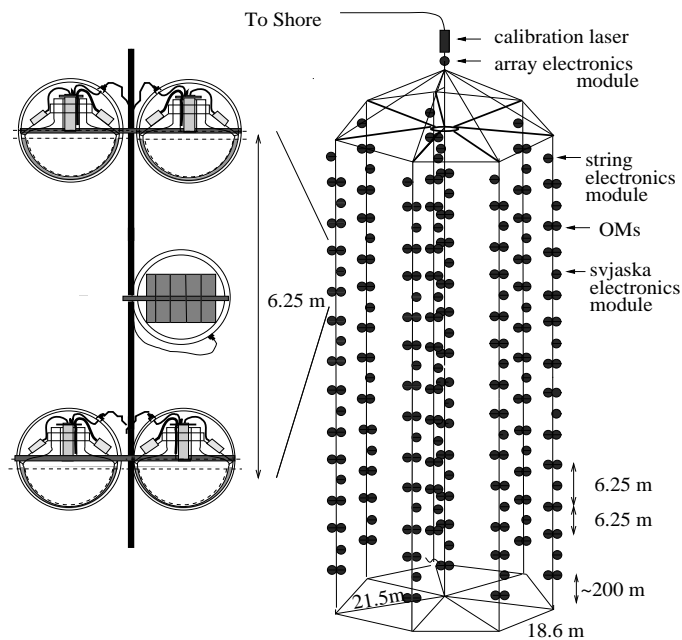}
\caption{\it Sketch of the NT-200 Baikal experiment. The expansion left-hand shows 2 pairs of optical modules (\textit{svjaska}) with the svjaska electronics module, which houses part of the readout and control electronics.}
\label{fig:baikal}
    \end{center}
\end{figure}

The Baikal Neutrino Telescope NT-200  was a second generation detector, deployed in Lake Baikal 3.6 km from shore at a depth of 1.1 km. It consists of 192 optical modules (OMs).  In April 1993, the first part of NT-200, the detector NT-36 with 36 OMs at 3 short strings, was set into operation. A 72-OMs array (NT-72) ran in 1995-96. In 1996 it was replaced by the four-string array (NT-96). Since April 1997  a six-string array with 144 OMs, take data in Lake Baikal (NT-144). NT-200 array was completed in April, 1998 (Fig. \ref{fig:baikal}). NT200 plus the new external strings form NT200+. 
An umbrella like frame carries 8 strings 72 m long, each with 24 pair of optical modules (OMs). The OMs contain 37 cm diameter PMT QUASAR­370 and are grouped in pairs along the strings. The PMs of a pair are switched in  coincidence  in order to suppress background from bioluminescence and PMT noise \cite{baik2}. 

The search strategy for high energy neutrinos relies on the detection of the Cherenkov light emitted by cascades, produced by neutrino interactions in a large volume below NT-200. The results of a search for high energy $\nu_e+\nu_\mu + \nu_\tau$  \cite{baik3} gives an upper limit of $E^{-2}d\Phi/dE <8.1 \times 10^{-7}$ GeV cm$^{-2}$  s$^{-1}$ sr$^{-1}$
in the energy range of 2$\times 10^4$ to $5\times 10^7$ GeV, see Fig. \ref{diffuse}.

\section{Detectors in the South Pole ice}\label{nutel}


An experiment at the South Pole, at the Amundsen-Scott station where the ice is about 2800 m deep, was pioneered by the AMANDA collaboration. They drilled holes in the ice using a hot water drill, and lowered strings of optical sensors before the water in the hole refreeze. 
The first AMANDA string was deployed in 1993, at a depth of 800-1000 m. It was quickly found that at that depth the ice had a very short scattering length, less than 50 cm (Fig. \ref{iceprop}). In 1995-96 AMANDA deployed 4 strings between 1500 and 2000 m deep. These detectors worked as expected, and AMANDA detected its first neutrinos \cite{amanda1}. This success led to AMANDA-II, which consisted of 19 strings holding 677 optical sensors. 

AMANDA was limited by its small size and some technological problems  \cite{ice1}. Its optical sensors consisted of photomultipliers (PMTs) with resistive bases in a pressure vessel. Not all of the optical modules survived the high pressures present when the water in the drill holes froze and AMANDA consumed considerable electrical power and required man\-po\-wer-intensive calibrations yearly.
High voltage was generated on the surface, and analogue signals were returned to the surface. Several transmission media were tried: coaxial cables, twisted pairs, and optical fibers. The 2.5 km long coaxial cables and twisted pairs dispersed the PMT pulses, and also the optical fibers (in roughly half of the OMs) had a very limited dynamic range. 
\begin{figure}[htb]
\begin{center}
\includegraphics[width = 8. cm]{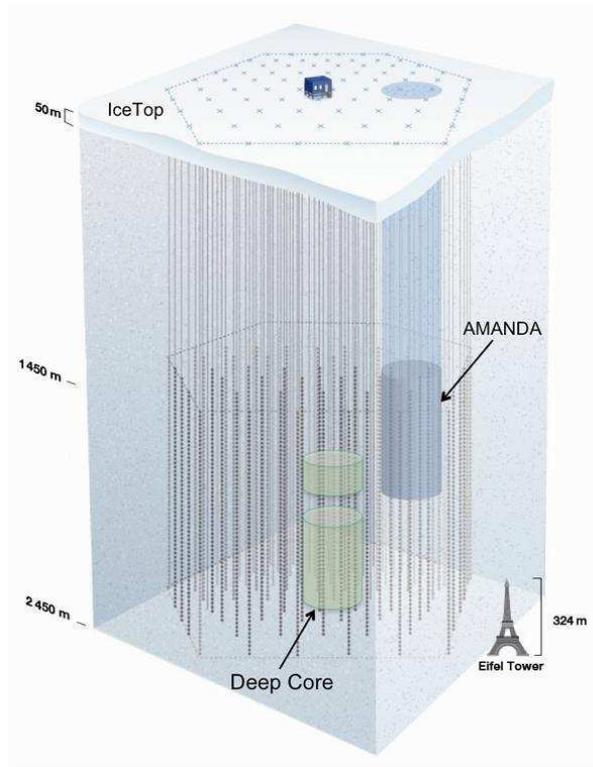}
\caption{\it The IceCube detector side view. Currently there are 59 buried InIce strings. The AMANDA detector appears in the right part of InIce. The IceTop surface array and the DeepCore (\S 9.4) are also shown.}
\label{fig:icecube}
    \end{center}
\end{figure}

IceCube was designed to avoid these problems and to be much simpler to deploy, operate and calibrate. 
When completed in 2011, it will consist on a deep detector (InIce) and a surface detector (IceTop), see Fig. \ref{fig:icecube}. The design of the main InIce part of the detector  \cite{ice13} consists of 80 strings, buried 1450 to 2450 meters under the surface of the ice, each bearing 60 Digital Optical Modules (DOMs), with 17 m spacing. The strings are placed on a 125 m hexagonal grid, providing a 1 km$^3$ instrumented volume. 
The surface electronics are in a counting house located in the center of the array. 

The IceTop surface air-shower  is an array of 80 stations \cite{ice14} for the study of extensive air showers. Each IceTop station, located above an IceCube string, consists of two tanks filled with ice. Each of those tanks contains two DOMs of same  design as the one used for the InIce part of the detector. The surface array can be operated looking for anti-coincidence with the InIce events to reject downgoing muons. It can also be used in coincidence, to provide a useful tool for cosmic ray composition studies. The array covers an area of about 1 km$^2$ as shown in Fig. \ref{fig:icecube}. The completed detector will be operated for 20 years. 

\begin{figure}[htb]
\begin{center}
\includegraphics[width = 8. cm]{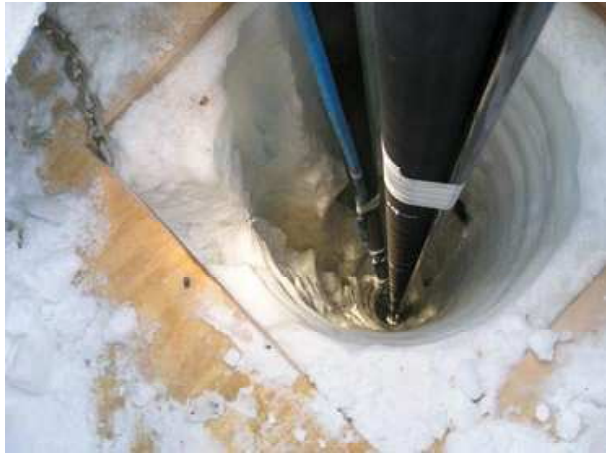}
\caption{\it The hot water hose and support cables disappear down one of one of boreholes drilled into the Antarctic ice to construct the IceCube Neutrino Observatory. Photo: Jim Haugen \cite{ice0}}
\label{fig:icehole}
    \end{center}
\end{figure}

Because of the Antarctic weather, high altitude and remote location of the South Pole, logistics is a key issue. The construction season lasts from November through mid-February (during the Austral summer). Everything needed must be flown to the Pole on ski-equipped LC-130 transports planes. 
The main task in IceCube construction is drilling holes for the strings, see Fig. \ref{fig:icehole}. This is done with a 5 MW hot-water drill, which generates a stream of $\sim$ 800 liters/minute of 88$^o$C water. This water is propelled through a 1.8 cm diameter nozzle, melting a hole through the ice. Drilling a 2500 m deep, 60 cm diameter hole takes about 40 hours. Deploying a string of DOMs takes about another 12 hours. 

Data acquisition with the partially finished IceCube detector is running smoothly and the detector is operating as expected. The detector began taking data in 2006 with a nine-string configuration (IC-9) and with a 22-string configuration in 2007 (IC-22)  \cite{ice4}. 
During the Austral summer 2007/08 18 more strings were buried, and 19 during the summer 2008/09. Currently (May 2009) 59 InIce strings with 3540 DOMS are deployed.

\subsection{The IceCube Data Acquisition System}
Each DOM used by IceCube comprises a 10" PMT (Hamamatsu R7081-02) housed in a glass pressure vessel and data acquisition (DAQ) electronics which  reads out, digitizes, processes and buffers the signals from the PMT. When individual trigger conditions are met at the DOM, it transfers fully digitized waveforms to a software-based trigger and event builder on the surface. 
The electronics acquire in parallel, on Analog Transient Waveform Digitizers (ATWDs) at 300 MHz, sampling over a 425 ns window. In addition the electronics also record the signal with a coarser 40 MHz sampling over a 6.4 $\mu$s window to record the late part of the signals. Two parallel sets of ATWDs on each DOM operate to reduce the dead-time: as one is active and ready to acquire, the other is read out. 

Data is transmitted to the surface via a single twisted copper cable pair, which also provides power. Each DOM consumes about 3.5 W. The cable also includes local coincidence circuitry, whereby DOMs communicate with their nearest neighbors. A more robust connector is used than in AMANDA, and a higher fraction of IceCube OMs survive the freezing of the ice. 
The main requirement for the IceCube hardware is high reliability without maintenance. Once deployed, it is impossible to repair a DOM. About 98\% of the DOMs survive deployment and freeze-in completely; another 1\% have lost their local coincidence connections, but they are usable. 
On the surface, the cables are connected to a custom PCI card in a PC; the remainder of the system is off-the-shelf. 

Each DOM also contains a 'flasher' board, which has 12 blue (405 nm) LEDs mounted around its edges. These LEDs are used for calibrations, to measure light transmission and timing between different DOMs, to check the DOM-to-DOM relative timing and study the optical properties of the ice.
The time calibration yields a timing resolution with a RMS narrower than 2 ns for the signal sent by the DOM to the surface  \cite{ice7}. The noise rate due to random hits observed for InIce DOMs is of the order of 300 Hz, which gives the possibility to monitor the DOM hit rates. 
This very low value makes the detector able to have a sensitivity to low energy (MeV) neutrinos from supernova core collapse in the Milky Way and in the Large Magellanic Cloud \cite{ice8}.

Data from the DOMs are time-sorted, combined into a single stream, and then monitored by a software trigger. The main trigger is based on multiplicity: it requires eight DOMs (with local coincidences) fired within 5 $\mu$s. This collects most of the neutrino events. Starting from 2008, a string trigger which improved sensitivity for low energy, requiring five out of seven adjacent DOMs fired within 1.5 $\mu$s, was added. When any trigger occurs, all data within the $\pm 10 \mu s$ trigger window is saved, becoming an event. If multiple trigger windows overlap, then all of the data from the ORed time intervals are saved as a single event.

The total trigger rate for 40 strings was about 1.4 kHz. The majority of the triggers (about 1 kHz) are due to atmospheric muons. A fast on-line filter system reduces the triggered events (6\% survives, for a data rate of $\sim$30 Gbytes/day), and selected events are transmitted via satellite to the Northern hemisphere. The rest of the data is stored on tapes at the South Pole station, and tapes are carried North during the Austral summer.

\subsection{Summary of the AMANDA-II results}
\begin{figure*}[htb]
\begin{center}
\includegraphics[width = 14.5 cm]{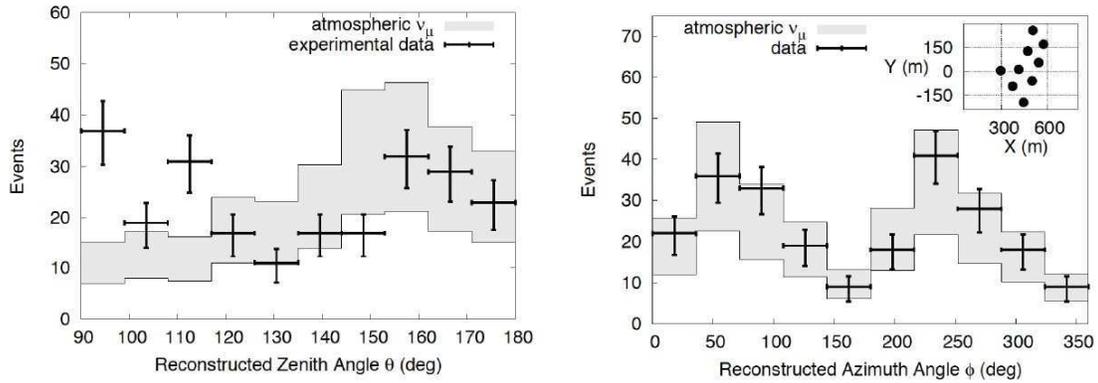}
\caption{\it Reconstructed zenith (on the left) and azimuth (on the right) angle distributions for the final sample of IC-9 events. A zenith angle of 90$^o$ (180$^o$) corresponds to horizontal (straight up-going) event. The shadowed area indicates expectations with systematic errors. The error bars are statistical only. The configuration of the IceCube strings seen from the top is also plotted (box on top of the right). The preferred axis of this configuration explains the features observed in the azimuth angle distribution.}
\label{fig:icecube2}
    \end{center}
\end{figure*}

IceCube has also integrated its predecessor, the AMANDA-II detector (the final configuration of the AMANDA detector as an independent entity). AMANDA, with 677 OMs distributed on 19 strings,  is now surrounded by IceCube (see Fig. \ref{fig:icecube}). For relatively low energy events, the dense configuration of AMANDA gives it a considerable advantage. Moreover, IceCube strings  surrounding AMANDA can be used as an active veto against cosmic ray muons, making the 
combined IceCube + AMANDA detector more effective for low energy studies than AMANDA alone. 

AMANDA-II has been taking data between 2000 and 2004 and has collected 4282 up-going neutrino candidates with an estimated background fraction of $\sim$ 5\% from wrongly reconstructed downward going muons. The analysis for point sources in the Northern hemisphere sky \cite{ice18}  yielded no statistically significant point source of neutrinos.  Assuming as usual a source of muon neutrinos with energy spectrum of $E^{-2}$, an upper limit was placed averaged over declination in the Northern hemisphere sky at 90\% confidence level: 
$E^{-2}d\Phi/dE <5.5\times 10^{-8}$ GeV cm$^{-2}$  s$^{-1}$  in the energy range from 1.6 TeV up to 2.5 PeV. 

Over the same period of time, a search for neutrino emission from 32 pre-selected specific candidate sources has been performed \cite{ice18}. No statistically significant evidence for neutrino emission was found, see Fig. \ref{astrolimits}. The highest observed significance, with 8 observed events compared to 4.7 expected background events, is at the location of the GeV blazar 3C273. 

In addition to searches for individual sources of neutrinos, AMANDA-II data taken between 2000 and 2003 have been used to set a limit on possible diffuse flux of neutrinos. As described in \S \ref{wb}, this diffuse flux can be distinguished from the background of atmospheric neutrinos due to its harder spectrum. This study relies on the number of triggered OMs which serve as an energy estimator. A limit of 
$E^{-2}d\Phi/dE <7.4\times 10^{-8}$ GeV cm$^{-2}$  s$^{-1}$ sr$^{-1}$ was placed (see Fig. \ref{diffuse}) on the diffuse muon neutrino in the energy range from 16 TeV to 2.5 PeV at 90\% confidence level \cite{icecubediffuse,ice26}. Additionally, AMANDA-II has searched for an all-flavor diffuse flux from the Southern sky, setting a limit of 
$E^{-2}d\Phi/dE <2.7 \times 10^{-7}$ GeV cm$^{-2}$  s$^{-1}$ sr$^{-1}$
in the energy range of $2 \times 10^5$ to $10^9$ GeV \cite{ice23}.

\subsection{First results from the IceCube 9 and 22 strings configuration} \label{icecube}

The IC-9 dataset has a total livetime of 137.4 days taken between June and November 2006. 234 neutrino candidates were identified on this data sample with 211$\pm 76_{syst}\pm 14_{stat}$ events expected from atmospheric neutrinos and less than 10\% pollution by the background of downward going muons \cite{ice24}. 
Zenith and azimuth angle distributions of these neutrino candidates are shown on Fig. \ref{fig:icecube2}. 
Agreement with simulation is good except for a discrepancy near the horizontal direction due to a residual contamination of down going muons. 
\begin{figure}[htb]
\begin{center}
\includegraphics[width = 12. cm]{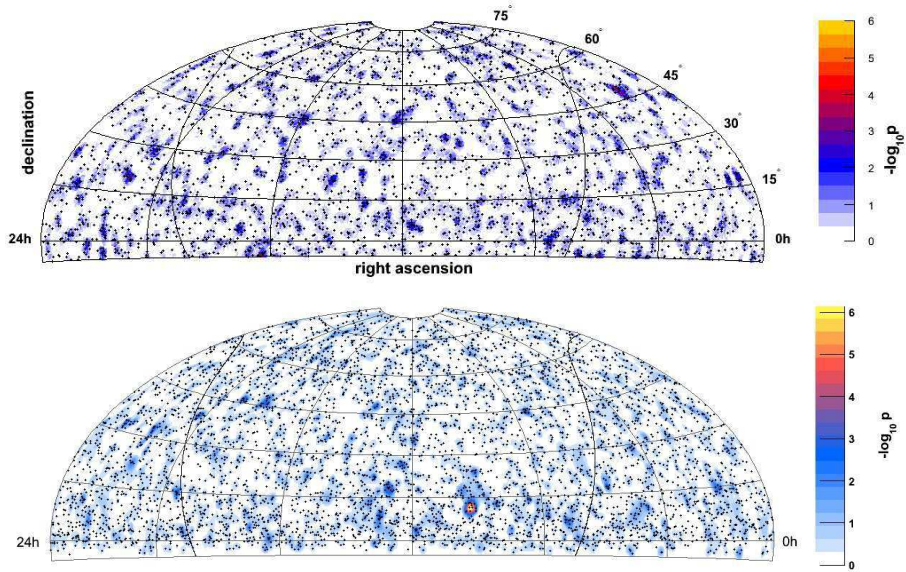}
\caption{\it IC-22 skymap with pre-trial p-values (in colors) and event locations (dots). Top: binned method. Bottom: unbinned method. Each method uses a different direction reconstruction technique \cite{ice2}.}
\label{fig:icesky}
    \end{center}
\end{figure}

The IC-22 detector took data in 2007-2008, with a lifetime of 276 days  \cite{ice2}. This data has been analyzed to search for extraterrestrial point sources of neutrinos using two methods: the binned and the unbinned maximum likelihood method. 
The binned method distinguishes a localized excess of signal from a uniform background using a circular angular search bin. The search bin radius depends on declination, and the mean value is $2.1^o$. 
The unbinned maximum likelihood method  \cite{ice3} constructs a likelihood function which depends on signal Probability Density Function (PDF) and background PDF, for a given source location and total number of data events.
Each analysis has followed its own event selection criteria, arriving at a final sample of 5114 (2956) events for the unbinned (binned) method. From simulation, a sky-averaged median angular resolution of $1.4^o$ is estimated for signal neutrinos with $E^{-2}$ spectrum.

The results from the all-sky search for both analyses are shown in Fig. \ref{fig:icesky}. The best sky-averaged sensitivity (90\% C.L.) is 
$E^{-2}d\Phi/dE <1.3 \times 10^{-8}$ GeV cm$^{-2}$  s$^{-1}$ to a generic E$^{-2}$ spectrum of $\nu_\mu$ over the energy range from 3 TeV
to 3 PeV. No neutrino point sources are found from the directions of a pre-selected catalogue nor in a search extended to the northern sky. Limits are improved by a factor of two compared to the total statistics collected by the AMANDA-II detector and by IC-9  \cite{ice25}.
A search for neutrinos coming from 26 galactic and extragalactic pre-selected objects has also been performed on this dataset, with null results.

\subsection{The future} 
During the coming years, IceCube will continue to grow starting from present configuration of 59 InIce strings. The capabilities of IceCube will be extended at both lower and higher energies in the near future. A compact core of 6 strings using IceCube DOM technology, called the DeepCore detector, will be deployed near the center of the main InIce detector, Fig. \ref{fig:icecube}. The inter-string spacing will be of the order of 72 m, allowing for the exploration of energies as low as 10-20 GeV. Surrounding IceCube strings will be used as an active veto to reduce the atmospheric muon background. The energy range that will be explored is very important for dark matter searches that were initiated with AMANDA  \cite{ice20,ice21}. 

Moreover, the ability to select contained events opens the search for downgoing astrophysical neutrino signals at low energies. This will allow looking above the current horizon of IceCube, even opening the possibility of looking at the galactic center or at the RX J1713.7-3946 source \cite{ice27}.
 
At EeV energies a possible extension of IceCube is also being studied in order to increase the detection volume and  to be sensible to the GZK neutrino using radio and/or acoustic detection of the signals generated by neutrino interacting in a huge volume of ice. With attenuation lengths of the order of a kilometer for 
acoustic (kHz frequency range) and for radio signals (MHz frequency range), a sparse instrumentation will be sufficient for this extension. Two projects are currently being explored: AURA (Askarian Underice Radio Array) for the radio signal  \cite{ice28} and SPATS (South Pole Acoustic Test Setup) for the acoustic signal. They are currently studying the polar ice and developing the hardware necessary to build a hybrid detector enclosing IceCube.

\section{The underwater neutrino projects in the Mediterranean Sea}\label{mediterranean}

\subsection{The ANTARES experiment}\label{antares}
The ANTARES project   \cite{coll} has been set up in 1996  \cite{anta_0}. Today it involves about 180 physicists, engineers and sea-science experts from 24 institutes of 7 European countries. ANTARES is at present the largest neutrino observatory in the Northern hemisphere, which represents a privileged sight of the most interesting areas of the sky like the galactic centre, where many neutrino source candidates are expected. 
\begin{figure*}[hbt]
\begin{center}
\includegraphics[width = 14.cm]{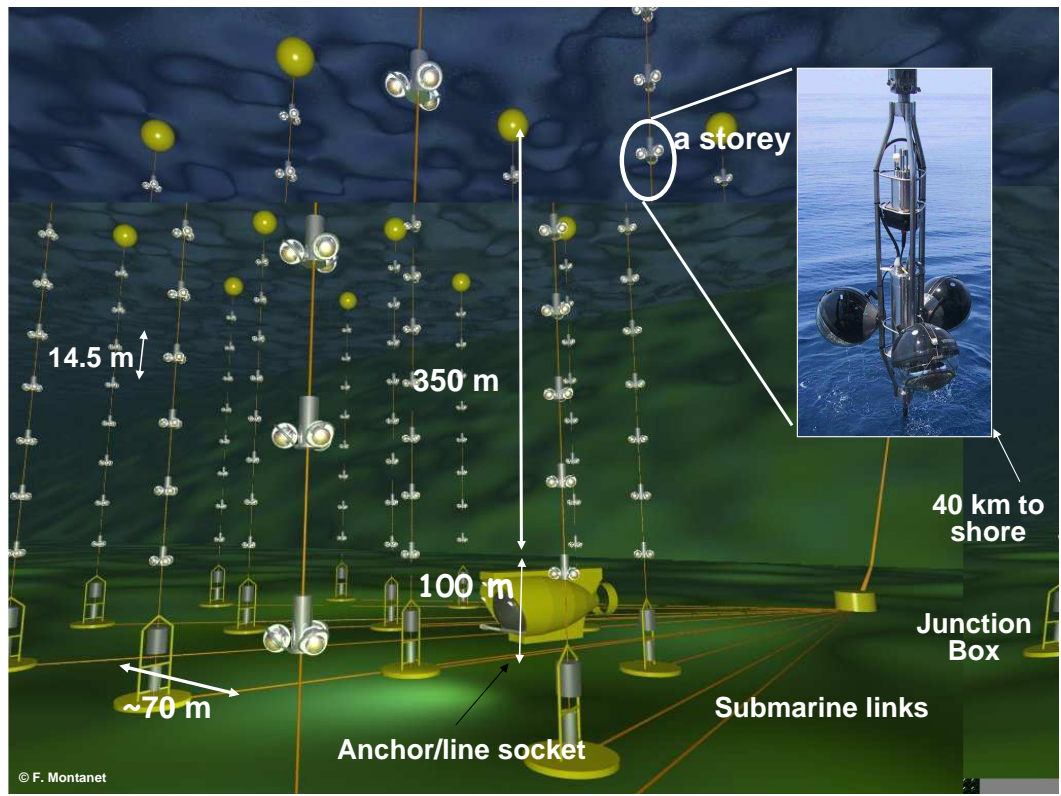}
  \caption{\it Schematic view of the ANTARES detector
    \label{detector} }
\end{center}
\end{figure*}

From 1996 to 1999 an extensive R\&D program has been successfully performed to prove the feasibility of the detector concept. Site properties have been studied such as: optical properties of the surrounding water  \cite{anta_water}; biofouling on optical surfaces  \cite{anta_sedi}; optical backgrounds due to bioluminescence and to the decay of the radioactive salts present in seawater  \cite{anta_om}; geological characteristics of its ground. These studies lead to the selection of the current site, 40 km off La Seyne-sur-Mer (France) at a 2475 m depth.  

A mini-instrumented line worked  in 2005   \cite{anta_mini}. The first detector line (also called \textit{string}) was connected in March 2006   \cite{anta_line1}, and the second line in September 2006. In July 2007, three more lines were connected, and data acquisition with five lines lasted up to December 2007. In this month, five more strings plus a dedicated instrumented line  for monitoring environmental quantities were connected. Finally, the detector was  completed with the connection of the last two strings on May, 30th  2008. The strings are made of mechanically resistant electro-optical cables anchored at the sea bed at distances of about 70 m one from each other, and tensioned by buoys at the top. Fig. \ref{detector} shows a schematic view of the detector array indicating the principal components of the detector.
Each string  has 25 storeys, each of them contains three optical modules (OM) and a local control module for the corresponding electronics. The OMs are arranged with the axis of the PMT 45$^o$ below the horizontal.  In the lower hemisphere there is an overlap in angular acceptance between modules, permitting an event trigger based on coincidences from this overlap.

On each string, and on the dedicated instrumented line, there are different sensors and instrumentation (LED beacons, hydrophones, compasses/tiltmeters) for timing and position calibration. The first storey is about 100 m above the sea floor and the distance between adjacent storeys is 14.5 m.
The instrumented volume corresponds to about 0.05 km$^3$. 

The basic unit of the detector is the optical module (OM), consisting of a photomultiplier tube, various sensors and associated electronics, housed in a pressure-resistant glass sphere   \cite{anta_om2}. Its main component is a 10" hemispherical photomultiplier model R7081-20 from Hamamatsu (PMT) glued in the glass sphere with optical gel. A $\mu$-metal cage is used to
shield the PMT against the Earth magnetic field. Electronics inside the OM are the PMT high voltage power supply and a LED system used for internal calibration. 

The total ANTARES sky coverage is 3.5$\pi$ sr, with an instantaneous overlap of 0.5$\pi$ sr with that of the IceCube experiment. The galactic centre will be observed 67\% of the day time.

\subsubsection{ The ANTARES Data Acquisition System}
The Data acquisition (DAQ) system of ANTARES is extensively described in  \cite{anta_daq}. 
The PMT signal is processed by an ASIC card (the Analogue Ring Sampler, ARS) which measures the arrival time and charge of the pulse. 
On each OM, the counting rates exhibit a baseline dominated by optical background due to sea-water $^{40}$K and bioluminescence coming from bacteria, as well as bursts of a few seconds duration, probably produced by
bioluminescent emission of macro-organisms, \S \ref{optbck}. Fig. \ref{rates_anta} shows the counting rates recorded by two OMs located on different storeys during the 2006-2008 runs. The average counting rate increases from the bottom to shallower layers. The baseline is normally between 50 to 80 kHz. 

Differently from the $^{40}$K background, bioluminescence suffers from seasonal and annual variations, see Fig. \ref{rates_anta}. There can be large variations of the rate, reaching hundreds of kHz in some periods.
Since September 2006 to December 2008 the mean counting rate is 75\% of the time below 100 kHz. A safeguard against bioluminescence burst is applied online by means of a high rate veto, most often set to 250 kHz.

The OMs deliver their data in real time and can be remotely controlled through a Gb Ethernet network. Every storey is equipped with a Local Control Module (LCM) which contains the electronic boards for the OM signal processing, the instrument readout, the acoustic positioning, the power system and the data transmission. Every five storeys the Master Local Control Module also contains an Ethernet switch board, which multiplexes the DAQ channels from the other storeys. 
At the bottom of each line, the Bottom String Socket is equipped with a String Control Module which contains local readout and DAQ electronics, as well as the power system for the whole line. Both the Master Local Control Modules and the String Control Modules include a Dense Wavelength Division Multiplexing (DWDM) system. The DWDM is used in data transmission to merge several 1Gb/s Ethernet channels on the same pair of optical fibres, using different laser wavelengths.
The lines are linked to the junction box by electro-optical cables which were connected using a unmanned submarine. A standard deep sea telecommunication cable links the junction box with a shore station where the data are filtered and recorded.

All OMs are continuously read out and digitized information ($hits$) sent to shore. In ANTARES, a $hit$ is a digitized PMT signal above the ARS threshold, set around 1/3 of the single photoelectron level (Level 0 hits, L0). On-shore, a dedicated computer farm performs a global selection of hits looking for interesting physics events (DataFilter). This on-shore handling of all raw data is the main challenge of the ANTARES DAQ system, because of the high background rates. 
The data output rate is from 0.3 GB/s to 1 GB/s, depending on background and on the number of active strings. 
Particular conditions define a subset of L0 for triggering purpose. This subset (called Level 1 hits, or simply L1) corresponds either to coincidences of L0 on the same OM triplet of a storey within 20ns, or to a single high amplitude L0 (typically $>$ 3 p.e.). 
The DataFilter processes all data online and looks for a physics event by searching a set of correlated L1 hits on the full detector on a $\sim 4 \ \mu s$ window. In case an event is found, all L0 hits of the full detector during the time window are written on disk, otherwise the hits are thrown away.

The trigger rate is between 1 to 10 Hz, depending on the number of strings in data acquisition. Most of the triggered events are due to atmospheric muons, successively reconstructed by track-finding algorithms. If ANTARES receives external GRB alerts  \cite{anta_daq}, all the activity of the detector is recorded for few minutes, \S \ref{grb}. In addition, untriggered data runs were collected on a weekly base. This untriggered data subset is used to monitor the relative PMT efficiencies, as well as to check the timing within a storey, using the $^{40}$K activity.

\begin{figure*}[tbh]
  \begin{center}
\includegraphics[width = 14. cm]{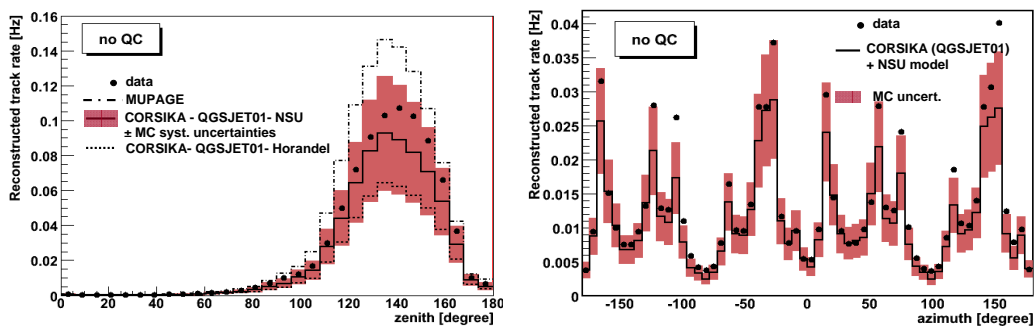}
\caption{ {\it (a) Zenith and (b) azimuth distributions of reconstructed tracks. Black points represent data. Lines refer to MC expectations, evaluated with two different simulation.  The shadowed  band (applied on one of the MC simulation) represents the systematic error ($\sim \pm 30$\%) due to environmental and geometrical parameters. No quality cuts (QC) are applied to reduce the wrongly-reconstructed upward going tracks in the region between 0$^o$ and 90$^o$ of zenith angle.}
\label{zenazi} }
\end{center}
\end{figure*}

\subsubsection{Time and positioning calibration systems}
Differently from the strings in ice, the ANTARES lines are flexible and move with the sea current, with displacements of the order of a few meters at the top for a typical sea current of 5 cm/s. 
The reconstruction of the muon trajectory is based on the differences of the arrival times of the photons between OMs. The ANTARES detector is expected to achieve an angular resolution of $< 0.3^o$ for muon events above 10 TeV, through timing measurements with precision of the order on the ns \cite{aguilarPHD}. This requires the knowledge of the OMs position with a precision of $\sim$ 10 cm (light travels 22 cm/ns in water). Pointing accuracy thus is limited by: $i)$ precision with which the spatial positioning and orientation of the OM is known; $ii)$ accuracy with which the arrival time of photons at the OM is measured; $iii)$ precision with which local timing of individual OM signals can be synchronized with respect to each other. 

The positions of the OMs are measured on a real-time, typically once every few minutes, with a system of acoustic transponders and receivers on the lines and on the sea bed, together with tiltmeters and compasses. The shape of each string is reconstructed by performing a global fit based on these information. Additional information needed for the line shape reconstruction are the water current flow and the sound velocity in seawater, which are measured using different equipments: an Acoustic Doppler Current Profiler; a Conductivity-Temperature-Depth sensors; a Sound Velocimeter. 

Relative time resolution between OMs is limited by the transit time spread of the signal in the PMTs (about 1.3 ns) and by the scattering and chromatic dispersion of light in seawater (about 1.5 ns for a light propagation of 40 m).
The electronics of the ANTARES detector is designed to contribute less than 0.5 ns to the overall time resolution. 

Complementary time calibration systems are implemented to measure and monitor the relative times between different components of the detector  within the ns level. These time calibrations are performed by: 

\noindent $i)$ the internal clock calibration system. It
consists of a 20 MHz clock generator on shore, a clock distribution system and a clock signal transceiver board placed in each LCM. The system also includes the
synchronization with respect to universal time, by assigning the GPS timestamp to the data. 

\noindent $ii)$  The internal Optical Module LEDs: inside each OM there is a blue LED attached to the back of the PMT. These LEDs are used to measure the relative variation of the PMT transit time using data from dedicated runs. 

\noindent $iii)$  The Optical Beacons   \cite{anta_beacon}, which allows the relative time calibration of different OMs by means of independent and well controlled pulsed light sources distributed throughout the detector. 

\noindent $iv)$ Several thousands of down-going muon tracks are detected per day. The hit time residuals of the reconstructed muon tracks (see Fig. \ref{dir_indir}) are used to monitor the time offsets of the OM,  enabling an overall space-time alignment and calibration cross-checks.

\subsubsection{Measurement of atmospheric muons and atmospheric neutrinos}

Atmospheric muons were an important tool to monitor the status of the detector and to check the reliability of the simulation tools and data taking. 
In ANTARES, two different Monte Carlo (MC) simulations are used to simulate atmospheric muons: one based on a full Corsika simulation  \cite{anna08}, and another based on a parameterization of the underwater muon flux  \cite{mau08}.

The  full MC simulation  \cite{brunner} is based on Corsika v.6.2, with the QGSJET  \cite{qgsjet} package for the hadronic shower development. Muons are propagated to the detector using the MUSIC  \cite{music} code, which includes all relevant muon energy loss processes.

The second MC data set is generated using parametric formulas  \cite{mupage0}, obtained with a full MC tuned in order to reproduce the underground MACRO flux  \cite{macro_flux1,macro_flux2}, energy spectrum  \cite{macro_ene1,macro_ene2} and distance between muons in bundle  \cite{macro_deco}. 
The characteristics of underwater muon events (flux, multiplicity, radial distance from the axis bundle, energy spectrum) are described with multi-parameters formulas in the range $1.5\div 5.0$ km w.e. and up to $85^\circ $ for the zenith angle. Starting from this parametrization, an event generator (called MUPAGE) was developed  \cite{mupage1} in the framework of the KM3NeT project  \cite{km3net} to generate underwater atmospheric muon bundles. 

In both simulations, muons that enter inside the surface of a virtual underwater cylinder (the {\it can}) are propagated using a GEANT-based program. 
The {\it can} defines the limit inside which charged particles in MC codes produce Cherenkov photons  \cite{brunner}).
Then, the background (extracted from real data) is added and the events are feed to a program which reproduces the DataFilter trigger logic. After this step, simulated data have the same format of the real ones.

\begin{figure}[tbh]
  \begin{center}
\includegraphics[width = 9. cm]{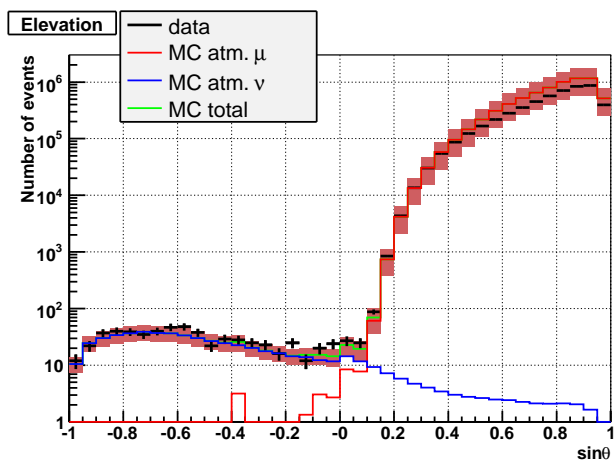}
  \caption{\it Distribution of the sinus of the zenith angle of reconstructed events (black crosses): 173 days of active time with 10 to 12 lines. Blue line: MC simulation of atmospheric neutrinos. Red line: atmospheric muons. The region between $-1<\sin\theta<0$ corresponds to upgoing particles. A contamination form atmospheric muons is present near the horizontal direction. From \cite{mau08}.}
\label{nuzenith}
\end{center}
\end{figure}

The main advantage of the full MC simulation (which is very large CPU time consuming) is the possibility of re-weight\-ing the events according to any possible
primary Cosmic Rays flux model.
The main advantage of the MUPAGE simulation is that a large sample is produced with a relatively small amount of CPU time (much less than the time needed to simulate the Cherenkov light inside the {\it can}), and it is particularly suited for the simulation of the background for neutrino events. 

Fig. \ref{zenazi} shows the zenith and azimuth distributions of reconstructed muon tracks. Black points represent experimental data. The solid   \cite{nsu} and the dotted   \cite{horandel} lines refer to Monte Carlo (MC) expectations obtained using the full MC simulation and two different CR composition models. The dashed-dotted line refers to the fast simulation   \cite{mupage1}.
The shadowed band gives an estimate of the systematic errors, due to the uncertainties on the environmental parameters, like water absorption and scattering lengths in the ANTARES site, and on the geometrical characteristics of the detector. In particular, given the fact that OMs are pointing downwards, at an angle of 135$^o$ w.r.t. the vertical, knowledge of the OMs angular acceptance at these large angles is critical for an accurate determination of the muon flux.

A different analysis is necessary when selecting neutrinos. A set of more severe quality cuts must be applied in order to remove  downward-going tracks wrongly reconstructed as upward. Data presented in Fig. \ref{nuzenith}  \cite{mau08} were collected during the 10-12 line configuration period, from December 2007 to December 2008. Atmospheric neutrino events are simulated using the Bartol flux  \cite{bartol}. Only events detected at least by two lines and with at least 6 floors are considered. Restricting to the upward-going hemisphere (neutrino candidates) the number of events are $3.4$ per day for data, and $2.9$ per day for simulations. The shadowed band represents the sum of theoretical  and systematic uncertainties.

\begin{figure}[tbh]
  \begin{center}
\vspace{8.0cm}
\includegraphics{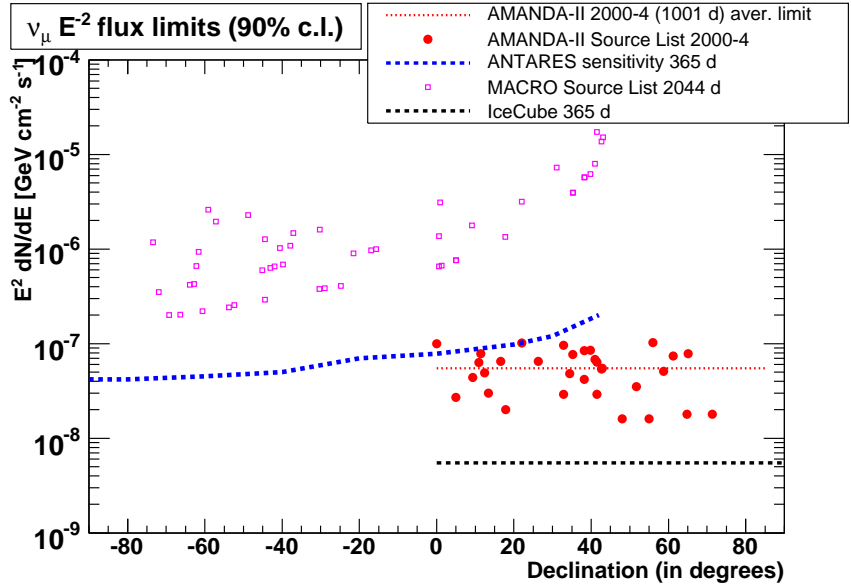}
  \caption{\it The sensitivity of the ANTARES experiment for point-like sources \cite{ciro}. The results equivalent to one year of data taking as a function of declination is presented as blue dotted line. Magenta squares are 90\% c.l. upper limits obtained by MACRO \cite{macro_astro}. Red circles are similar limits by AMANDA-II \cite{ice18}, while the red dotted line is the average limit. The black dotted line shows the one year expected sensitivity for IceCube.}
\label{astrolimits}  
\end{center}
\end{figure}

Upper flux limit from the direction of selected  candidate sources were also evaluated using still more stringent criteria for the selection of upward-going muons  \cite{ciro}. The data with 5 lines were used and 140 active days. Even with less than half a detector, these limits are the best ones for experiments looking at the Southern hemisphere; they are shown as a function of the declination of the sources, and are compared with other experiments in Fig. \ref{astrolimits}.

The energy of the crossing muon or of secondary particles generated by neutrino interactions inside the instrumented volume is estimated from the amount of light deposited in the PMTs. Several estimators based on different techniques were developed  \cite{romeyer}. MC studies show that this resolution  is between $log_{10}(\sigma_E/E)=0.2\div 0.3$ for muons with energy above 1 TeV. 
The event energy measurement is a mandatory requirement for the study of the diffuse flux of high energy neutrinos. 
MC simulations indicate that after 3 years of data taking ANTARES can set an upper limit for diffuse fluxes of $E^2\Phi<3.9\times 10^{-8}$ GeV cm$^{-2}$ s$^{-1}$ sr$^{-1}$ (see Fig. \ref{diffuse}).

\subsection{ The NEutrino Mediterranean Observatory}\label{nemo}

The NEutrino Mediterranean Observatory (NEMO) is a project   \cite{nemoc,migneco} of the Italian National Institute of Nuclear Physics (INFN).
The activity has been mainly focused on the search and characterization of an optimal site for the detector installation;  on the
development of key technologies for the km$^3$ underwater telescope; on a feasibility study of the km$^3$ detector, which included the analysis of all construction and installation issues and optimization of the detector geometry by means of numerical simulations \cite{sapienza}.

The validation of the proposed technologies via an advanced R\&D activity, the prototyping of the proposed technical solutions and their relative validation in deep sea is carried out with two pilot projects NEMO Phase-1 and Phase-2. 

Since  1998,  the  NEMO  collaboration  conducted  more  than  20  sea campaigns for the  search and the characterization of  an optimal site where to install an  underwater neutrino telescope. 
A deep site  with proper features in terms of  depth and water optical
properties  has been  identified at  a  depth of  3500 m  about 80  km
off-shore from Capo Passero (36$^\circ$ 16' N 16$^\circ$  06' E), see \S \ref{water}   \cite{bib@siti_capone}. 

\begin{figure}[ht]
\begin{center}
\includegraphics[height=12.cm]{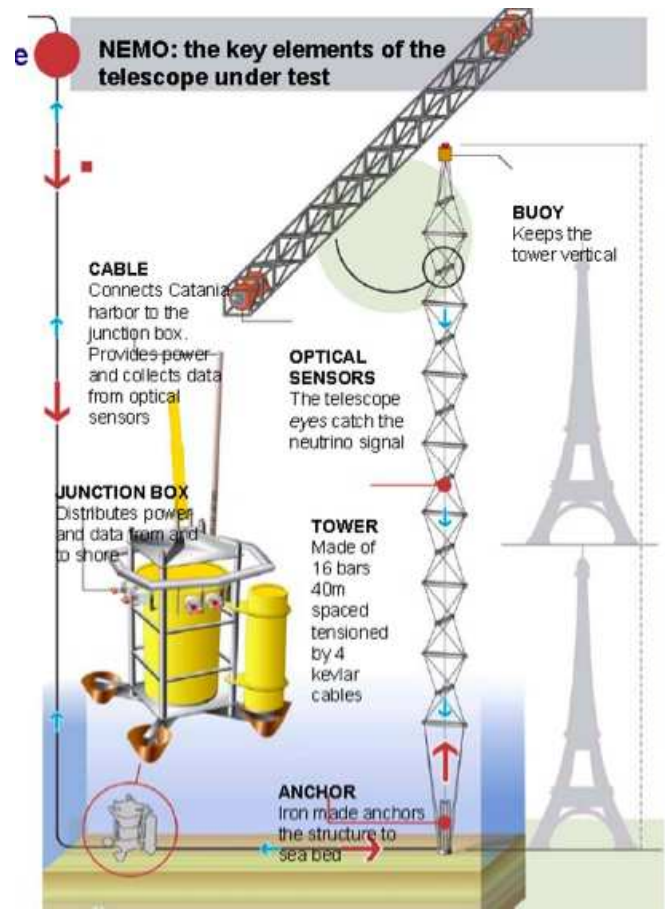}
\caption{\it The key-elements of the NEMO proposed version for the km$^3$ detector: the Junction Box, for power and data distribution, and the Tower with the bars and OM.
 \label{fig@NEMOKM3detector} }
\end{center}
\end{figure}

The main feature of a km$^3$ telescope  is its modularity.   The proposed NEMO basic element is the instrumented {\em NEMO-tower} (see Fig. \ref{fig@NEMOKM3detector}): it is
about 700 m high,  and it is composed of 16 floors, 40 m spaced;
each floor  is rotated  by 90$^\circ$, with  respect to the  upper and
lower  adjacent ones,  around  the  vertical axis  of  the tower.  Each floor is equipped  with two OMs (one down-looking and  one horizontally looking) at both extremities.  In addiction  to the OMs, the tower hosts several
environmental  instruments  plus  the  hydro\-pho\-nes  for  the  acoustic positioning system.   The tower structure  is anchored at the  sea bed and it is kept vertical by an appropriate buoyancy on the top.

The  NEMO  Phase-1 project  allowed  a  first validation  of the  technological  solutions proposed  for the  km$^3$
detector    \cite{migneco}.  The apparatus  included prototypes
of all the  critical elements: the Junction Box  and a reduced version
(one fourth) of the tower, called the {\em mini-tower}.  On December
2006, both the Junction Box and the mini-tower were deployed and successfully activated at a test  site at  2000  m depth near the Catania harbour.  The  underwater  detector  was connected  to the  shore station  via a  28 km  electro-optical cable.

The  NEMO  Phase-1  Junction Box  was  built following  the  concept  of  double
containment.   Pressure resistant  steel vessels  were hosted  inside a large fiberglass  container, which was  filled with silicon oil to compensate the external pressure.   This solution has the advantage of decoupling  the  two  problems  of pressure  and  corrosion  resistance. The  electronic components  capable of  withstanding high pressures were installed directly in the oil bath.

The mini-tower  was equipped with sixteen 10"  Hamamatsu R7081-SEL PMTs,
mounted on 15 m long floors. The  floors were spaced 40 m one from the
other,  with an  additional spacing  of  150 m  from the  base.

\begin{figure}[ht]
\begin{center}
\includegraphics[height=7.cm]{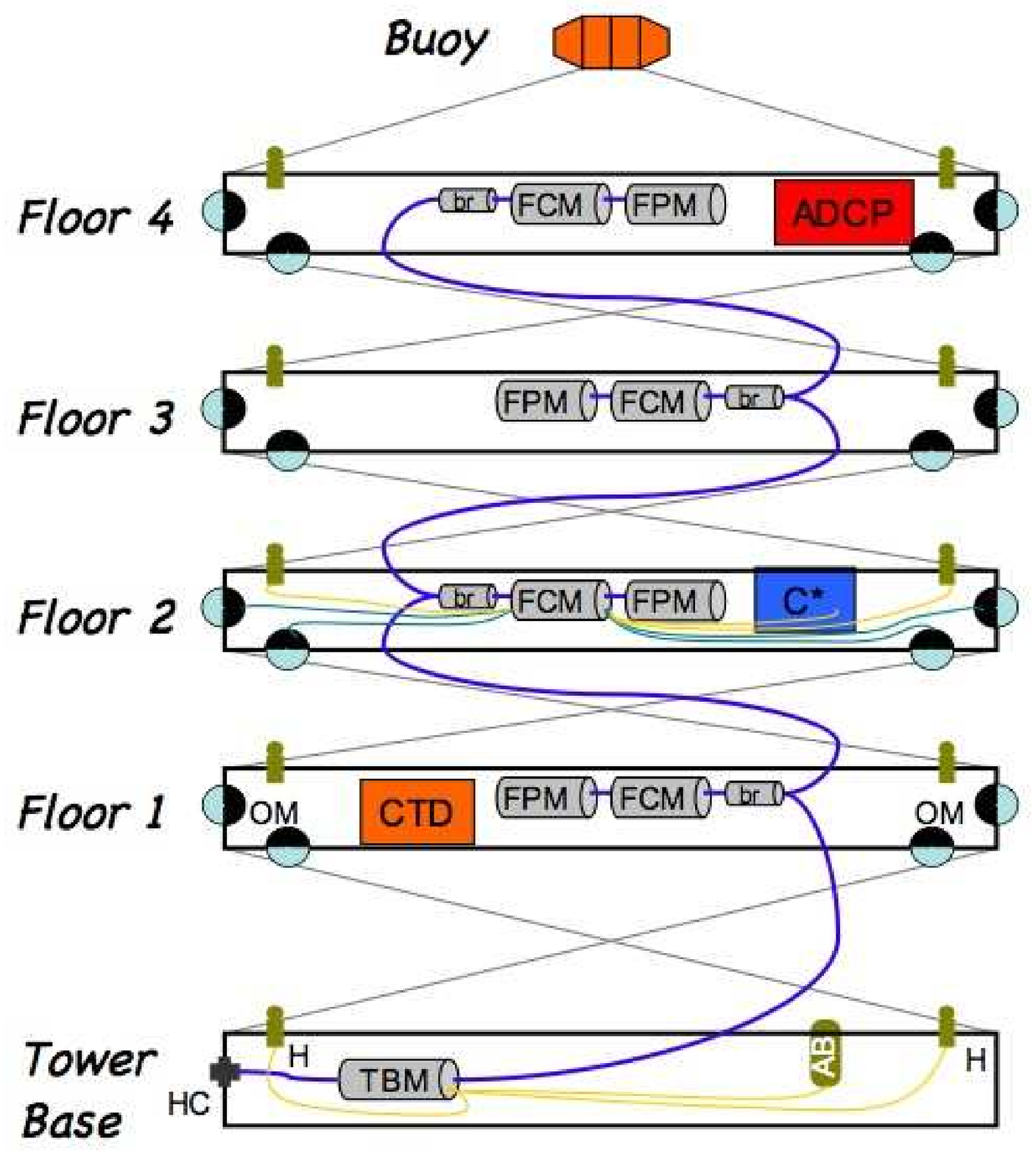}
\includegraphics[height=5.cm]{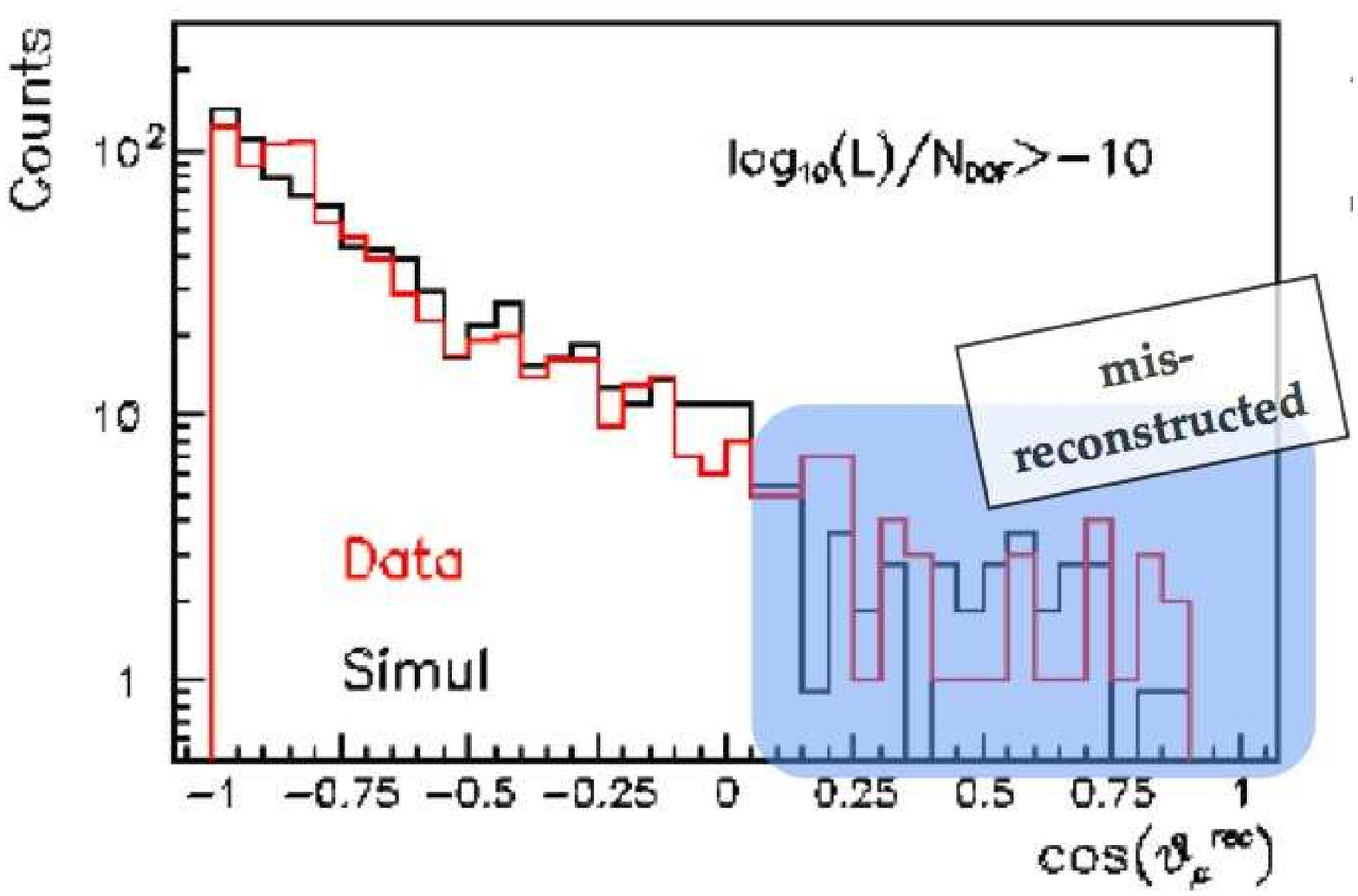}
\caption{\it Up: sketch of  the Phase-1 mini-tower: the OMs on the floors
and other  environmental instruments: the Acoustic Doppler Current Profiler (ADCP), for  profiling the sea current  velocities;   the  C*,   for  measuring  the   water  optical properties; the CTD, for salinity, temperature and density of water.  
Bottom: Distribution of  the reconstructed  events as  a  function  of   the  cosine  of  the  zenith  angle, $\cos(\vartheta_\mu^{rec})$, compared with the MUPAGE MC \cite{bib@vlvnt_amore}. 
 \label{fig@MINITORRE} }
\end{center}
\end{figure}  

In addition  to the  OMs, the  instrumentation
included several sensors for calibrations and environmental monitoring
(see Fig. \ref{fig@MINITORRE}, upper part).  In particular two hydrophones
were mounted on  the tower base and at the  extremities of each floor.
These, together with  acoustic beacons placed at  the tower base and
on   the  seabed,  were  used  for  the determination of the OM positions with a precision of about 10 cm.  Moreover,  a time calibration system was  linked to the OMs  through external  optical  fibers.  The time
offsets were determined with the precision of 1 ns.

One important technical choice of the design of  the data acquisition system for  NEMO Phase-1 is the scalability to a much bigger apparatus
  \cite{bib@daq_ameli}.  The  electric signal from the  PMTs was sampled with 2 Flash-ADC (100 MHz each) staggered by 5 ns, for a total 200 MHz sampling  and a  low power  consumption.  Each  PMT  generating an over threshold  pulse (\textit{hit})  was   characterized  by  its  time-stamp,  total integrated charge and sampled  signal waveform. The latter allowed an off-line  reconstruction  of  the  hit time  with  precision of  $\sim$1 ns   \cite{bib@vlvnt_simeone}.   The hits from the four PMTs of each floor were continuously collected by
the Floor Control Module  (FCM) boards, converted into optical signals by an electro-optical transceiver and  sent to shore through one of the optical-fiber  of  the 28  km  cable  by  using the  Dense  Wavelength Division Multiplex  protocol.  On shore, twin FCM boards de-multiplexed the incoming signal and distributed the data to the on-line trigger for a first raw selection of  data.  The trigger was based on coincidences occurred on near OMs within 20 ns and on large amplitude single hits. When a trigger seed was found, all hits occurred within a time window of $\pm 2$ $\mu$s centered on the  seed time were recorded.

A data analysis  was done on  a small sample  of selected
events,  recorded   during  23$^{rd}$  and   24$^{th}$  January  2007,
corresponding  to  a  livetime  of  11.3 hours.   From  the  
analyzed data  set, 2260  atmospheric muon  events were reconstructed  and their angular  distribution  measured   \cite{bib@vlvnt_amore} (see  Fig. \ref{fig@MINITORRE}, bottom).

The  Phase-1  project  provided  a fundamental test of the  technologies proposed for the realization and  installation of the  detector. Some problems occurred in Phase-1 after some months of functioning. Buoyancy of the tower decreased with the time (due to the construction process of the buoy), producing a lowering of the tower position.
Another problem was related to a malfunction inside the JB that required the recovery for a full diagnosis, which pointed out a malfunctioning of the  optical penetrator. 

The Phase-2 was planned to validate the new solutions at the depths  of the site of Capo Passero.  In July 2007, a 100 km  Alcatel electro-optical cable was laid on the seabed linking the  3500 m deep sea site to shore.   The cable is  a  10 kV DC,  along a  single electrical conductor, allowing  a power  transport larger than  50 kW.  The DC/DC converter, which converts  the high voltage  coming from shore into the 400 V required for the detector, is produced by Alcatel and will be deployed by the end of 2009.   The  data transmission  is provided through 20 single mode optical fibers   \cite{bib@cavo_sedita}.
The shore  station, located inside  the harbor area of  Portopalo di Capo Passero, was completed during 2009 in an ancient {\em winery}  building.

A  complete mechanical tower, a fully equipped NEMO mini-tower and a reduced version of an ANTARES string are planned to  be installed on  the Capo Passero  Site by the end of 2009.   Some features of the tower and mini-tower were modified according  to the experience obtained from  Phase-1: ({\em i}) floor length  is reduced  from  15 m  to 12  m;  ({\em  ii})  higher simplification  of the  floor  electric connections;  ({\em iii})  new time-calibration flashers  embedded directly into the  OMs; ({\em iv}) new electronic components which  allow a lower power consumption; ({\em v})  a  new  acoustic   positioning  system  with  special   road-band hydro\-pho\-nes able  to measure the  environmental acoustic background noise at 3500 m depth.   

The cable and the shore station were proposed to the KM3\-Net consortium as well suited to host the Mediterranean km$^3$ detector.

\subsection{NESTOR}\label{nestor}
 

The Neutrino Extended Submarine Telescope with Oceanographic Research (NESTOR) collaboration has developed an approach to operating a deep‐sea station, in the Southern Ionian Sea off the coast of Greece at depths exceeding 3500 m, permanently connected to shore by an in‐situ bidirectional cable,  for multi‐disciplinary scientific research.

\begin{figure*}[ht]
\begin{center}
\includegraphics[width=15cm]{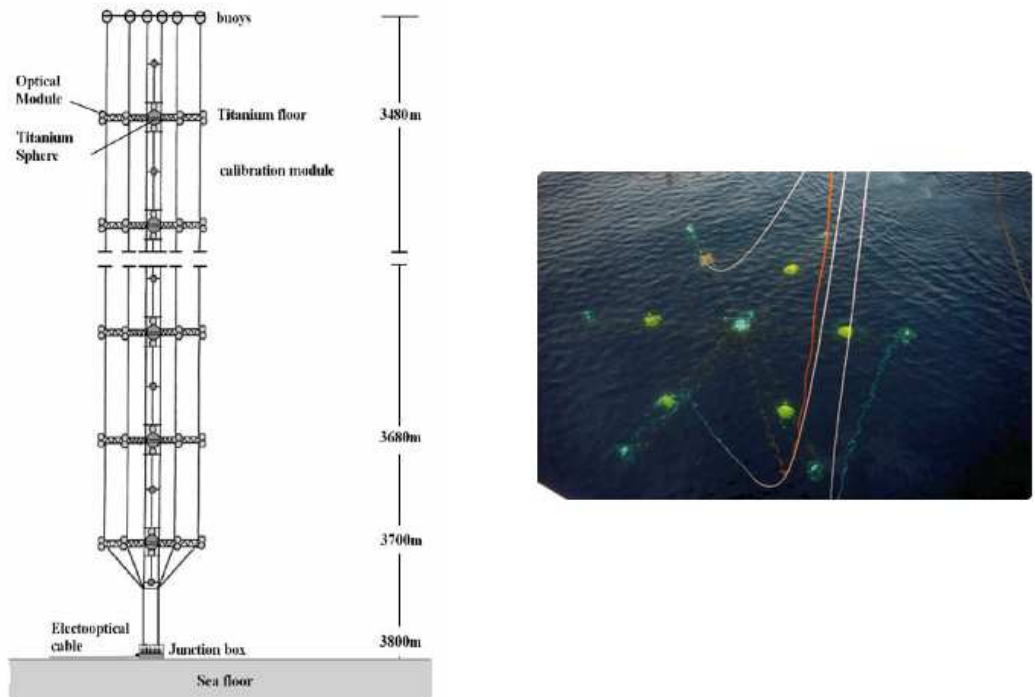}
\caption{\it Left: the proposed full NESTOR Detection Unit. Right: one  floor of the NESTOR tower during its deployment for the first test in March 2003.}
\label{fig@NestorTower}
\end{center}
\end{figure*}

The basic element of the proposed NESTOR  detector is a 32 m diameter hexagonal floor ($star$). A central casing supports  a  1 m diameter  spherical  titanium  pressure housing  which contains   the  data   acquisition   electronics,  power   converters, monitoring, control and data  transmission equipment.  Attached to the
central  casing there are  six  arms built  from  titanium tubes  to form  a lightweight but rigid  lattice girder structure. The arms  can also be
folded for transport and deployment.  Two OMs are installed at the two end of each arm, one facing upwards and the other  downwards:  OMs are  also  installed above  and below  the central casing making a total of 14 units per floor.  Using the OMs in pairs gives 4$\pi$ coverage, enhancing discrimination between upward and downward  going particles. In the NESTOR version, the tower will consist of 12 of such floors, spaced vertically 20-30m (see Fig. \ref{fig@NestorTower}, left). 

In  January 2002  a prototype  was completed and  deployed at a depth  of 4100 m  (project LAERTIS). The  station  transmitted  the
acquired data  to shore  from temperature  and pressure sensors,  compass, light
attenuation   meter,  water   current  meter   and  an   ocean  bottom
seismometer. After a period of some months, the station was recovered. 

In March  2003, the NESTOR  collaboration successfully deployed  a test
floor of the  detector tower, fully equipped with  12 OMs,
final    electronics     and    associated    environmental    sensors
 \cite{bib@Anassontzis2002}  (see Fig.  \ref{fig@NestorTower}, right).   The  detector  continuously operated for  more than a month. For  about 1.1\% of the total experimental time,  bioluminescent activity was observed around the   detector.   This   caused   about   1\%   dead   time.    
The prolonged period of running under stable operating conditions made it 
possible to measure the cosmic ray muon flux as a function of zenith angle and to derive the deep intensity relation \cite{bib@NestorFlux}.

\section{The KM3NeT Consortium}\label{km3net}
\begin{figure*}[tbh]
\begin{center}
\vspace{12.0cm}
  \includegraphics{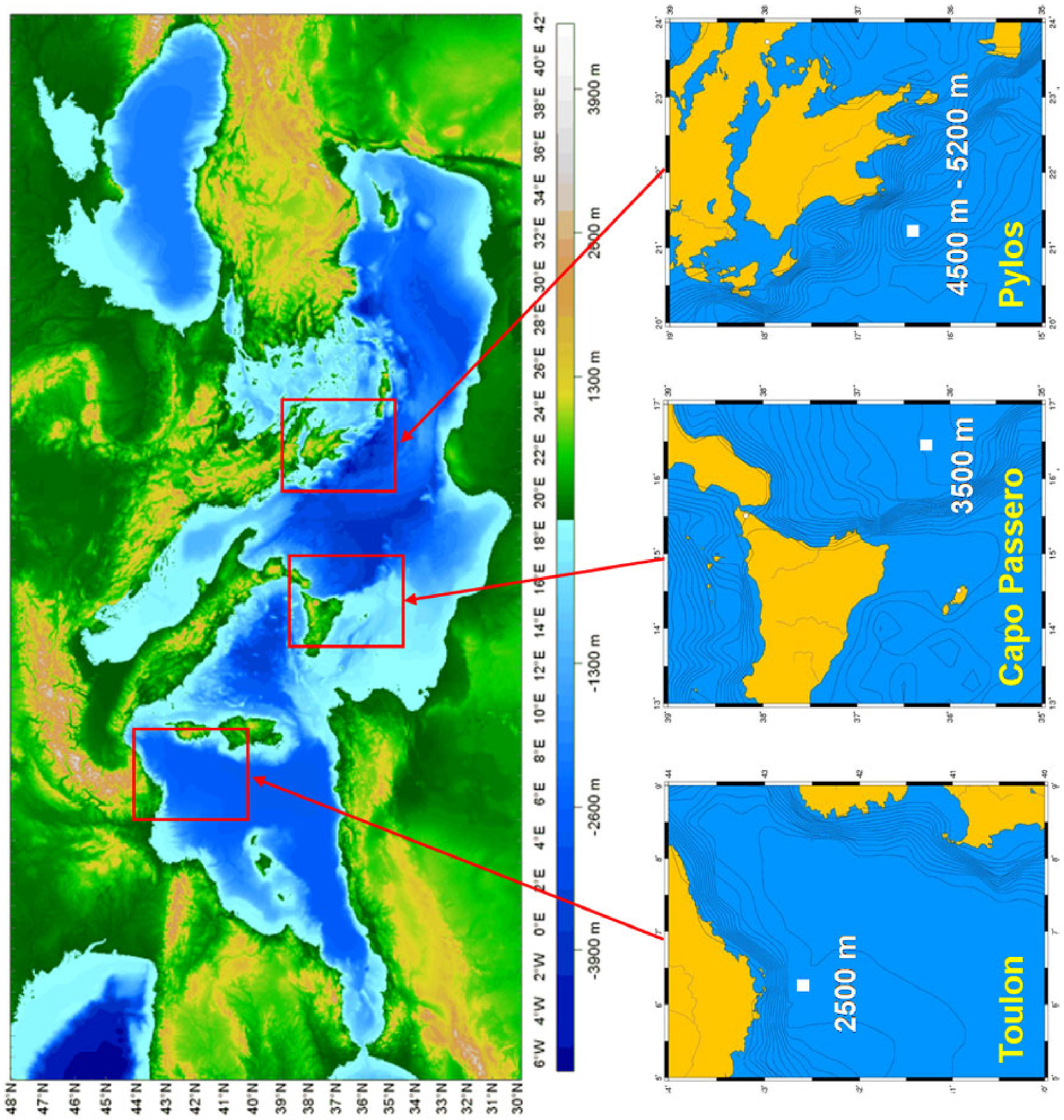}
  \caption{\it The three sites candidate to host  the KM3Net telescope.
    \label{locationNT} }
\end{center}
\end{figure*}

KM3NeT is a future deep-sea research infrastructure planned to host a neutrino telescope with a volume of at least one cubic kilometre to be constructed in the Mediterranean Sea. In February 2006, the Design Study for the infrastructure, funded by the EU FP6 framework, started. The KM3NeT research infrastructure has been singled out by ESFRI (the European Strategy Forum on Research Infrastructures) to be included in the European Roadmap for Research Infrastructures. The primary objective of the Design Study is the development of a cost-effective design for a cubic-kilometre sized deep-sea infrastructure housing a neutrino telescope with unprecedented neutrino flux sensitivity at TeV energies and providing long-term access for deep-sea research.
In April 2008 the Conceptual Design Report for the KM3NeT infrastructure was made public  \cite{km3}.

\begin{figure*}[htb]
\begin{center}
  \includegraphics[width=14cm]{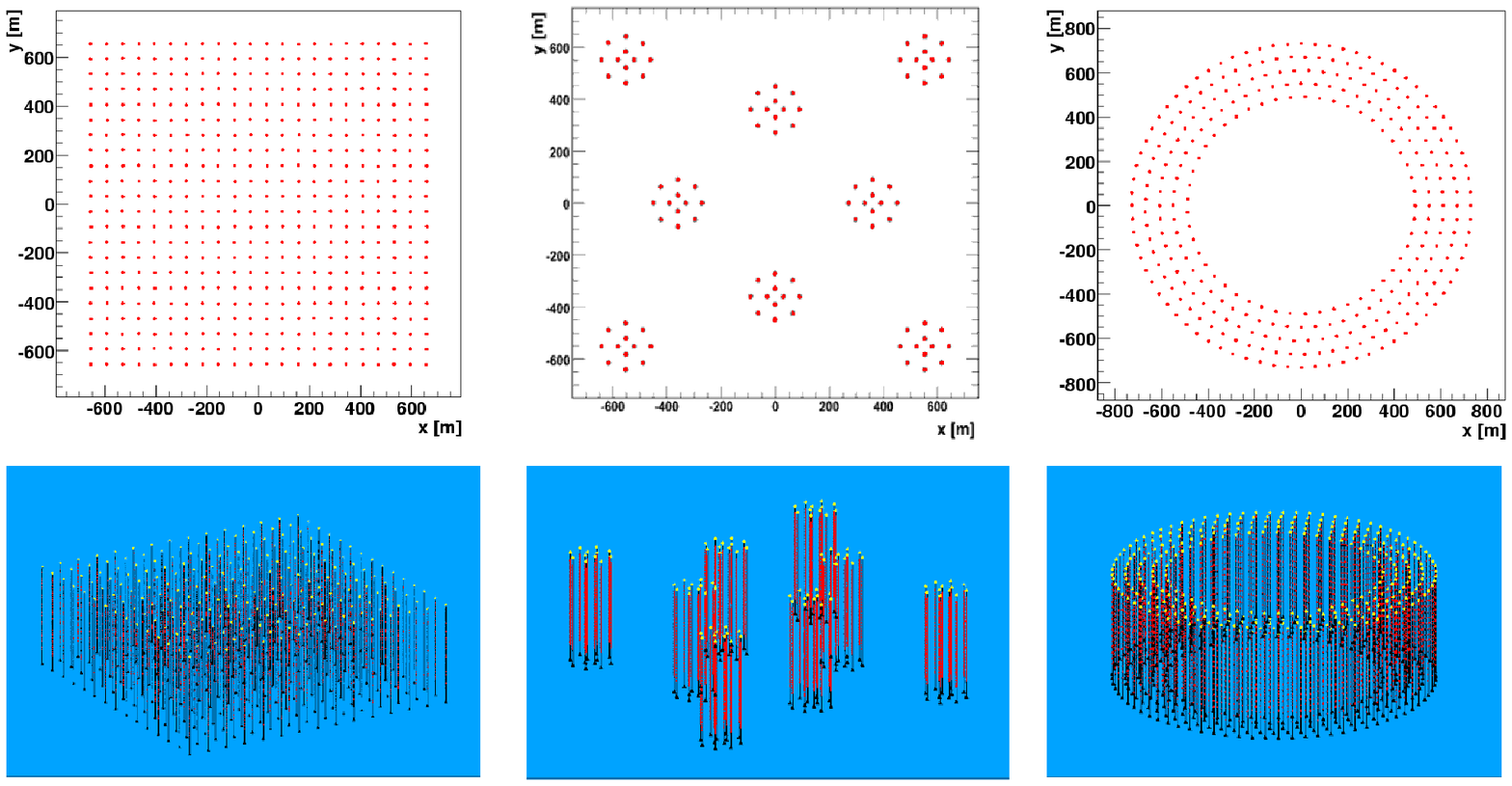}
  \caption{\it Three geometry layouts for the KM3NeT detector: left, a
    squared grid; center, clustered Detection Units; right, ring configuration. }
\label{fig@GeometryLayout}
\end{center}
\end{figure*}

The Preparatory Phase of the infrastructure, funded by the EU FP7 framework, started in March 2008. The primary objective of the KM3NeT Preparatory Phase is to pave the way to political and scientific convergence on legal, governance, financial engineering, aspects concerning the choice of the site and of the infrastructure and to prepare rapid and efficient construction once it gets approved. Reconciliation of national and regional political and financial priorities with scientific and technological considerations will be a major issue, as has become apparent in the KM3NeT Design Study. The construction of the KM3NeT infrastructure is foreseen to start after the three year Preparatory Phase, which has been organised in work packages. Each work package has its own coordinator and executive committee.

Design, construction and operation of the KM3NeT neutrino telescope will be pursued by a consortium formed around the institutes currently involved in the ANTARES, NEMO and NESTOR pilot projects (see    Fig. \ref{locationNT}). Based on the leading expertise of these research groups, the development of the KM3\-NeT telescope is envisaged to be achieved, after the Preparatory Phase, within a period of about four years for construction and deployment. 

The  KM3NeT facilities will  provide support to  scientific,  long term  and real-time  measurements, also  to a  wide range  of other {\em  associated }  earth and marine sciences, like oceanology, geophysics and marine biology.

The KM3Net detector is expected to exceed IceCube in  sensitivity  by  a  substantial factor,  exploiting  the  superior
optical properties of  seawater as compared to  the Antarctic ice and
an increased overall PMT area.  In Table 1, we summarize the required angular and energy resolution needed by the future KM3NeT detector, according to different types of astrophysical neutrino sources. 
{\small
\begin{table*}[ht]
\begin{center}
\begin{tabular}{|c|c|c|c|}
\hline 
Source & $E_\nu$ range & channel & KM3NeT requisites\tabularnewline
\hline
\hline 
Steady point Source & $10-10^{3}$ TeV & $\nu_{\mu}\;N\rightarrow \mu\;X$ & Angular res. $\sim 0.1^\circ$\tabularnewline
\hline 
Transient point Source &&  & Angular res. $\sim 0.1^\circ$ \\ 
 & $10-10^{3}$ TeV &$\nu_{\mu}\;N\rightarrow \mu\;X$&+ time coincidence with a\\
(e.g. GRBs)&&&  GRB Coordination Network\tabularnewline
\hline 
Diffuse Flux & $>10^2$ TeV & $\nu_{l}\;N\rightarrow l\;X $ &Energy res.$\sim 0.3$ in $\log\,E$ \\
&&$\nu_{l}\;N\rightarrow \nu_l\;X $&
\tabularnewline
\hline
\end{tabular}
\label{tab@km3net_goals}
\caption{\it Target sources, neutrino energy range, interaction channels
  and resolution constraints for the KM3NeT telescope.}
\end{center}
\end{table*}
}

The KM3NeT neutrino  telescope will be composed of a number of vertical structures (detection units) which are  anchored to  the sea  bed  and usually  kept vertical  by one  or several  buoys at  their top.   Since  there  is  still  a  variety  of  viable  design  options,  corresponding simulation  studies are rather  generic, concerning both assumed neutrino fluxes and detector properties. 
Fig. \ref{fig@GeometryLayout} shows three  possible layouts:  a  squared grid  of  Detection Units  (left), clustered (middle) and ring (right) configurations.  Another configuration  could consist  of arranging  the Detection Units  in a homogeneous hexagon.
\begin{figure}[htb]
\begin{center}
\includegraphics[width=9cm]{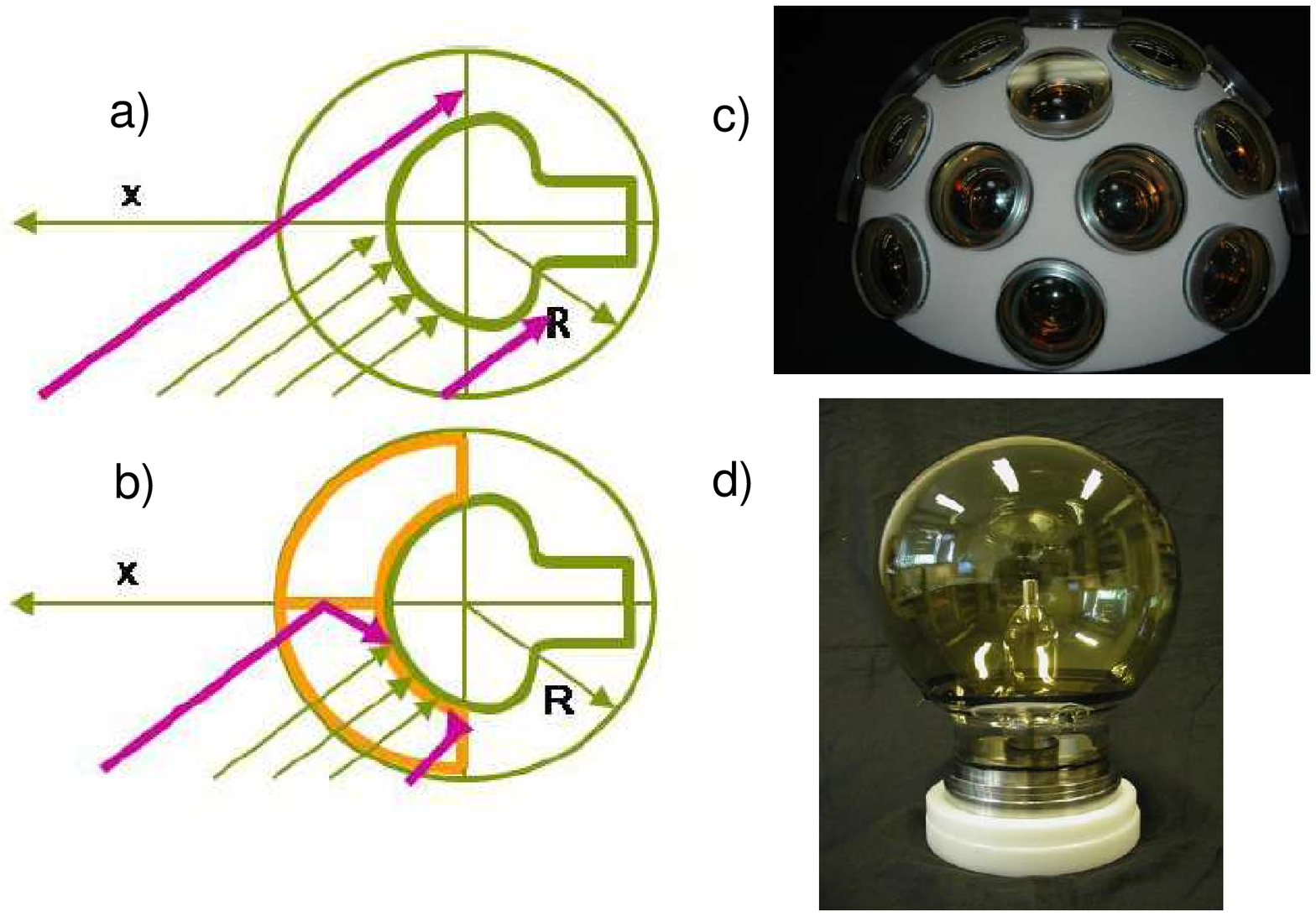}
\caption{\it The studied optical modules: (a) One single 10'' PMT in a
  ``benthos'' sphere; (b) Multi-cathode PMT, with mirror separations to
  subdivide the PMT acceptance; (c) Multi-PMT optical module made by 20
  PMTs, 3'' each; (d) Spherical geometry X-HPD (8'' prototype). }
\label{fig@OMs}
\end{center}
\end{figure}

Each  detection  unit will  carry  the  photo-sensor and  possibly
further  devices  for calibration  and  environmental measurements  on
mechanical structures  which can be  like ANTARES  {\em storeys},
NEMO's {\em floors} or NESTOR's {\em stars}.  Such structure
will support the necessary sensors, supply interfaces, data lines
and  electronic   components  where  applicable.    The  basic photo-sensor unit remains the Optical Module (OM)  \cite{bib@ANTARESsoft,bib@Carr_soft}, which can host one or several PMTs, their high-voltage bases  and their interfaces to an acquisition system of nanosecond-precision.

Whereas  all of the  current neutrino  telescope projects  use OMs composed of  a single large (typically 10'')  standard PMT per OM, alternative solutions are also under investigation for KM3NeT.  In addition to the {\em  classical} solution described above (see Fig. \ref{fig@OMs},   case  a),   various  tests   with   multi-cathode  PMTs  (see Fig. \ref{fig@OMs}, case b),
multi-PMT OMs (see Fig. \ref{fig@OMs}, case c), and large spherical hybrid PMTs (see Fig. \ref{fig@OMs} case d)  \cite{bib@HybridPMT} are performed together with computer simulation for studying the telescope response accordingly. 

\begin{figure}[ht]
\begin{center}
\includegraphics[width=9cm]{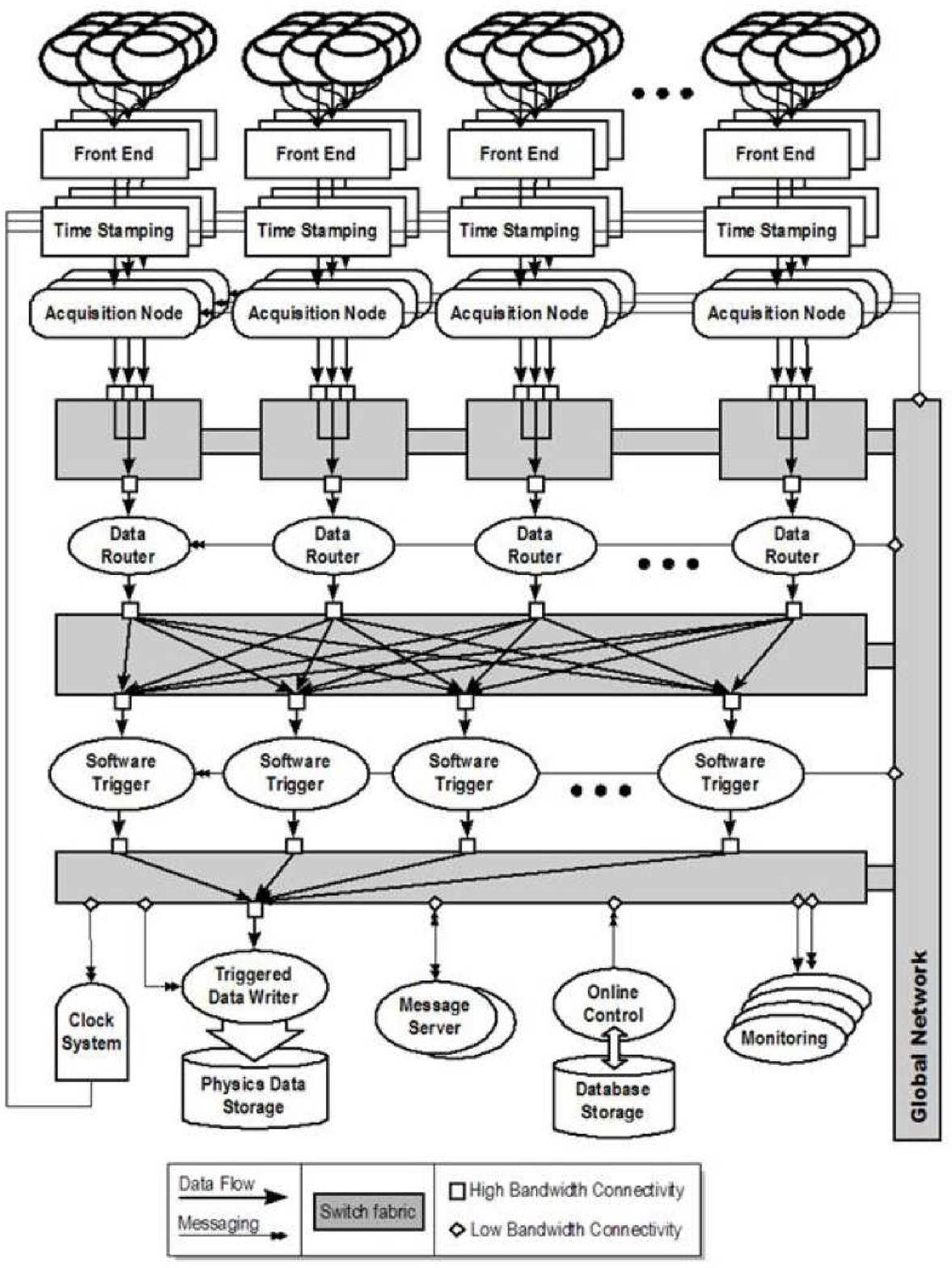}
\caption{\it The Trigger and Data Acquisition System  (TRIDAS) for the
  KM3NeT project.}
\label{fig@Tridas}
\end{center}
\end{figure}

The data transport devices and power harness of each Detection Unit is planned to be connected via the anchor to a deep-sea cable network. This
network  can contain one  or more  junction boxes  and one  or several
electro-optical  cables  to shore.  It  also  provides  power  and
slow-control  communication  to  the  detector. On  shore,  a  station
equipped with  appropriate computing  power is required  for collecting
the data, applying online  filter algorithms and transmitting the data
to mass storage devices (see Fig. \ref{fig@Tridas} where the trigger
and DAQ system is sketched). 

The  deployment  of the  Detection  Units on  the  sea  bed and  their
maintenance along  the years of  the telescope live time  require the
development of appropriate machines and infrastructures. 
For Detection Units deployment, the NESTOR Institute has 
developed a  central-well,  ballasted   platform   called  {\em   Delta
  Berenike}. For completing  the detector construction (junctions between
underwater  connectors) and maintenance down  in the deep-site,
underwater remotely operated robotic devices will be necessary.

\section{Conclusion}\label{conclusion}

In this work, we have reviewed most of the phenomenological aspects  of neutrino astrophysics, the status and future prospectives of neutrino Cherenkov detectors.

Neutrino telescopes will cover the high-energy part of the neutrino spectrum from astrophysical or cosmogenic origin. 
The history is started \cite{davis} with the detection of neutrinos ($E_\nu \sim 1-10$ MeV) from the Sun: it  corresponds to the direct observation of the nuclear reactions occurring in the core of stars \cite{bah-sun1,bah-sun2}. Different experimental technologies were used to study solar neutrinos: radiochemical \cite{home,gallex,sage}, water \cite{sk-sol} and heavy-water \cite{sno}  Cherenkov detectors and liquid scintillator \cite{borex} detectors. All experiments agree with the  hypothesis of neutrino flavor oscillations. 
In addition to dedicated solar $\nu$ experiments, other underground apparatus  in the eighties were motivated by the proton decay prediction of most grand unified theories (GUTs); they were also MeV-GeV neutrino detectors \cite{koshiba}. Those in acquisition during 1987 were lucky enough to catch neutrinos from the supernova 1987A in the Large Magellanic Cloud. The measurement of neutrinos emitted by SN1987A provided a direct observation of the core-collapse of a star \cite{raffelt}. 
The size of these neutrino experiments (also the biggest ones, as Super-Kamiokande and MACRO) is too small to have the hope to discover high energy neutrinos from cosmic sources. With the advent of neutrino telescopes, only the very low energy relic  neutrinos from cosmological origin  seems to be far to be recognized.

Some fundamental questions are still open after a century from the discovery of Cosmic Rays: which are the astrophysical engines in our Galaxy able to accelerate CRs up to $\sim 10^6$ TeV and which are those in the Universe able to accelerate CRs up to more than $\sim 10^8$ TeV?
IACT telescopes have found a variety of galactic and extragalactic TeV $\gamma$-ray sources whose spectrum extends up to some tens on TeV, where the intrinsic sensitivity of the present generation of instruments fall down.

At present, no definitive proofs exist that hadrons are also accelerated at the sites where TeV $\gamma$-ray are observed. If hadrons are present, a strong relationship between the CR energy spectrum and that of secondary $\gamma$-rays and neutrinos must exist. Compelling evidence for CR accelerators would be the detection of cosmic TeV neutrinos.
In particular, no exponential cut\-off is observed in the CR spectrum above $\sim$ 10 TeV, despite what is reported by TeV $\gamma$-ray  experiments for some galactic sources (see Fig. \ref{gammatonu}).
The exponential cut\-off would suppress the expected flux of neutrinos above 10 TeV: in this energy range, the detectors have a large neutrino effective area (Fig. \ref{km3effarea}) and the $\nu \rightarrow \mu$ kinematic production allows a measurement of the neutrino direction within $0.2^o$ in water (Fig. \ref{fig:nu_mu}). With this angular resolution, the atmospheric neutrino background can be largely reduced.
However, galactic CRs can reach $10^6$ TeV: the exponential cut\-off observed in the TeV gamma ray spectrum of some sources can be due to an instrumental/lack of statistic effect, to gamma absorption effects at sources, or because the sources of CRs above few tens of TeV are hidden to gamma rays. In any case, engines that accelerate hadrons up to 10$^6$ TeV exist in the Galaxy, and they are presumably accompanied by high energy neutrinos.  

Only a neutrino telescope in the Mediterranean Sea, with effective area $> 25$ times larger than that of the ANTARES detector, can probably solve this puzzle. 
The next years will thus be decisive for neutrino astronomy. 
While the IceCube experiment at the South Pole is in an advanced stage of construction, the technological challenges to build a huge neutrino telescope in deep sea have been surmounted by experiments like ANTARES, NE\-MO and NESTOR, which have motivated the approval of the design study of the kilometer-scale version (the KM3NeT consortium). 
These new generation of neutrino telescope experiments will achieve effective volumes which will be able to explore the Northern sky (the IcuCube experiment in the South Pole) and the Southern sky (the underwater Mediterranean experiment) in a way never seen before.

The advantage of the telescope in the Mediterranean Sea with respect to the IceCube detector is a better exposure to the galactic center region (Fig. \ref{gammasky}), and a much better angular resolution on the measurement of the neutrino direction . The latter has two consequences: the association possibility with known celestial objects; the reduction of the background from atmospheric neutrinos, which increases as the square of the search-window opening angle. 
On the other hand, the IceCube experiment has a better chance to observe the diffuse flux of neutrinos of extragalactic origin, being the theoretical upper limits accessible within few years (Fig.    \ref{diffuse}). 
Concerning the possibility that the detector in the Mediterranean Sea (if national and regional political and financial priorities cannot be easily conciliated with scientific and technological considerations) 
 will be distributed in more than one site (the \textit{multisite option}), this will not affect the sensitivity for the CC $\nu_\mu$ channel but reduces the confinement of the neutrino induced electromagnetic and hadronic showers and the possibility to detect with high efficiency the $\nu_\tau$ channel.

Identification  of neutrino from cosmic accelerators has not been claimed so far; neutrino flux can only be evaluated using models and this sets the scale of the detector instrumented volume. The value of $\sim$ 1 km$^3$ seems the minimum in order to observe  neutrinos both of galactic and extragalactic origin. The hunt for the first high energy neutrino of cosmic origin has started.

\vskip 1.0cm
\noindent \textbf{Acknowledgments}
\vskip 0.5cm

The attention of the high energy physics community is concentrated  to the beginning of the LHC operations. One of the hidden motivation for this work is to present some of the most interesting open questions in astroparticle physics, particularly to young researchers and post-graduate students. For this reason, we primarily acknowledge many young colleagues and PhD students, mainly from the ANTARES and NEMO collaboration and from the KM3NeT consortium, whose analyses and figures where used and quoted in this work. 
The authors wishes to thank many colleagues for useful discussion about the content of this paper from all the ``water and ice'' neutrino community and in particular our colleagues in Bologna (M. Bazzotti, S. Biagi, G. Carminati, S. Cecchini, G. Giacomelli and  A. Margiotta).
We thank E. Guzman De La Hoz for reading the final version. Finally, 
the authors are grateful for useful comments and suggestions from the anonymous referees.  
This work was partially supported (T.C.) by a grant of the Preparatory Phase of KM3NeT infrastructure, funded by the EU FP7 framework.

\end{document}